\documentclass[aps,prd,preprintnumbers,showpacs,twocolumn,groupedaddress,nofootinbib]{revtex4}
\usepackage{graphicx}
\usepackage{latexsym}
\def\beq{\begin{equation}}
\def\eeq{\end{equation}}
\def\bey{\begin{eqnarray}}
\def\eey{\end{eqnarray}}

\def\lsim{\mathrel{\raise.3ex\hbox{$<$\kern-.75em\lower1ex\hbox{$\sim$}}}}
\def\gsim{\mathrel{\raise.3ex\hbox{$>$\kern-.75em\lower1ex\hbox{$\sim$}}}}

\begin{document}

\title{Stringent Constraints on the Dark Matter Annihilation Cross Section From the Region of the Galactic Center}  
\author{Dan Hooper$^{1,2}$, Chris Kelso$^{1,3}$, and Farinaldo S.~Queiroz$^{1,4}$}
\affiliation{$^1$Center for Particle Astrophysics, Fermi National Accelerator Laboratory, Batavia, IL 60510, USA}
\affiliation{$^2$Department of Astronomy and Astrophysics, University of Chicago, Chicago, IL 60637, USA}
\affiliation{$^3$Department of Physics, University of Chicago, IL 60637, USA}
\affiliation{$^4$Departamento de F\'{i}sica, Universidade Federal da Para\'{i}ba, Caira Postal 5008, 58051-970, Jo\~{a}o Pessoa, PB, Brasil}

\date{\today}

\begin{abstract}

For any realistic halo profile, the Galactic Center is predicted to be the brightest source of gamma-rays from dark matter annihilations. Due in large part to uncertainties associated with the dark matter distribution and astrophysical backgrounds, however, the most commonly applied constraints on the dark matter annihilation cross section have been derived from other regions, such as dwarf spheroidal galaxies. In this article, we study Fermi Gamma-Ray Space Telescope data from the direction of the inner Galaxy and derive stringent upper limits on the dark matter's annihilation cross section. Even for the very conservative case of a dark matter distribution with a significant ($\sim$kpc) constant-density core, normalized to the minimum density needed to accommodate rotation curve and microlensing measurements, we find that the Galactic Center constraint is approximately as stringent as those derived from dwarf galaxies (which were derived under the assumption of an NFW distribution). For NFW or Einasto profiles (again, normalized to the minimum allowed density), the Galactic Center constraints are typically stronger than those from dwarfs.

\end{abstract}

\pacs{95.85.Pw,95.55.Ka,95.35.+d; FERMILAB-PUB-12-505-A}
\maketitle

\section{Introduction}

A major scientific mission of the Fermi Gamma-Ray Space Telescope (FGST) is the search for gamma-rays from dark matter annihilations. To date, Fermi data has been used to search for dark matter annihilation products from dwarf spheroidal galaxies~\cite{dwarffirst,dwarf}, galaxy clusters~\cite{clusters,clusters2,clusters3}, the Galactic Halo~\cite{Anderson:2010hh}, galactic subhalos~\cite{Ackermann:2012nb}, and from among the isotropic gamma-ray background~\cite{cosmo,cosmo2}. The most stringent of these constraints are beginning to probe annihilation cross sections at or around the value predicted for a simple thermal relic ($\sigma v \sim 3\times 10^{-26}$ cm$^3$/s)\footnote{We define a ``simple thermal relic'' as a particle which annihilates during the era of thermal freeze-out through processes which are not strongly dependent on velocity and whose abundance is not primarily depleted through coannihilations with other particles. The actual value of the annihilation cross section required in order for a simple thermal relic to be produced with the measured dark matter abundance is actually slightly smaller than the commonly used canonical value; approximately $2.2\times 10^{-26}$ cm$^3$/s for dark matter particles heavier than about 10 GeV~\cite{Steigman:2012nb}.}, at least for dark matter particles with masses below a few tens of GeV and which annihilate to final states which result in significant fluxes of gamma-rays. And while the particle (or particles) that make up the dark matter may possess an annihilation cross section that is well below this value, the range of cross sections that is presently being probed by Fermi represents a very well motivated and theoretically significant benchmark.

Observations of gamma-rays from the inner region of the Milky Way are of particular interest for indirect dark matter searches. Due to its relative proximity and the high densities of dark matter present in this region, the Galactic Center is predicted to be the single brightest source of dark matter annihilation products in the sky. For dark matter distributed according to the commonly used NFW (Navarro-Frenk-White) profile, for example, one predicts a flux of gamma-ray annihilation products from the Galactic Center that is roughly four orders of magnitude higher than from even the most promising of the known dwarf spheroidal galaxies (Segue 1, Ursa Major, etc.). Even if the dark matter distribution in the inner Galaxy has a significant constant-density core, the Galactic Center will significantly outshine all other sources of dark matter annihilation products. Tempering this advantage is the fact that astrophysical backgrounds from the Inner Galaxy are fairly bright, and are not particularly well understood. So while even a very faint gamma-ray signal from one or more dwarf galaxies could possibly provide us with evidence of annihilating dark matter, a much higher flux of annihilation products would be required from the inner Galaxy before it could be identified as such.

In previous studies, a number of independent groups have studied Fermi data from the direction of the Galactic Center~\cite{Hooper:2011ti,HG2,Abazajian:2012pn,aharonian,Boyarsky:2010dr,HG1} (preliminary results from the Fermi Collaboration have also been presented~\cite{Vitale:2011zz,Morselli:2010ty,Vitale:2008zz}). In particular, in Refs.~\cite{Hooper:2011ti,HG2,HG1}, a bright gamma-ray source was found to be present at the Galactic Center, with spectral and morphological characteristics consistent with those predicted from annihilating dark matter. This conclusion was recently confirmed by Ref.~\cite{Abazajian:2012pn}. A debate regarding the origin of this emission is ongoing and has focused on a number of possibilities, including that these gamma-rays largely result from cosmic ray collisions with gas~\cite{Hooper:2011ti,Linden:2012iv,Abazajian:2012pn,HG2,aharonian,Linden:2012bp}, from a population of $\sim 10^3$ millisecond pulsars~\cite{Hooper:2011ti,Abazajian:2012pn,HG2,pulsars,Wharton:2011dv}, or from annihilations of $\sim$$7$\,--\,$40$ GeV dark matter particles with an annihilation cross section on the order of $\sigma v \sim 10^{-26}$ cm$^3$/s~\cite{Hooper:2011ti,HG2,HG1,Abazajian:2012pn}. Throughout this paper, we will remain agnostic as to the origin of this emission, and use the observed spectral and spatial distribution of the observed gamma-rays to derive upper limits on the dark matter annihilation cross section.

We find that even for very conservative choices for the dark matter distribution (such as cored halo profiles), we derive constraints on the dark matter's annihilation cross section which are comparable to the strongest constraints found from other regions of the sky. In particular, our most conservative constraints from the Galactic Center (assuming a distribution with a kiloparsec-scale constant-density core, normalized to the minimum value compatible with the Milky Way's rotation curve) are comparable to the constraints derived by the Fermi collaboration from the combination of all dwarf spheroidal galaxies (under the assumption of NFW halo profiles). If instead we adopt an NFW, Einasto, or contracted profile to describe the dark matter distribution in the Galactic Center, the constraints we derive on the dark matter annihilation cross section are more stringent than those from any other gamma-ray observations.

\section{Analysis Procedure}
\label{analysis}

We begin our analysis by generating contour maps of the gamma-ray flux from the region surrounding the Galactic Center. These maps were generated using the latest (corresponding to the analysis software update of April 18, 2012) data release from the Fermi-LAT taken over the time period between August 4, 2008 and June 19, 2012. Due to their superior point spread function, we use only front-converting events from the Pass 7 ultraclean class and, as recommended by the FGST collaboration, we include only events with zenith angles smaller than 100 degrees and do not include events recorded while the Fermi satellite was transitioning through the South Atlantic Anomaly or while the instrument was not in survey mode.

\begin{figure*}[t]
\centering
\includegraphics[angle=0.0,width=1.86in]{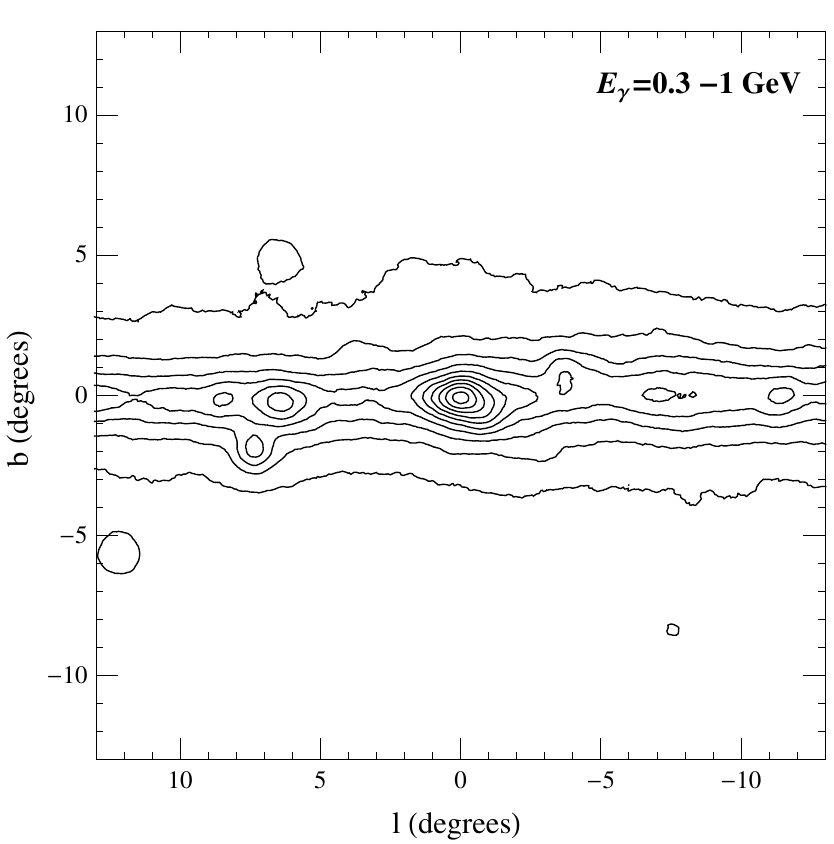}
\includegraphics[angle=0.0,width=1.86in]{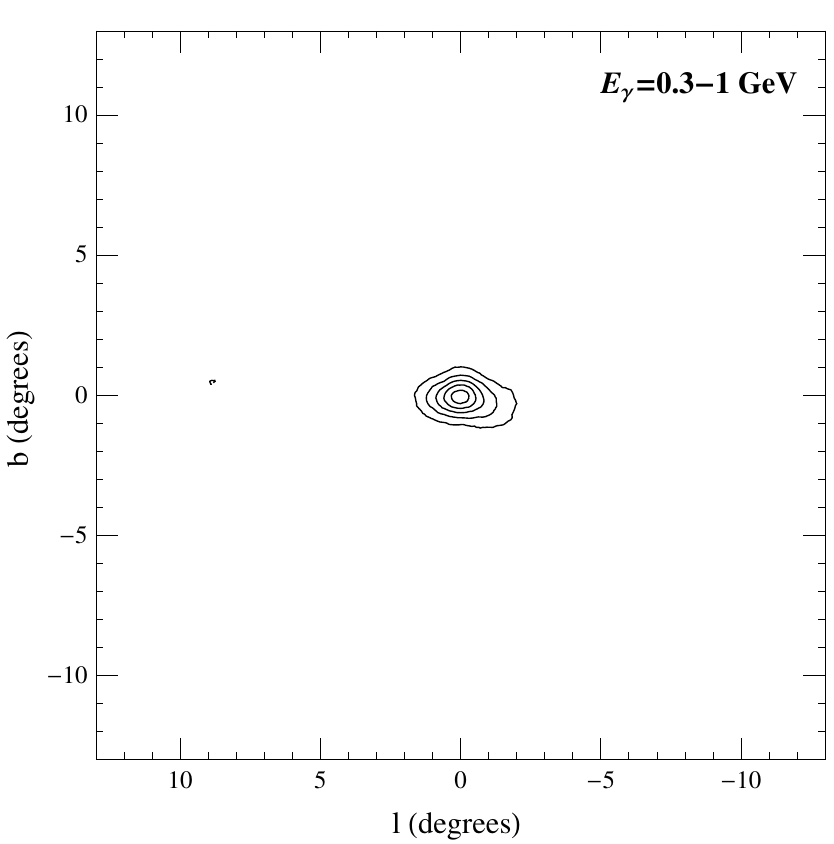}\\
\includegraphics[angle=0.0,width=1.86in]{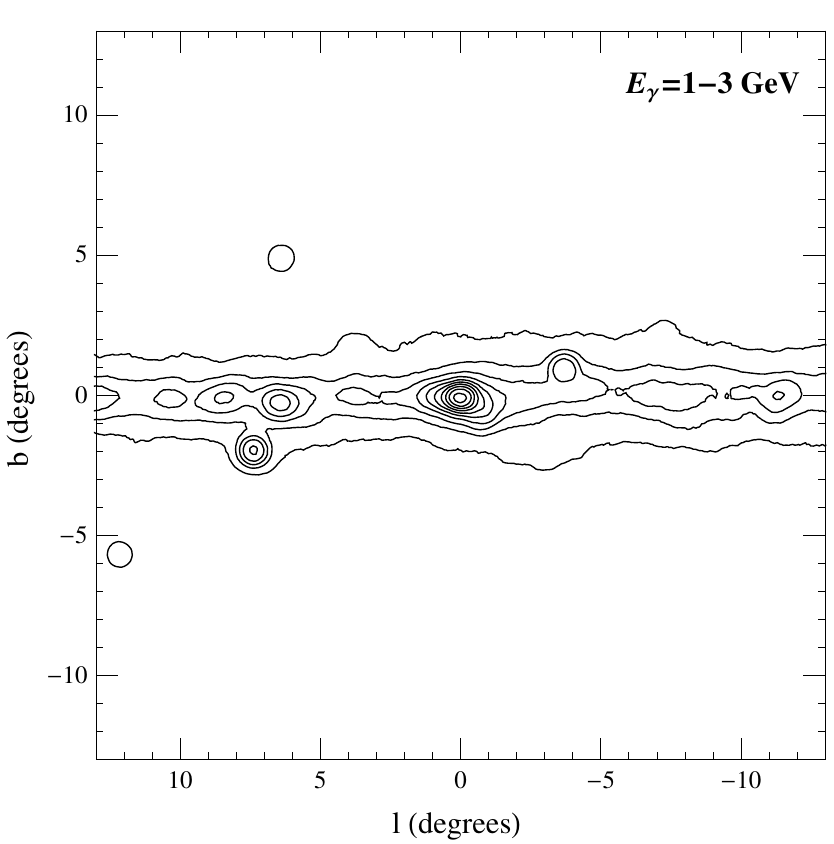}
\includegraphics[angle=0.0,width=1.86in]{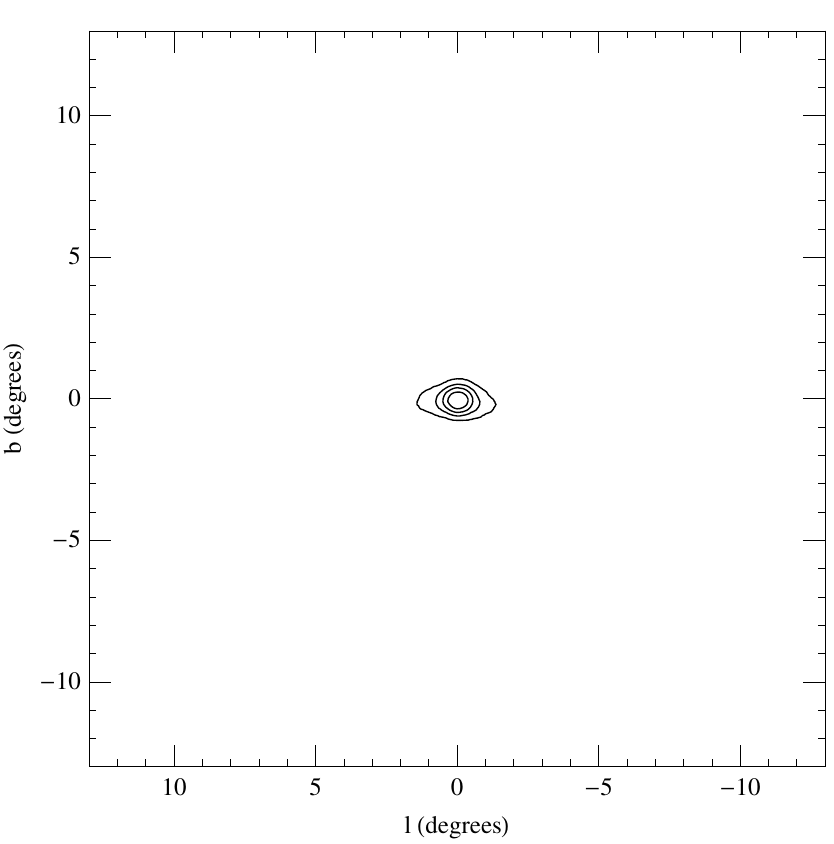}\\
\includegraphics[angle=0.0,width=1.86in]{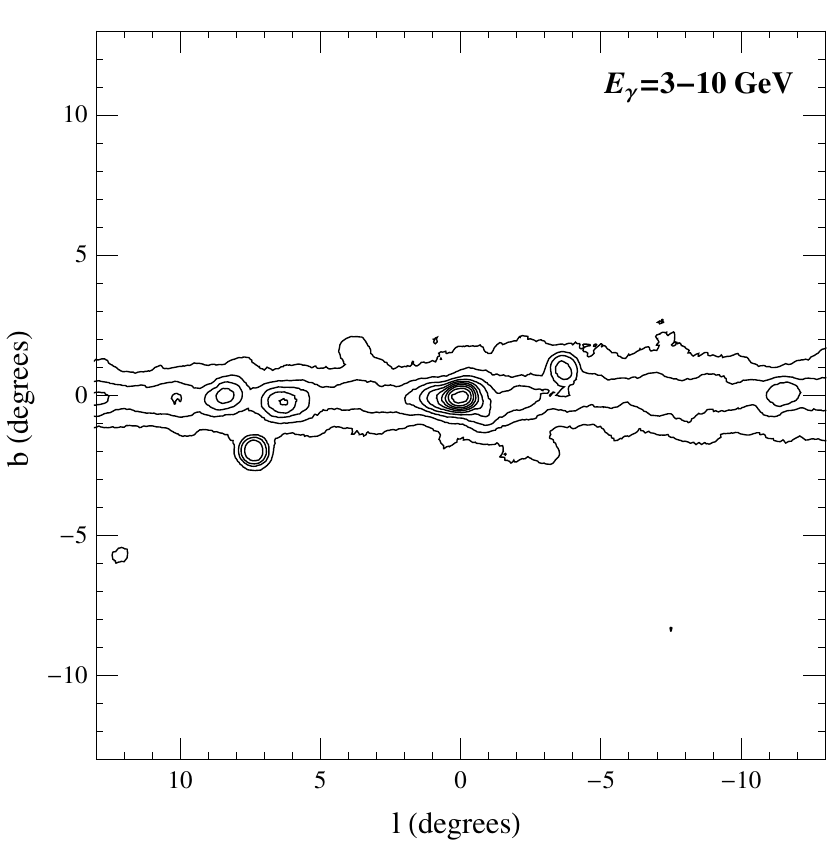}
\includegraphics[angle=0.0,width=1.86in]{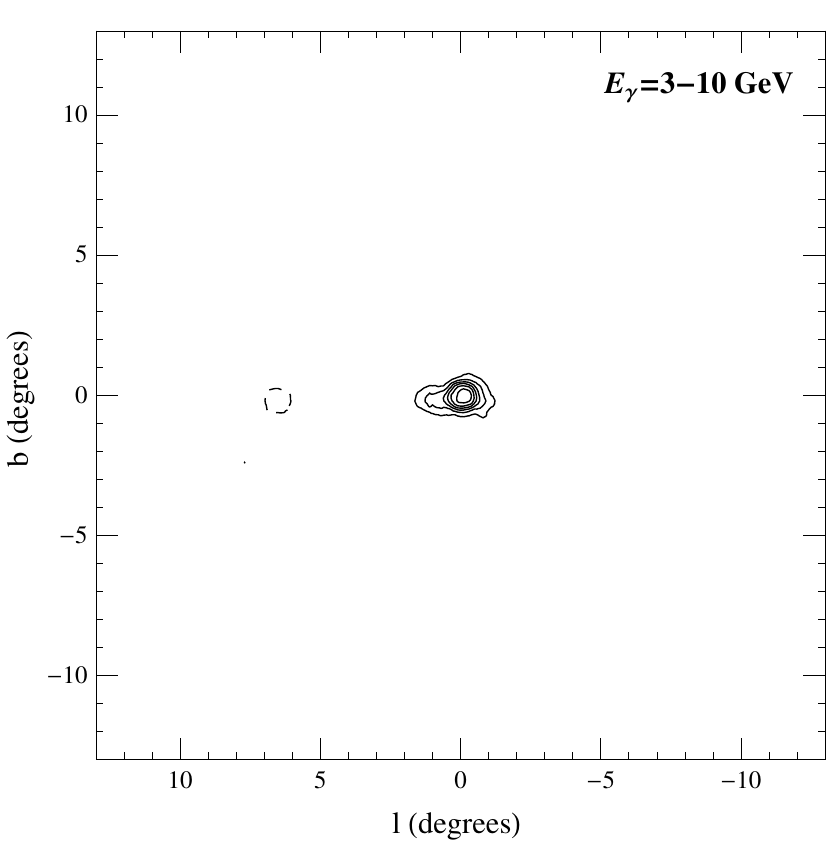}\\
\includegraphics[angle=0.0,width=1.86in]{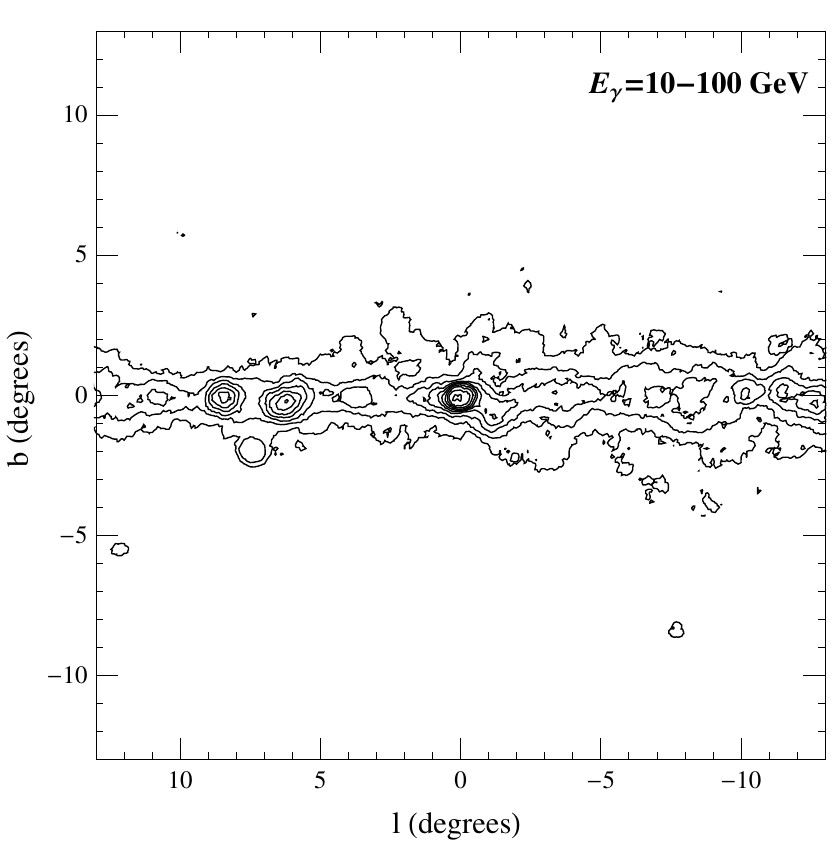}
\includegraphics[angle=0.0,width=1.86in]{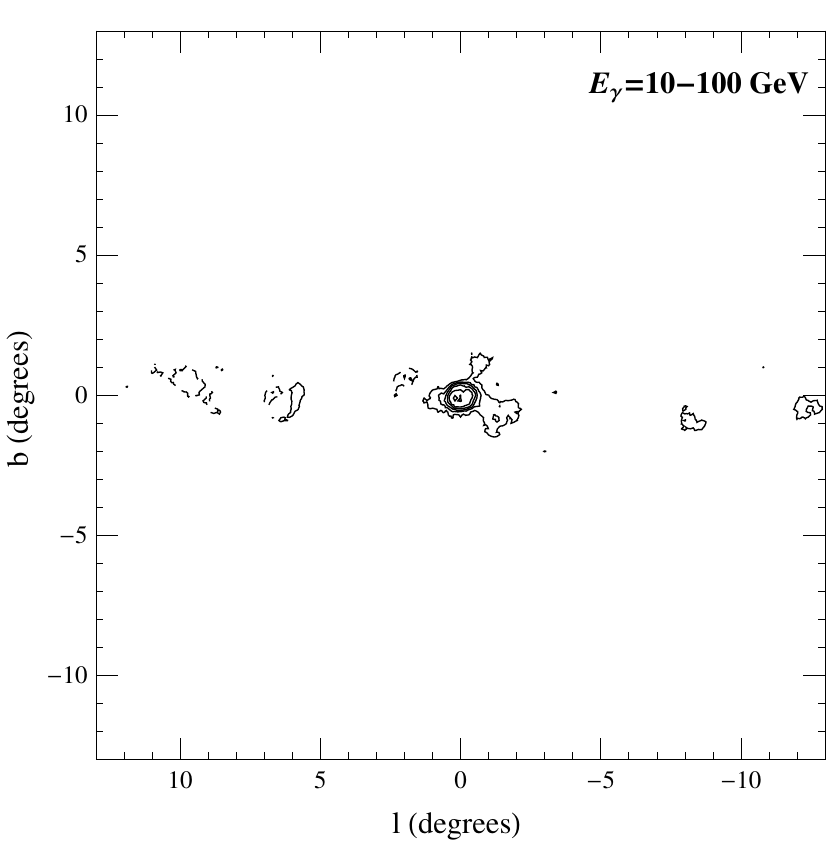}
\caption{Contour maps of the gamma-ray flux from the region surrounding the Galactic Center, as observed by the Fermi Gamma-Ray Space Telescope. The left frames show the raw maps, while the right frames show the maps after subtracting the emission from the sources found in the Second Fermi Source Catalog (not including the central source) and from the Galactic Disk. All maps have been smoothed over a scale of 0.5 degrees. See text for more details.}
\label{maps}
\end{figure*}

Our raw maps are shown in the left frames of Fig.~\ref{maps} for four different energy ranges between 300 MeV and 100 GeV. In each map, ten contours are shown, distributed linearly between $2.45\times 10^{-8}$ and $2.45\times 10^{-7} {\rm cm}^{-2}\, {\rm s}^{-1}$ sq deg$^{-1}$ (300-1000 MeV), $1.06\times 10^{-8}$ and $1.06\times 10^{-7} {\rm cm}^{-2}\, {\rm s}^{-1}$ sq deg$^{-1}$ (1-3 GeV), $2.60\times 10^{-9}$ and $2.60\times 10^{-8} {\rm cm}^{-2}\, {\rm s}^{-1}$ sq deg$^{-1}$ (3-10 GeV), and $3.60\times 10^{-10}$ and $3.60\times 10^{-9} {\rm cm}^{-2}\, {\rm s}^{-1}$ sq deg$^{-1}$ (10-100 GeV). We have smoothed each of the maps at a scale of 0.5 degrees (the contour maps thus represent the average flux observed within a 0.5 degree radius of a given direction in the sky). 

In the right frames of Fig.~\ref{maps}, we show the maps after subtracting the emission from known point sources and from the Galactic Disk, following the approach of Ref.~\cite{Hooper:2011ti}. In particular, we have subtracted a template map including all of the sources in the region contained within the Fermi Second Source Catalog~\cite{catalog} (adopting central values for the intensity and location of each source, as reported in the catalog), with the exception of the bright central source, which cannot be easily disentangled from dark matter annihilation products in the case of a strongly cusped or contracted halo profile. To account for emission from the disk, we subtract a template with a morphology derived from the line-of-sight gas densities~\cite{gas,gas2} as a function of galactic latitude (this template is only very mildly dependent on galactic longitude in the region of the sky being studied here). For details, see Ref.~\cite{Hooper:2011ti}. As the morphologies of these subtracted backgrounds are not at all like those predicted from dark matter annihilation, we are in no danger of unknowingly absorbing any would-be dark matter signal. Throughout this paper, dashed lines are used to denote negative contours of the same magnitude as assigned to the solid lines (resulting from oversubtraction).

After subtracting the contributions from the disk and from known point sources, central residuals remain in each energy range (outside of the inner $\sim$2$^{\circ}$, this subtraction leaves only very modest residuals, typically on order of 10\% or less of the residual flux in the innermost region). This central residual emission almost certainly includes some degree of contributions from the central supermassive black hole~\cite{hesspoint,others}, unresolved point sources, and from cosmic ray interactions with gas (such as is observed at higher energies by HESS~\cite{ridge}, for example). To be conservative, we do not subtract any such components from this residual and derive all of our limits assuming that the entire residual flux could potentially be the products of annihilating dark matter.

\section{Dark Matter Annihilation in the Inner Galaxy}

The flux of gamma-rays in a direction, $\psi$, from dark matter annihilations is given by:
\begin{equation}
\Phi(\psi) = \frac{\sigma v}{8\pi m^2_{\rm DM}}\frac{dN_{\gamma}}{dE_{\gamma}} \int_{\rm los} \rho^2(l) dl,
\end{equation}
where $\sigma v$ and $m_{\rm DM}$ are the annihilation cross section and mass of the dark matter particle, respectively. $dN_{\gamma}/dE_{\gamma}$ is the gamma-ray spectrum produced per annihilation (as calculated using PYTHIA~\cite{pythia}), and the square of the dark matter density profile is integrated over the observed line-of-sight (los). This integral is often written in terms of the dimensionless function, $J$:
\begin{equation}
J(\psi) \equiv \frac{1}{8.5 \, {\rm kpc}} \bigg(\frac{1}{0.3 \, {\rm GeV}/{\rm cm}^3}\bigg)^2 \,  \int_{\rm los} \rho^2(l) dl.
\label{J}
\end{equation}
In this section, we begin by considering a dark matter distribution which follows the well-known NFW (Navarro-Frenk-White) profile~\cite{nfw}:
\begin{equation}
\rho(r) \propto \frac{1}{(r/R_s) [1+(r/R_s)]^2},
\end{equation}
where $R_s=20$ kpc is the scale radius of the halo. 

And while the NFW profile is motivated by the results of numerical simulations~\cite{Navarro:2008kc}, there is some (relatively mild) variation in the profiles that are predicted by such approaches. The results of the Via Lactea II simulation, for example, favor profiles with a somewhat steeper inner slope than NFW ($\rho \propto r^{-1.2}$)~\cite{vialactea}, while the Aquarius Project finds that the density slope varies with $r$~\cite{aquarius}. The results of the Aquarius simulation have been used to motivate the commonly used Einasto profile:
\begin{equation}
\rho(r) \propto \exp\bigg[\frac{-2.0}{\alpha}\bigg\{\bigg(\frac{r}{R_s}\bigg)^{\alpha}-1\bigg\} \bigg],
\end{equation}
where $R_s=20$ kpc is the scale radius of the halo. We will take a value of 0.17 for the parameter $\alpha$.

Neither the NFW nor the Einasto profile take into account the potentially important impact of baryonic physics on the distribution of dark matter in the inner Milky Way. Generally speaking, Milky Way-sized dark matter density profiles are expected to be contracted as a result of dissipating baryons, leading to the steepening of the inner profile~\cite{ac}. The degree to which this effect is manifest depends on the fraction of the baryons that dissipate slowly by radiative cooling.

Several state-of-the-art hydrodynamical simulations (performed by different groups, using different codes) have found that Milky Way-sized halos become significantly contracted by baryonic processes, increasing the density of dark matter in their inner volumes relative to that predicted by dark matter-only simulations (see Ref.~\cite{Gnedin:2011uj} and references therein). These simulations, which include the effects of gas cooling, star formation, and stellar feedback, predict a degree of adiabatic contraction which typically steepens the inner slopes of dark matter density profiles from $\gamma \approx 1.0$ to $\gamma \approx 1.2-1.5$ within the inner $\sim$10 kpc of Milky Way-like galaxies~\cite{Gnedin:2011uj,mac}.  The resolution of such simulations is currently limited to scales larger than $\sim$100 parsecs~\cite{100pc}. In contrast, it has also been argued that strong feedback could result in the flattening of the inner slopes of galactic dark matter profiles~\cite{governato}.  This appears particularly likely in the case of low mass galaxies ({\it ie.} dwarf galaxies), although it is less clear that such effects are likely to dominate in larger, Milky Way-sized, systems. 

In an effort to consider a wide range of possible baryonic effects, we will derive limits using both contracted and cored halo profiles. To account for the possible effects of baryonic contraction, we adopt a generalized form of the NFW profile:
\begin{equation}
\rho(r) \propto \frac{1}{(r/R_s)^{\gamma} [1+(r/R_s)]^{(3-\gamma)}},
\end{equation}
where $\gamma$ is the inner slope of the profile ($\gamma=1$ recovers the NFW form), and $R_s=20$ kpc is again the scale radius of the halo. In the following section, we will adopt values of $\gamma=1.2$ and 1.4, which we consider to represent mildly and significantly contracted profiles (although even larger values of the inner slope are sometimes found in hydrodynamical simulations~\cite{Gnedin:2011uj}, we will not consider such profiles here). Alternatively, some simulations have found that resonant interactions between the stellar bar and the dark matter can flatten the density of the central cusp~\cite{Kuhlen:2012qw,Weinberg:2001gm} (see also, however, Refs.~\cite{Sellwood:2002vb,Valenzuela:2002np,Colin:2005rr}). With this in mind, we will also consider NFW-like profiles with flat, constant density cores within a radius of either 0.1 kpc or 1 kpc. 

Although profiles with inner slopes less steep than NFW ($\gamma=1$) are sometimes found in hydronamical simulations with strong feedback, constant-density cores ($\gamma=0$) are are rather extreme.  Profiles with softened cusps ($\gamma \sim 0.5-0.8$, for example) exhibit much milder density suppression that is encapsulated in our constant-density cored profiles. Our results using these cored profiles should thus be thought of as quite conservative, leading to less bright dark matter annihilation signals than might otherwise be anticipated.

\begin{figure*}[!t]
\centering
\includegraphics[angle=0.0,width=7.in]{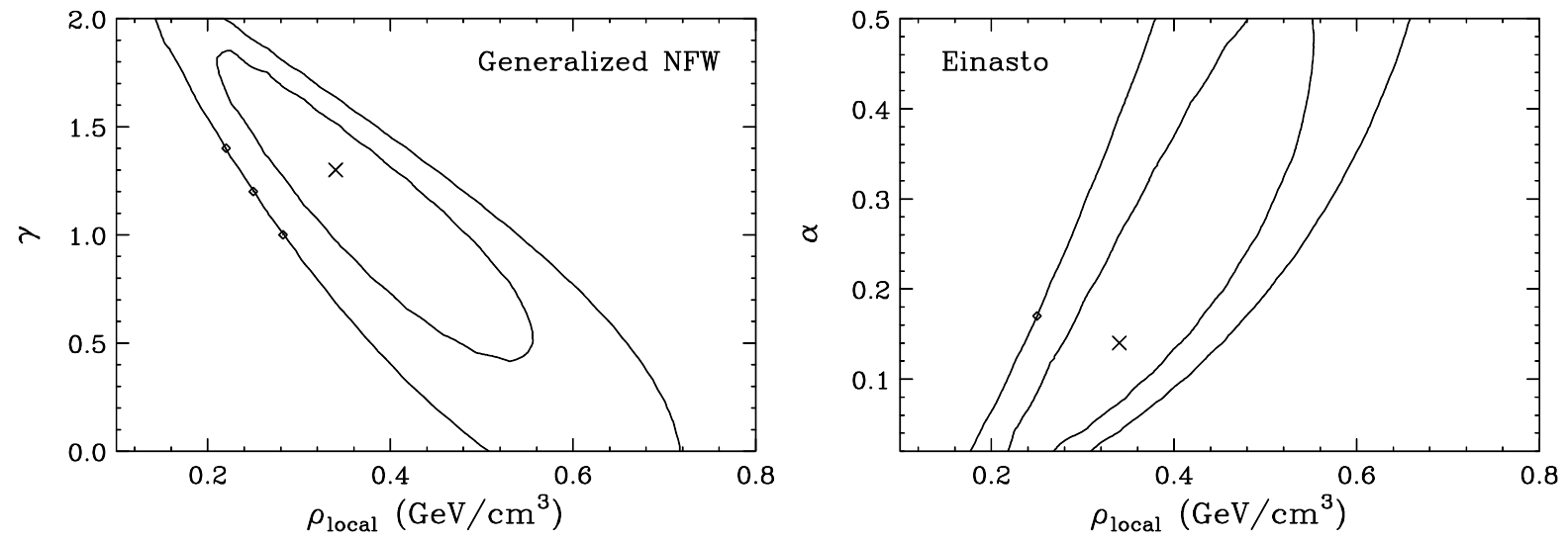}
\caption{Constraints on the Milky Way's dark matter distribution from the combination of dynamical (rotation curves) and microlensing observations, as presented in Ref.~\cite{local}. The crosses in each frame denote the best-fit point. When deriving constraints on the dark matter annihilation cross section, we adopt halo profile normalizations which are at the lower $2\sigma$ boundary of these constraints (shown as small diamonds). We thank the authors of Ref.~\cite{local} for providing the numerical values for the contours shown.}
\label{bertone}
\end{figure*}

At present, the distribution of dark matter in the inner Milky Way is not very strongly constrained by observations. In particular, the gravitational potential of the Milky Way's innermost kiloparsecs is dominated by baryons (stars and gas), which makes it difficult to strongly constrain the subdominant dark matter distribution in this region. In Fig.~\ref{bertone}, we show the results of a recent study of microlensing and dynamical observations of our galaxy~\cite{local}. While this study finds that these observational constraints are compatible with an NFW profile ($\gamma=1$), they also allow for much steeper ($\gamma \sim 1.8$) or shallower ($\gamma \sim 0.5$) distributions.   

In each frame of Fig.~\ref{bertone}, the best-fit profile is marked by a cross. To be conservative in the deriving of annihilation constraints, for each halo model we will normalize the dark matter density using the minimum (the lower $2\sigma$ boundary) value found in Ref.~\cite{local}. These points are denoted by small diamonds in Fig.~\ref{bertone}.

\begin{figure*}[t]
\centering
\includegraphics[angle=0.0,width=7.in]{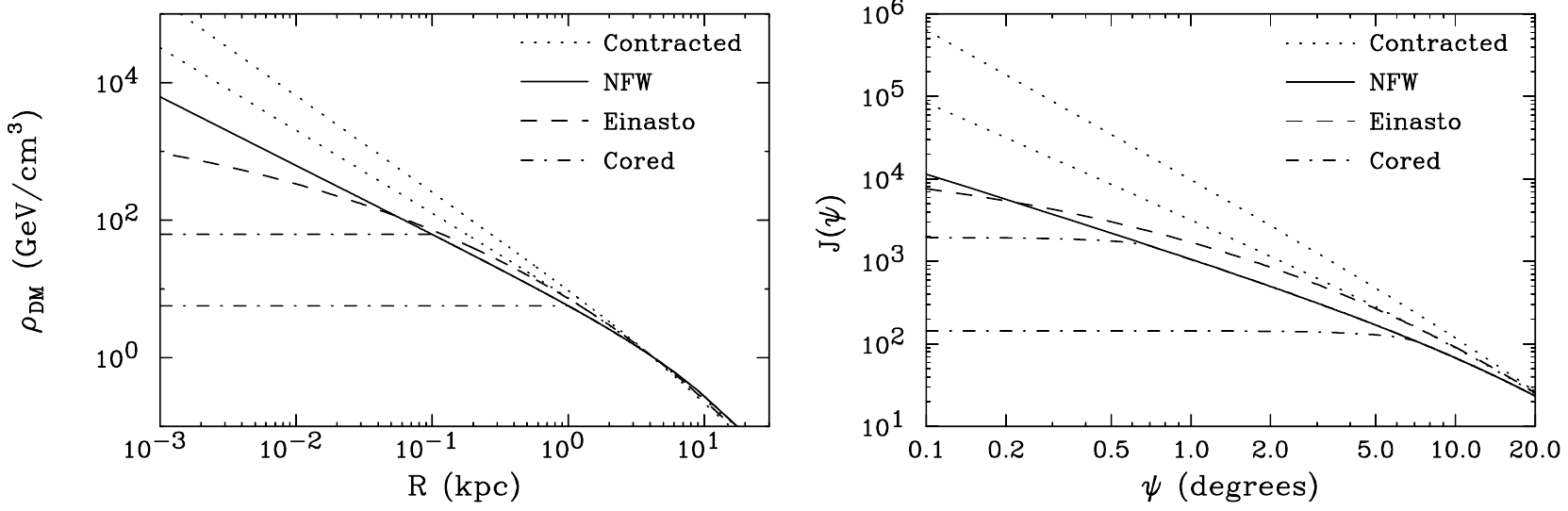}
\caption{Left: Dark matter density profiles used in this study. Right: The value of the density squared line-of-sight integral, $J$, as a function of the angle from the Galactic Center, $\psi$. The contracted profiles are generalized NFW profiles with inner slopes of 1.2 and 1.4. The cored profiles are NFW profiles with constant densities within either 0.1 or 1.0 kpc. See text for more details.}
\label{profiles}
\end{figure*}

In this study, we will derive dark matter annihilation constraints using six halo profile models. In the left frame of Fig.~\ref{profiles}, we plot each of these density profiles (each normalized to the best-fit value as shown in Fig.~\ref{bertone}). Although these profiles are all very similar outside of the innermost few kiloparsecs of the Milky Way, they can vary considerably within the inner kiloparsec or so. In the right frame of this figure, we show the values of the density squared line-of-sight integral, $J$ (see Eq.~\ref{J}), as a function of the angle away from the Galactic Center, $\psi$, for each of these profiles. 

Upon first glance of this plot, one may expect the flux of gamma-rays from the innermost fraction of a degree around the Galactic Center to depend dramatically on the halo profile that is adopted. In reality, however, the finite angular resolution of Fermi (about 1$^{\circ}$ for a front-converting, 1 GeV photon) mitigates this variation to a significant extent.

\section{Constraints on the Dark Matter Annihilation Cross Section}

\begin{figure*}[t]
\centering
\includegraphics[angle=0.0,width=1.86in]{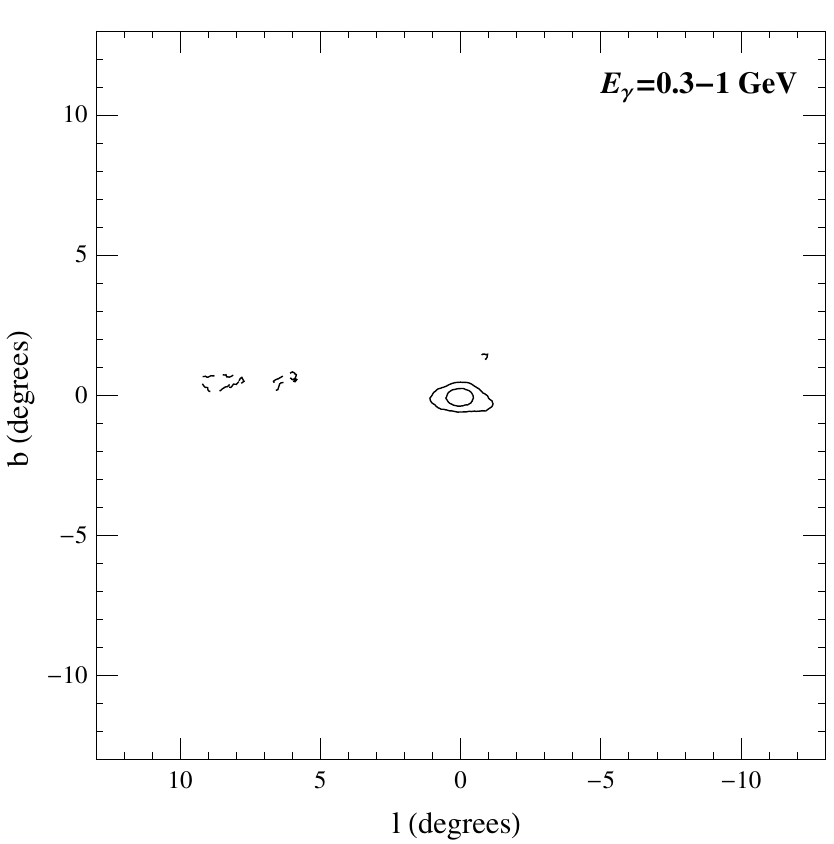}
\includegraphics[angle=0.0,width=1.86in]{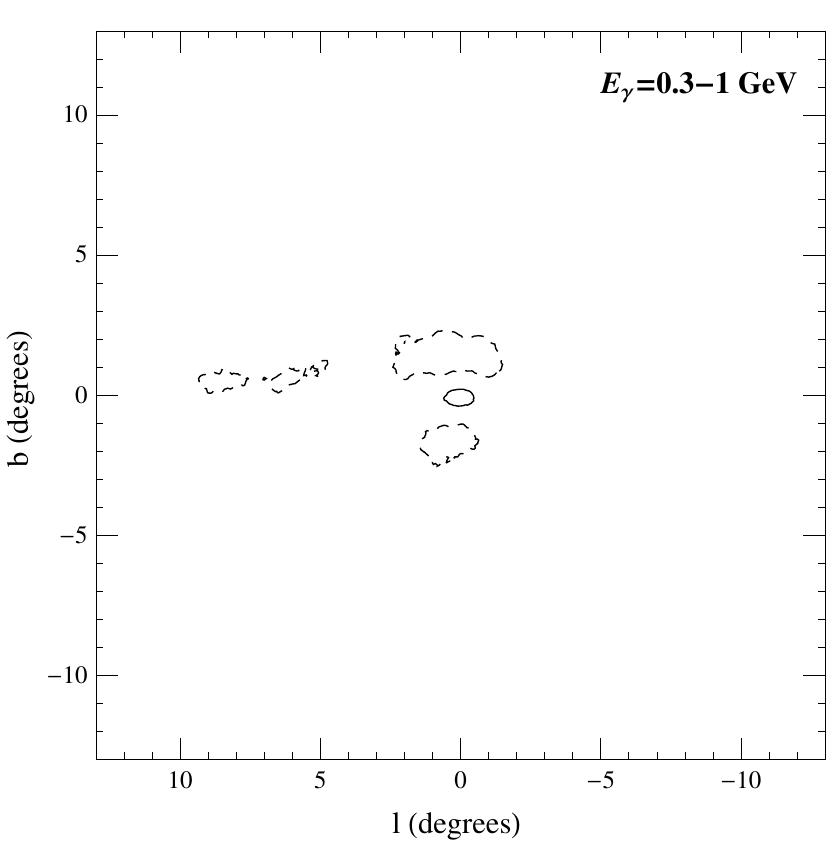}
\includegraphics[angle=0.0,width=1.86in]{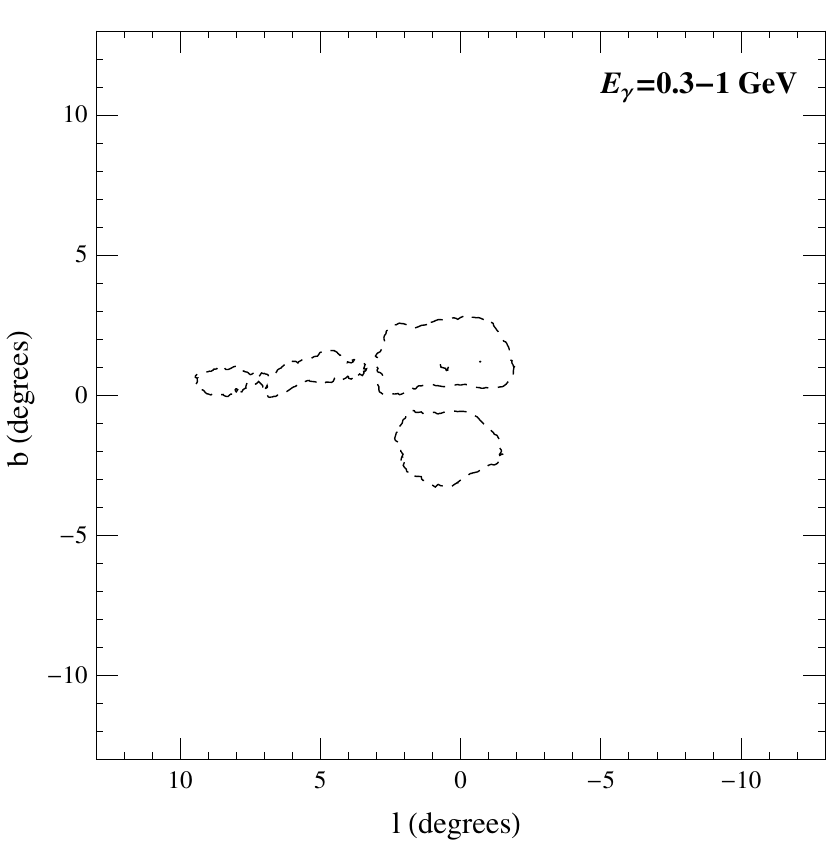}\\
\includegraphics[angle=0.0,width=1.86in]{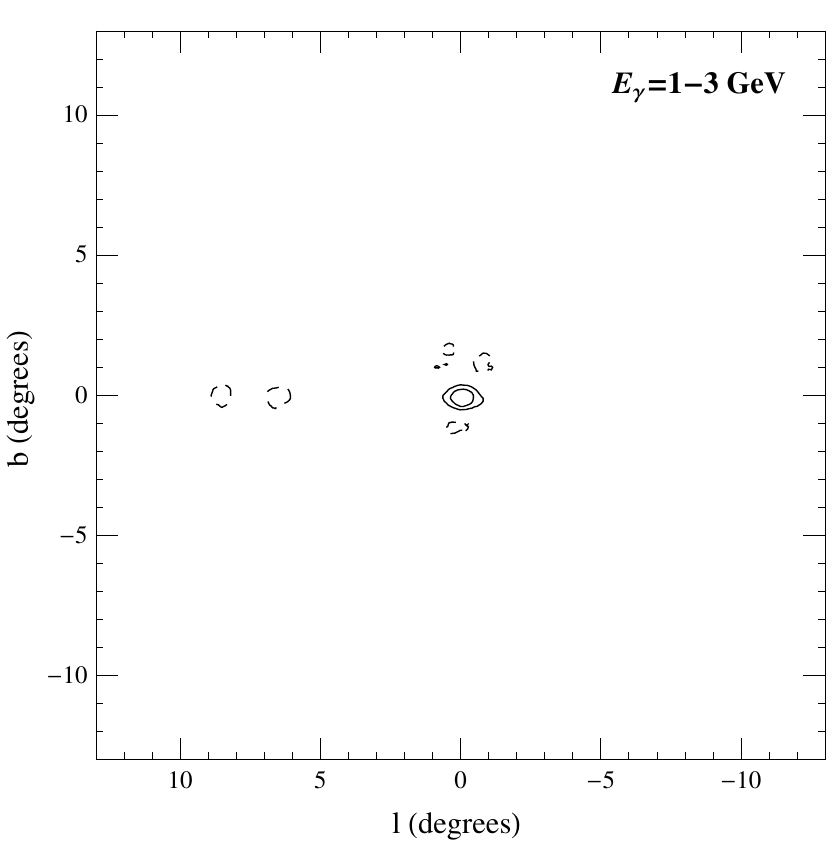}
\includegraphics[angle=0.0,width=1.86in]{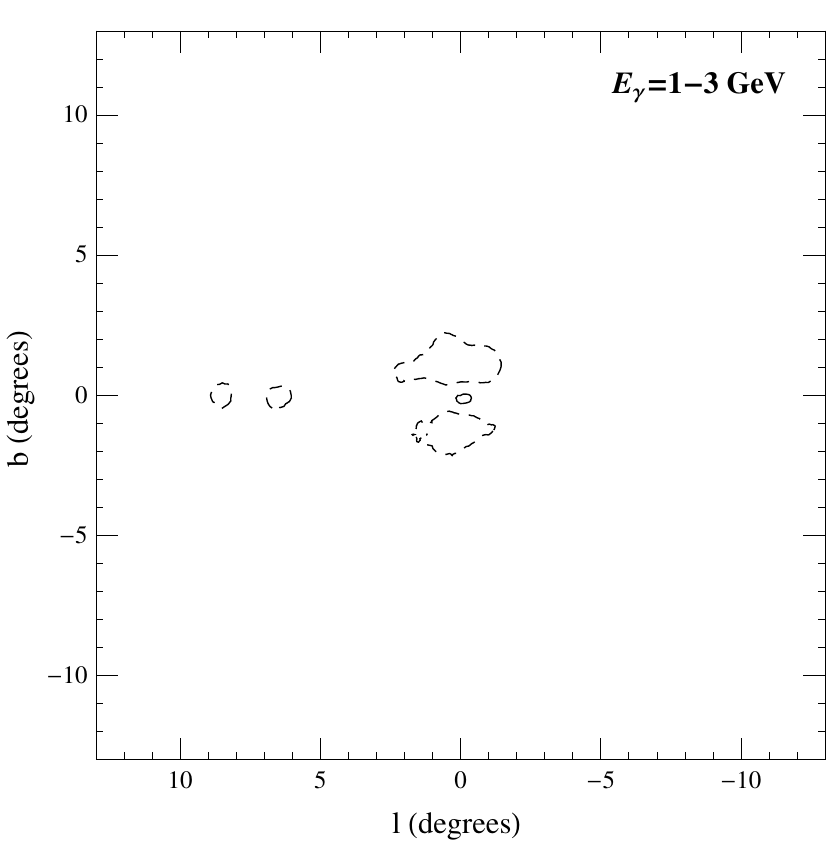}
\includegraphics[angle=0.0,width=1.86in]{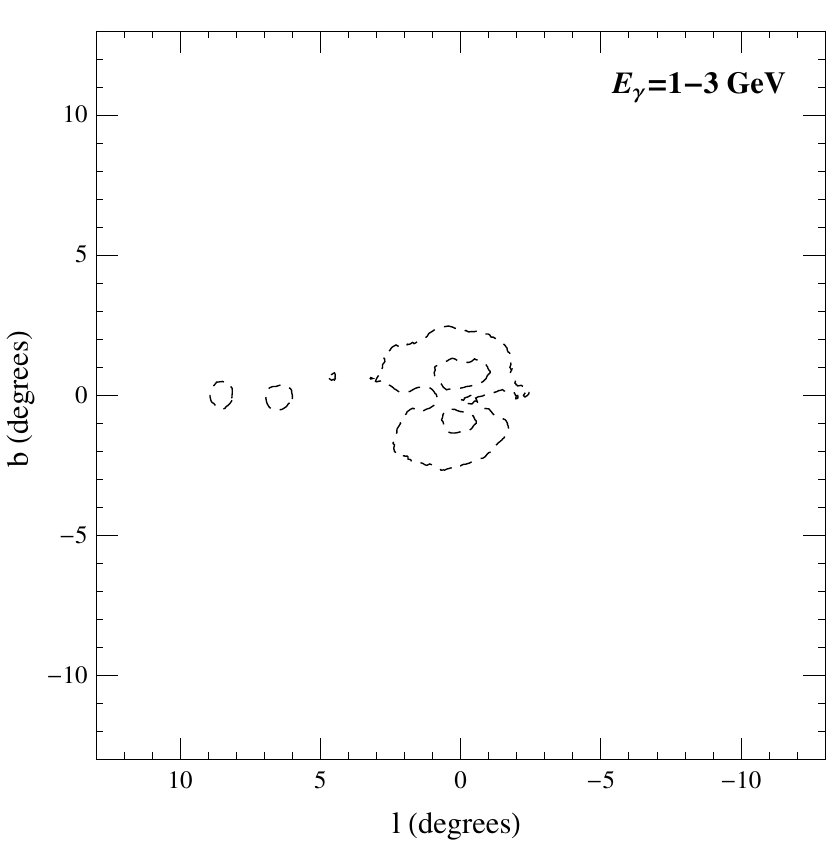}\\
\includegraphics[angle=0.0,width=1.86in]{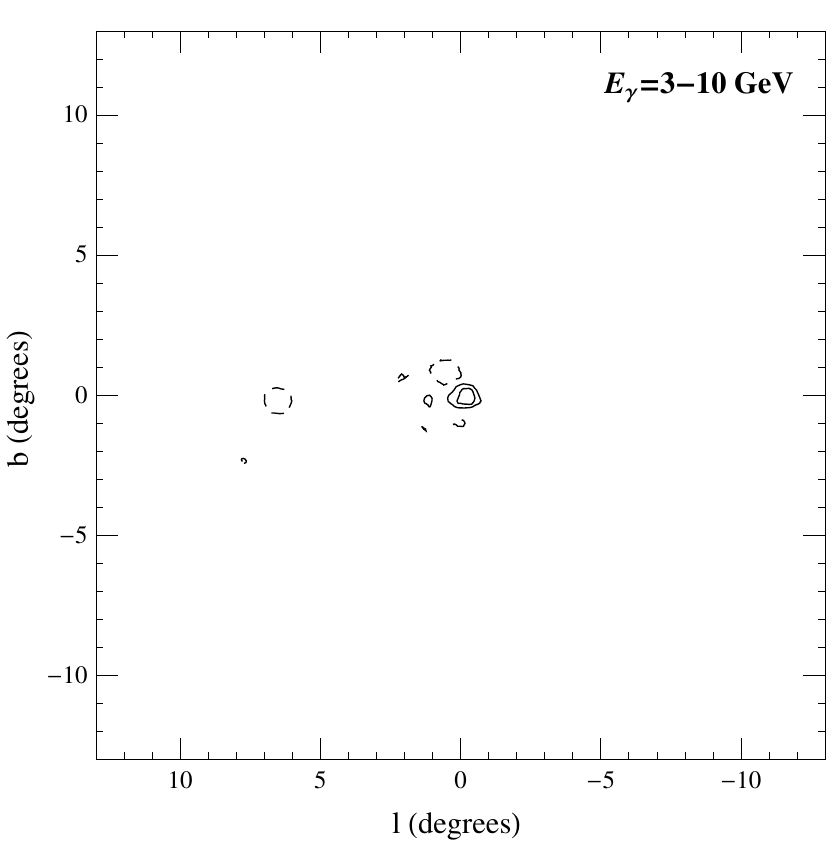}
\includegraphics[angle=0.0,width=1.86in]{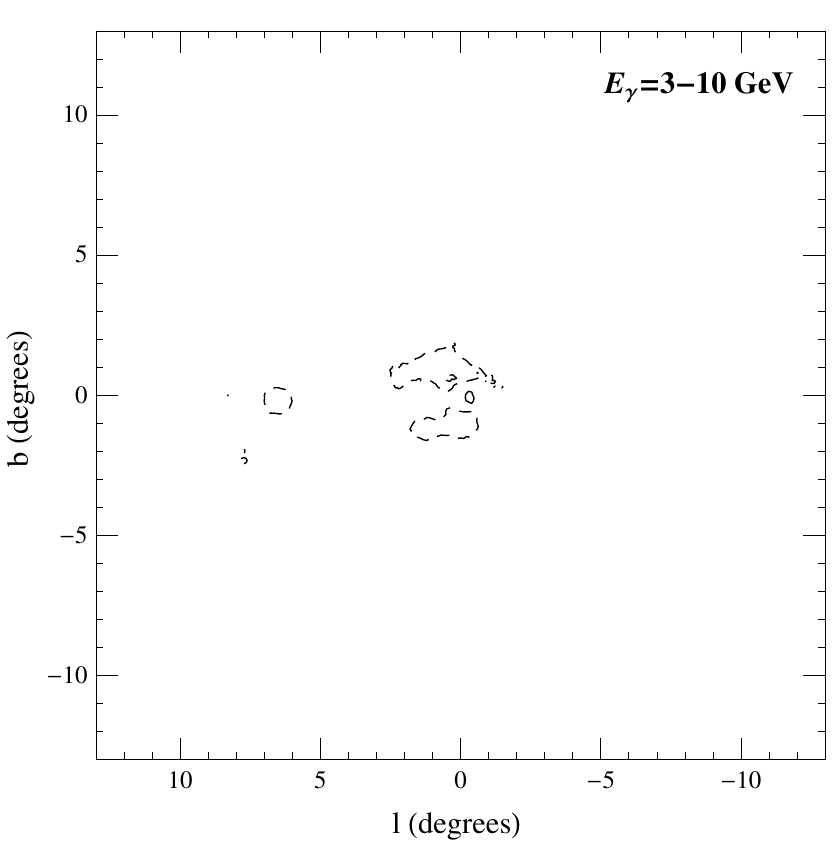}
\includegraphics[angle=0.0,width=1.86in]{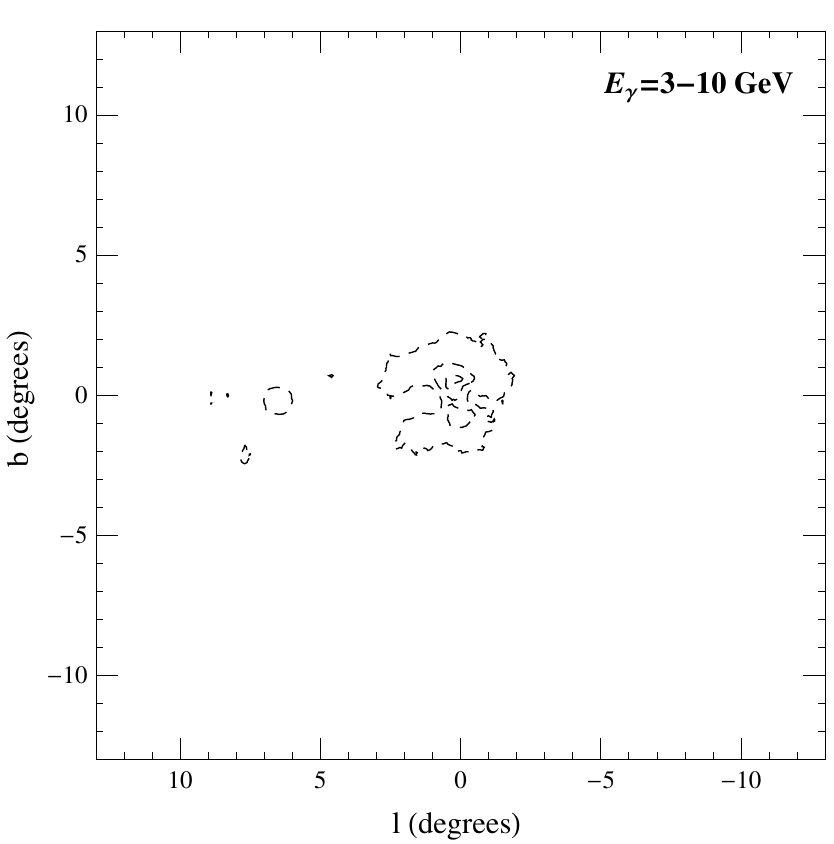}\\
\includegraphics[angle=0.0,width=1.86in]{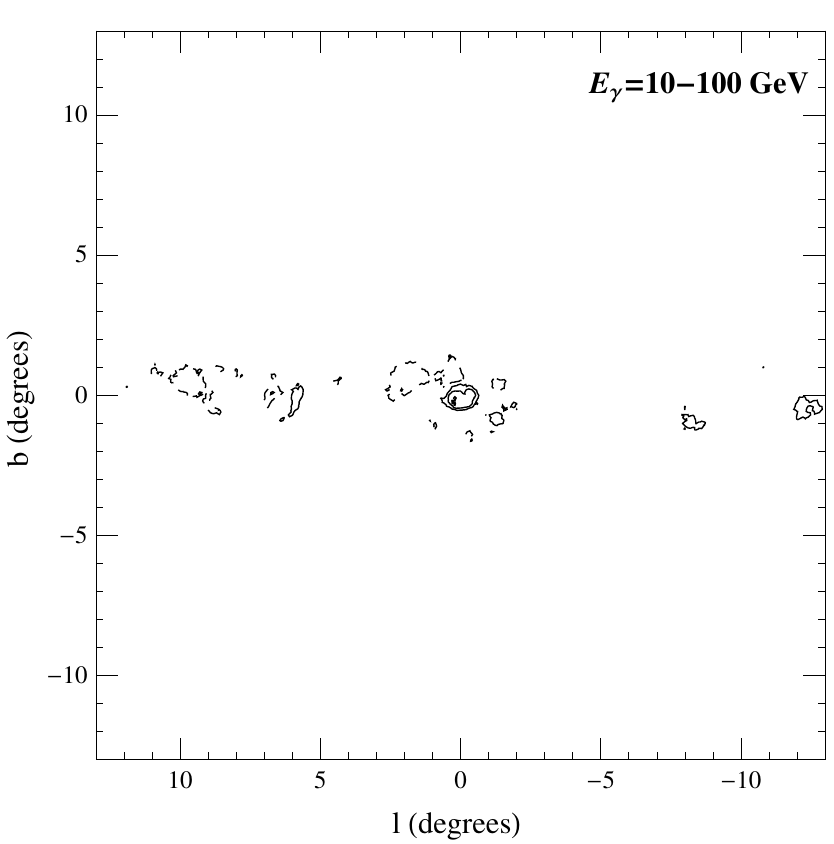}
\includegraphics[angle=0.0,width=1.86in]{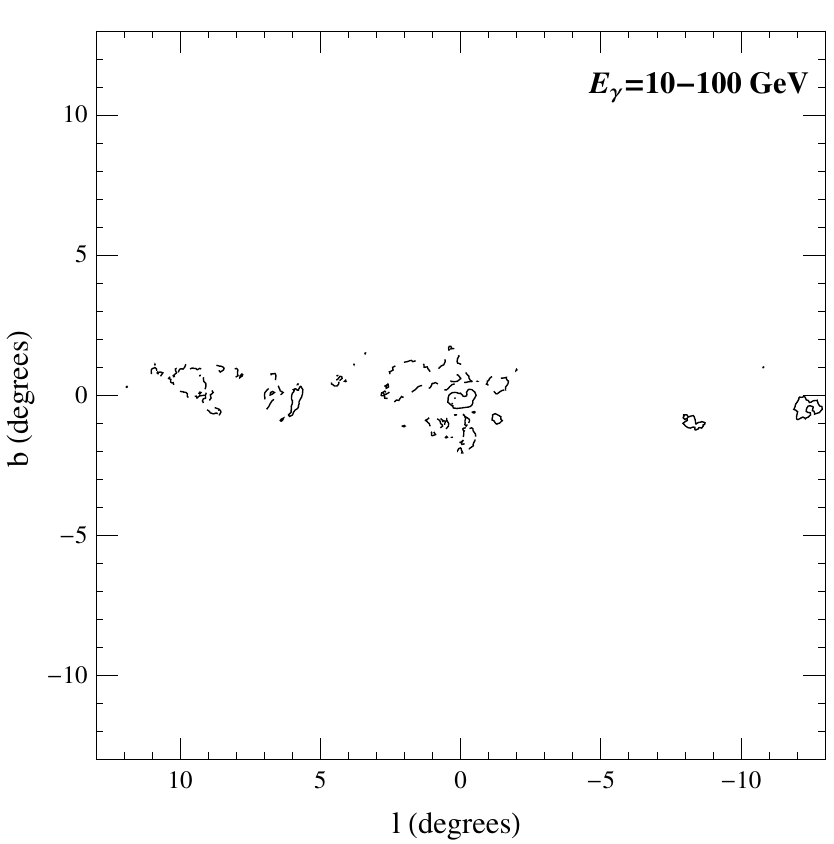}
\includegraphics[angle=0.0,width=1.86in]{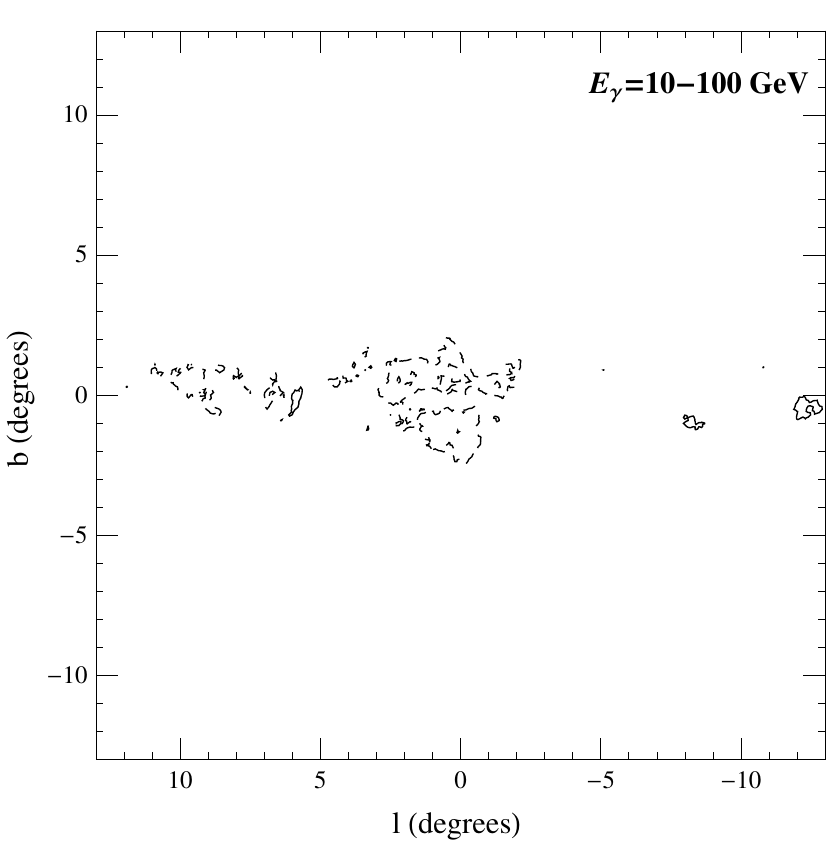}
\caption{Contour maps of the gamma-ray flux from the region surrounding the Galactic Center, after subtracting varying degrees of emission from dark matter distributed according to an NFW profile. As the flux of dark matter annihilation products is increased (moving from left-to-right), regions of the maps become increasingly oversubtracted (denoted by dashed contours). In this case of an NFW distribution, this occurs most noticeably in the regions approximately 1-2$^{\circ}$ north and south of the Galactic Center.}
\label{maps2}
\end{figure*}

\begin{figure*}[t]
\centering
\includegraphics[angle=0.0,width=1.86in]{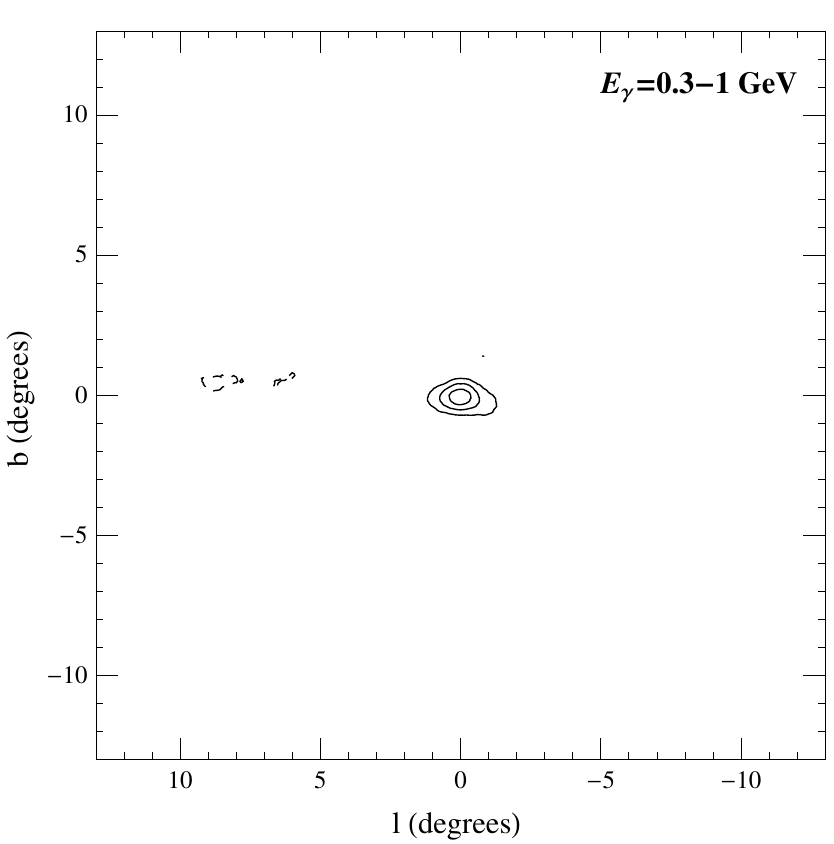}
\includegraphics[angle=0.0,width=1.86in]{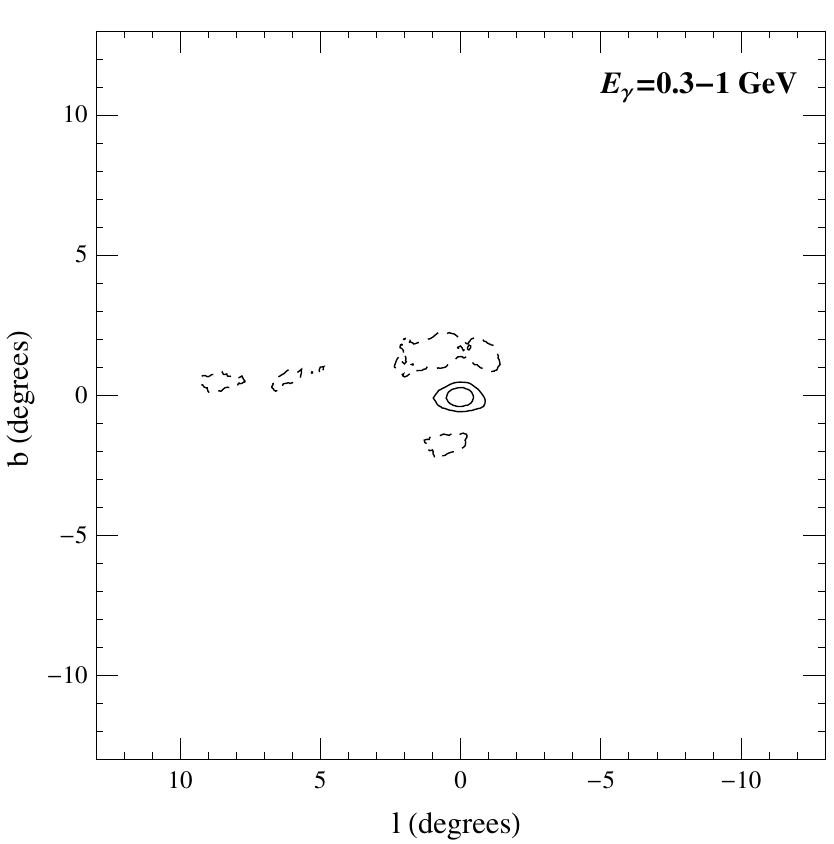}
\includegraphics[angle=0.0,width=1.86in]{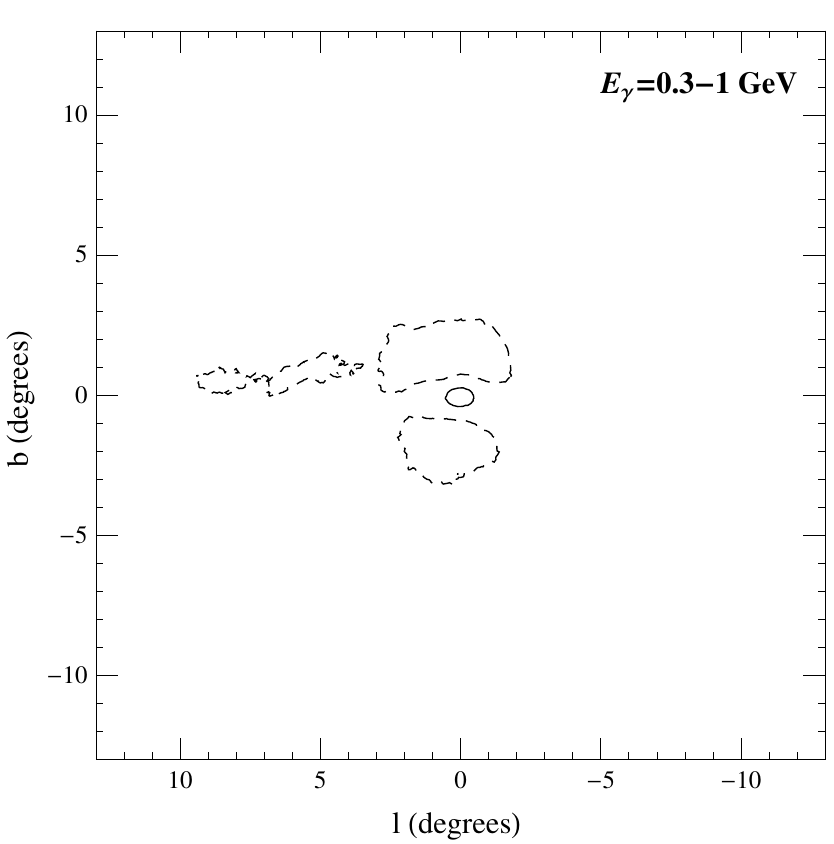}\\
\includegraphics[angle=0.0,width=1.86in]{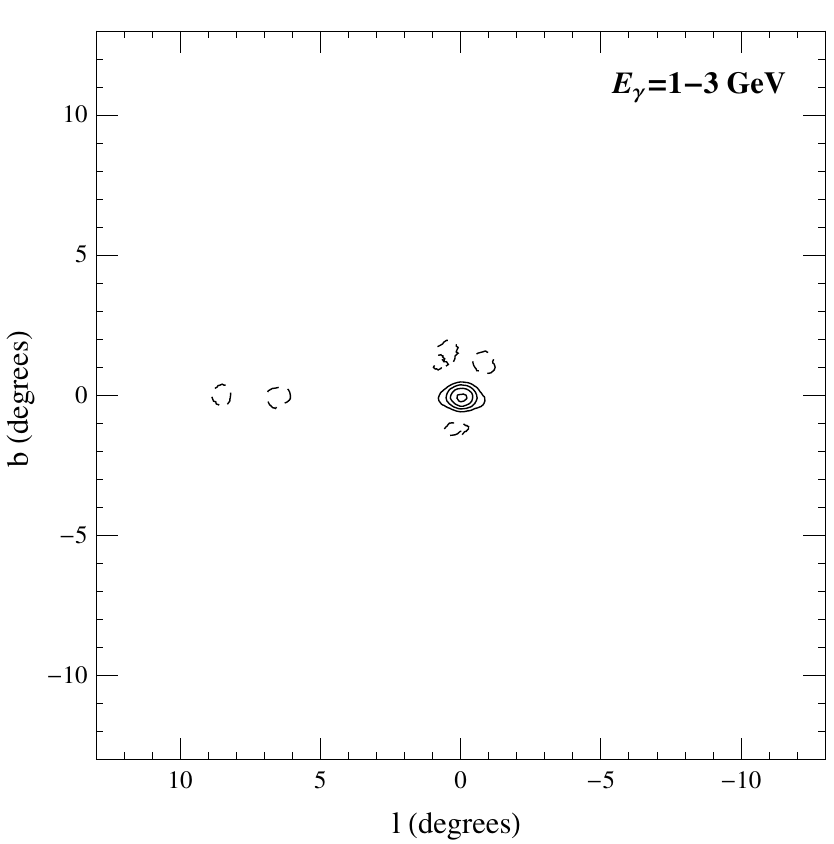}
\includegraphics[angle=0.0,width=1.86in]{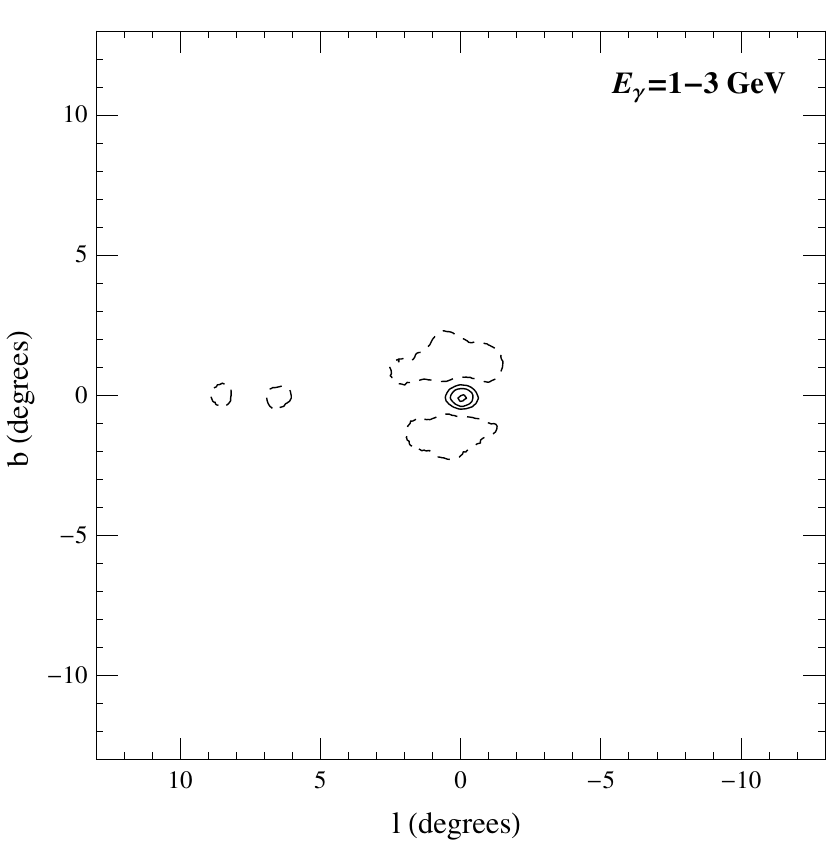}
\includegraphics[angle=0.0,width=1.86in]{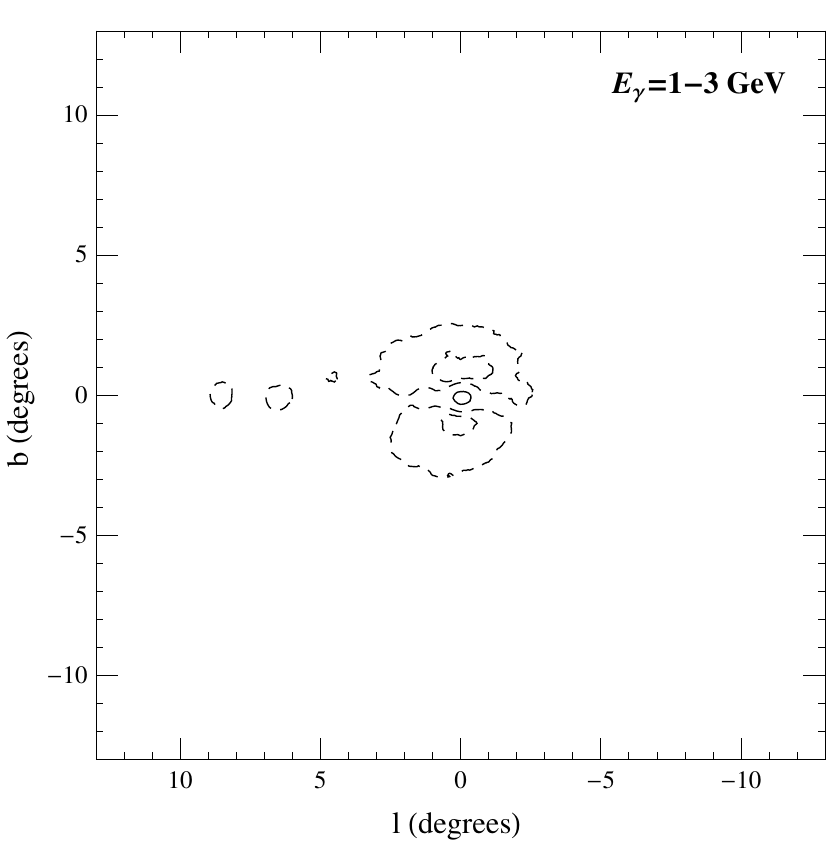}\\
\includegraphics[angle=0.0,width=1.86in]{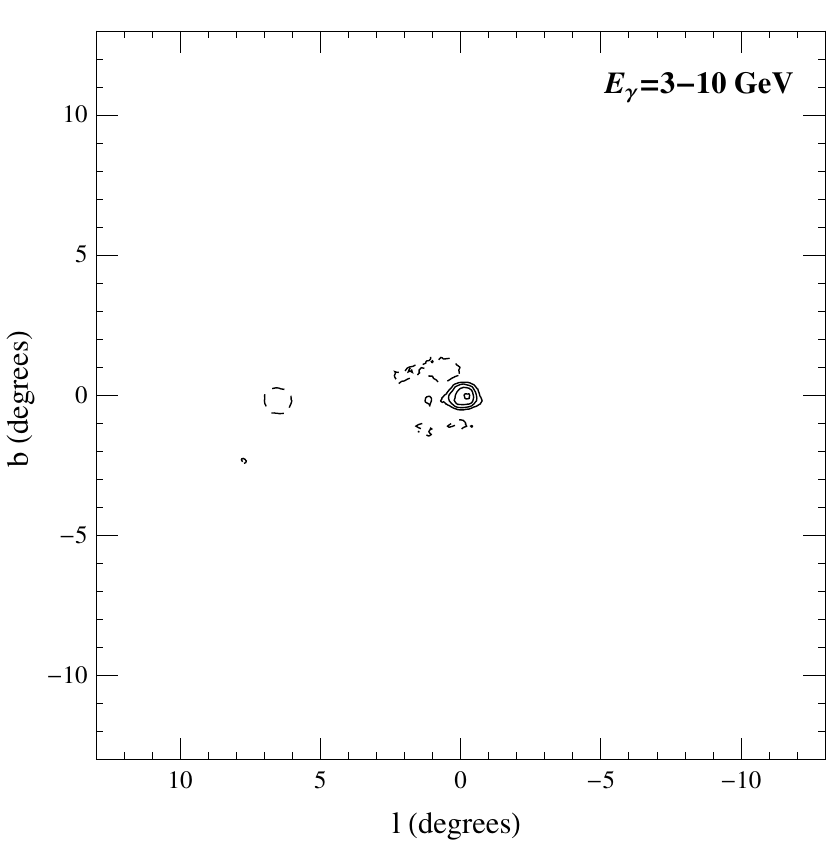}
\includegraphics[angle=0.0,width=1.86in]{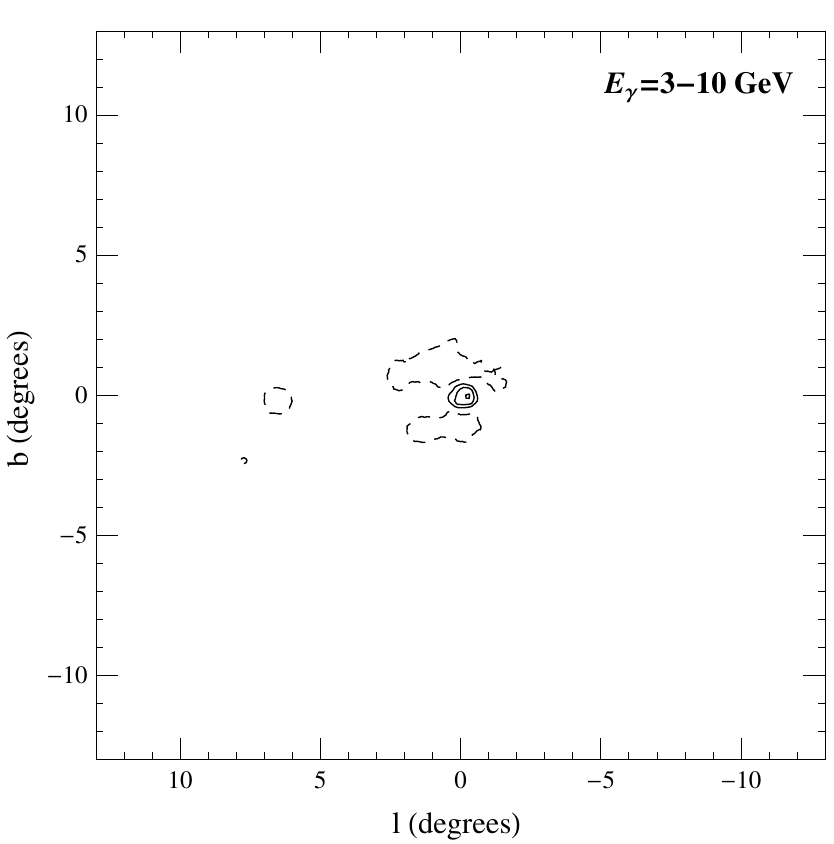}
\includegraphics[angle=0.0,width=1.86in]{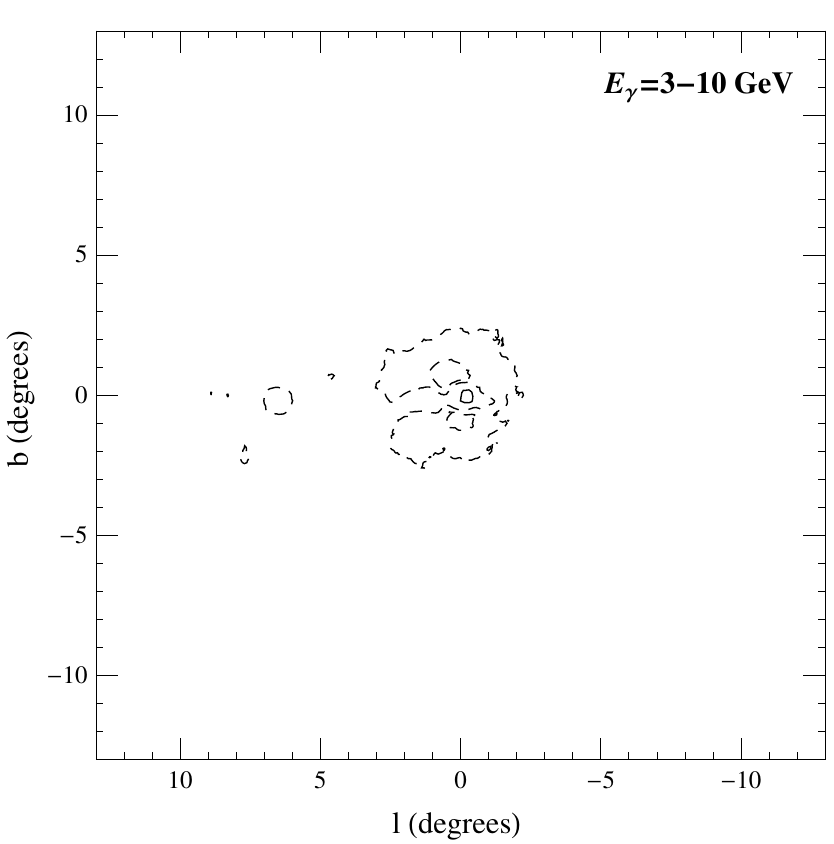}\\
\includegraphics[angle=0.0,width=1.86in]{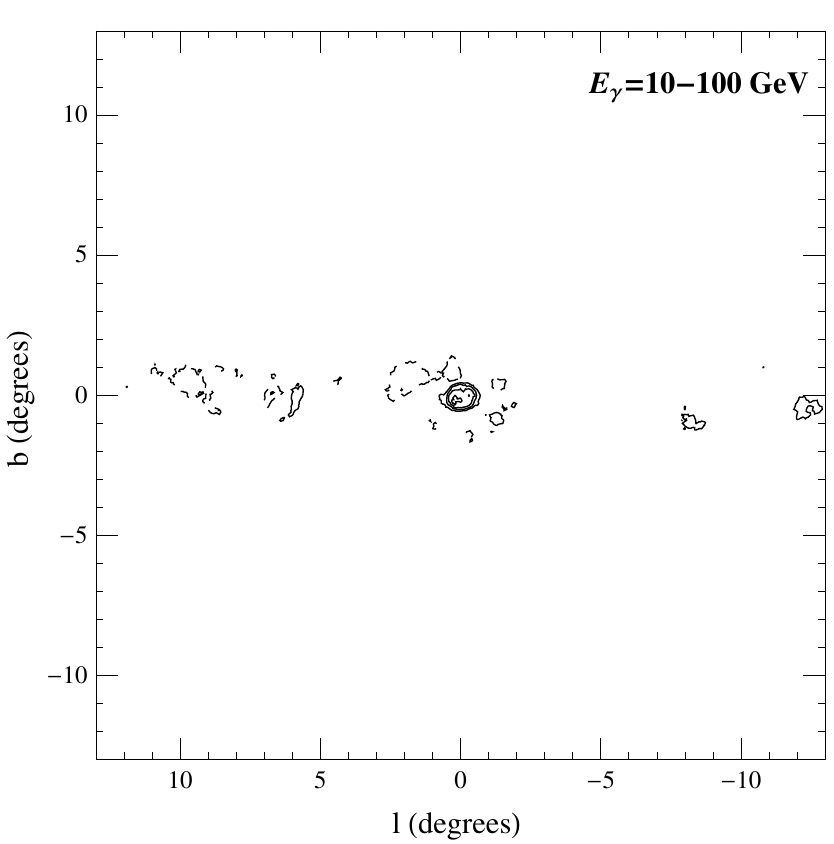}
\includegraphics[angle=0.0,width=1.86in]{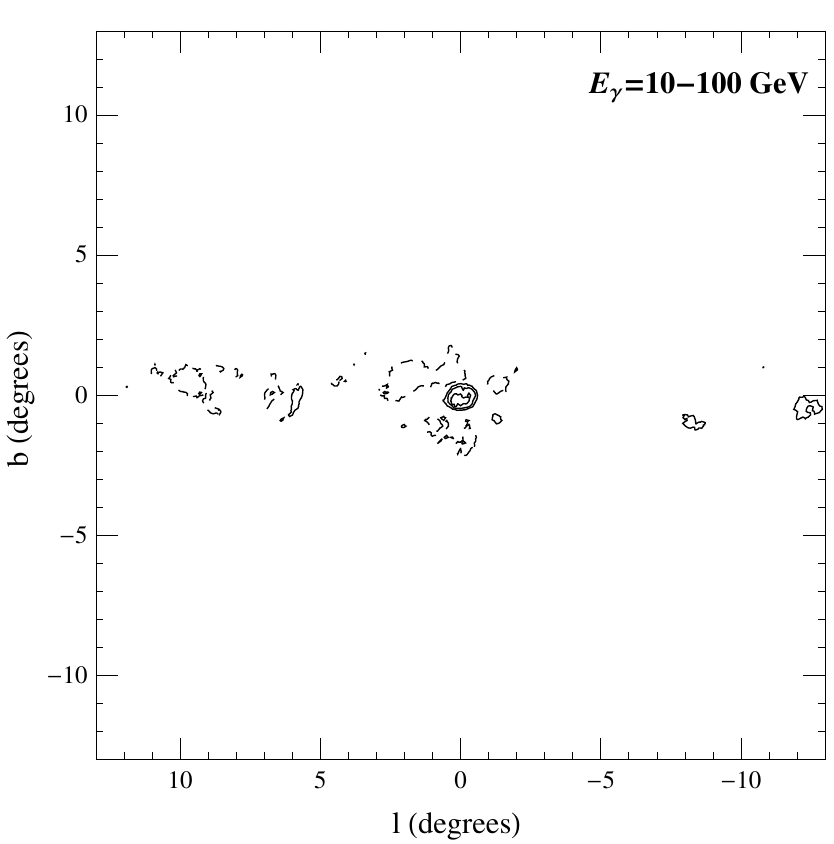}
\includegraphics[angle=0.0,width=1.86in]{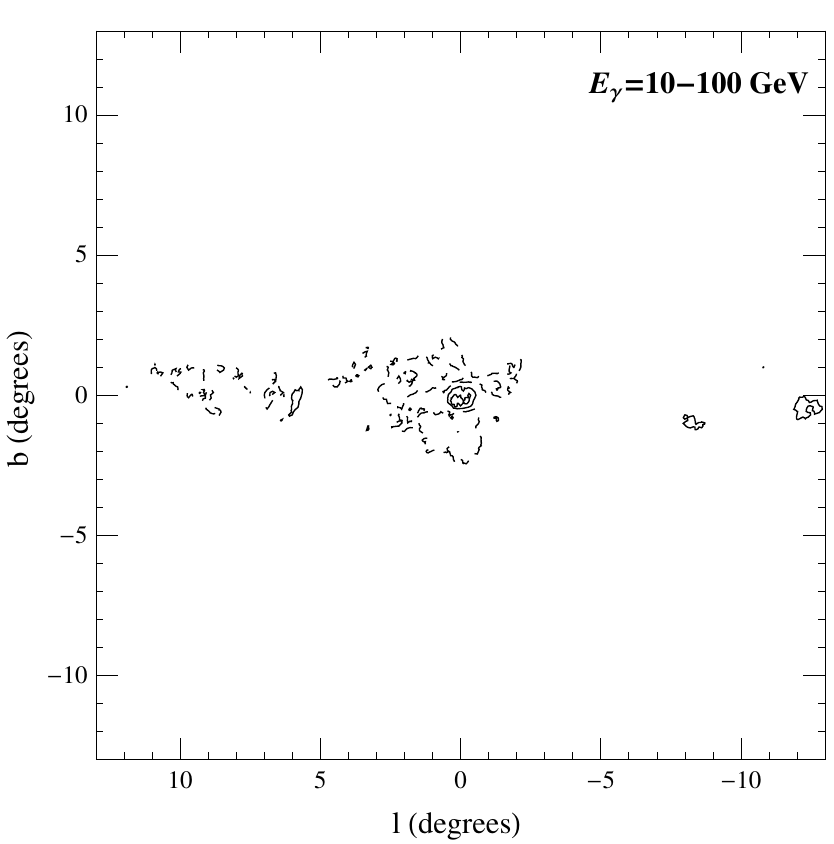}
\caption{Contour maps of the gamma-ray flux from the region surrounding the Galactic Center, after subtracting varying degrees of emission from dark matter distributed according to an Einasto profile. As the flux of dark matter annihilation products is increased (moving from left-to-right), regions of the maps become increasingly oversubtracted (denoted by dashed contours). In this case of an Einasto distribution, this occurs most noticeably in the regions approximately 1-2$^{\circ}$ north and south of the Galactic Center.}
\label{mapseinasto}
\end{figure*}

\begin{figure*}[t]
\centering
\includegraphics[angle=0.0,width=1.86in]{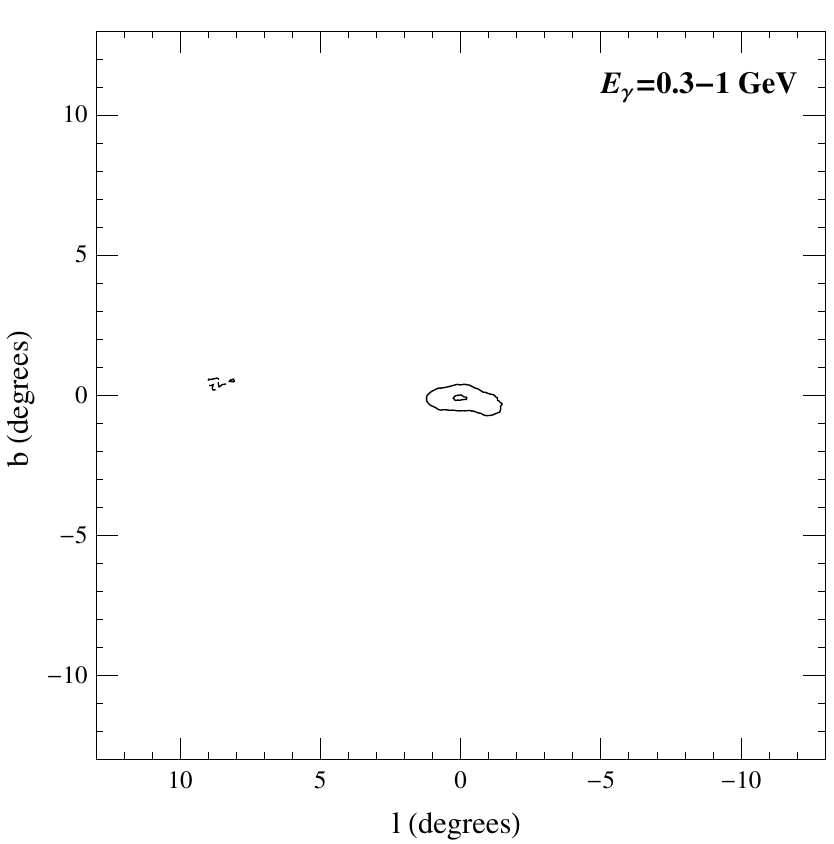}
\includegraphics[angle=0.0,width=1.86in]{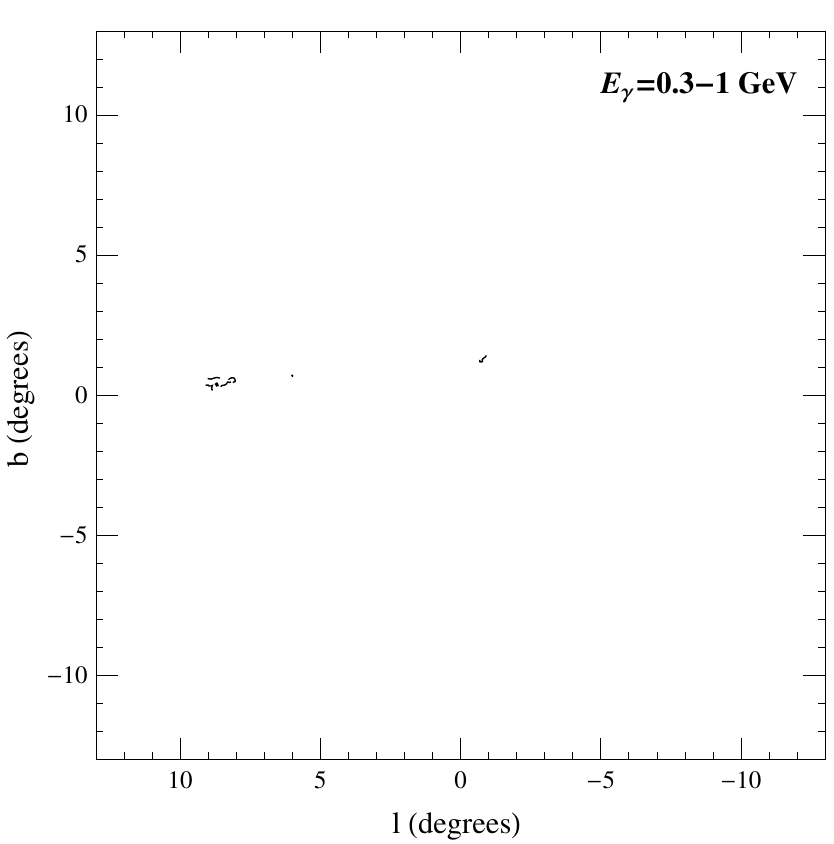}
\includegraphics[angle=0.0,width=1.86in]{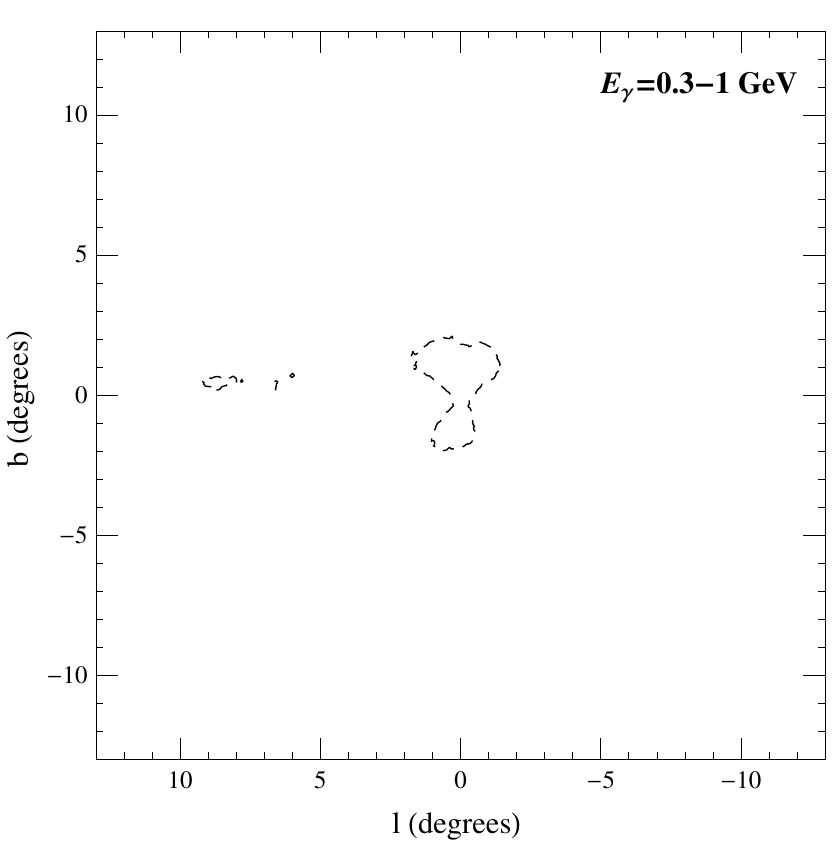}\\
\includegraphics[angle=0.0,width=1.86in]{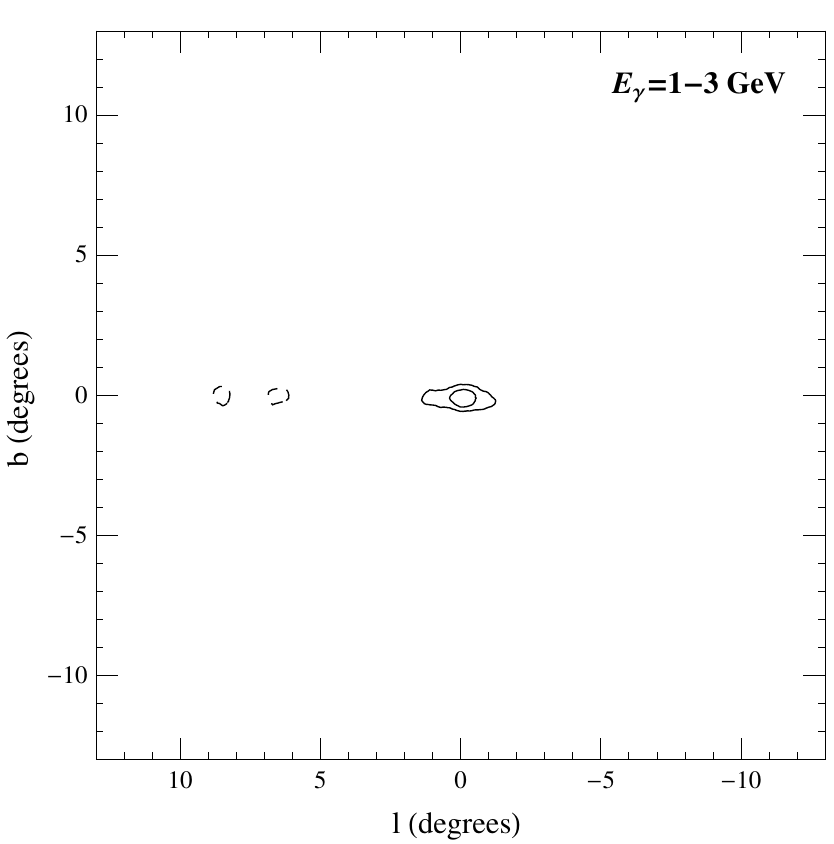}
\includegraphics[angle=0.0,width=1.86in]{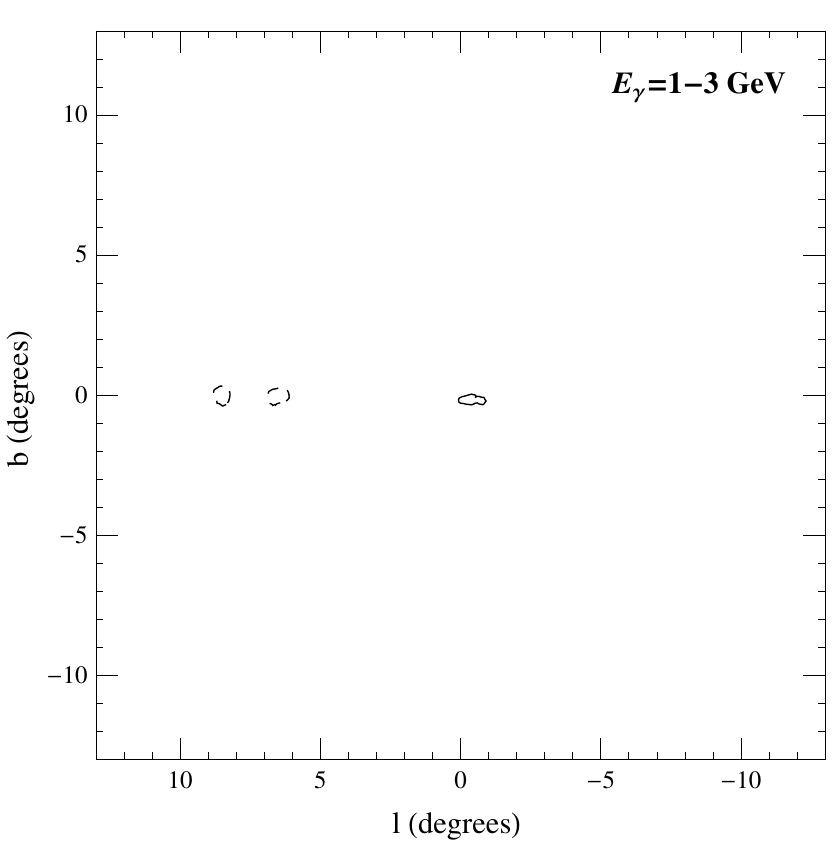}
\includegraphics[angle=0.0,width=1.86in]{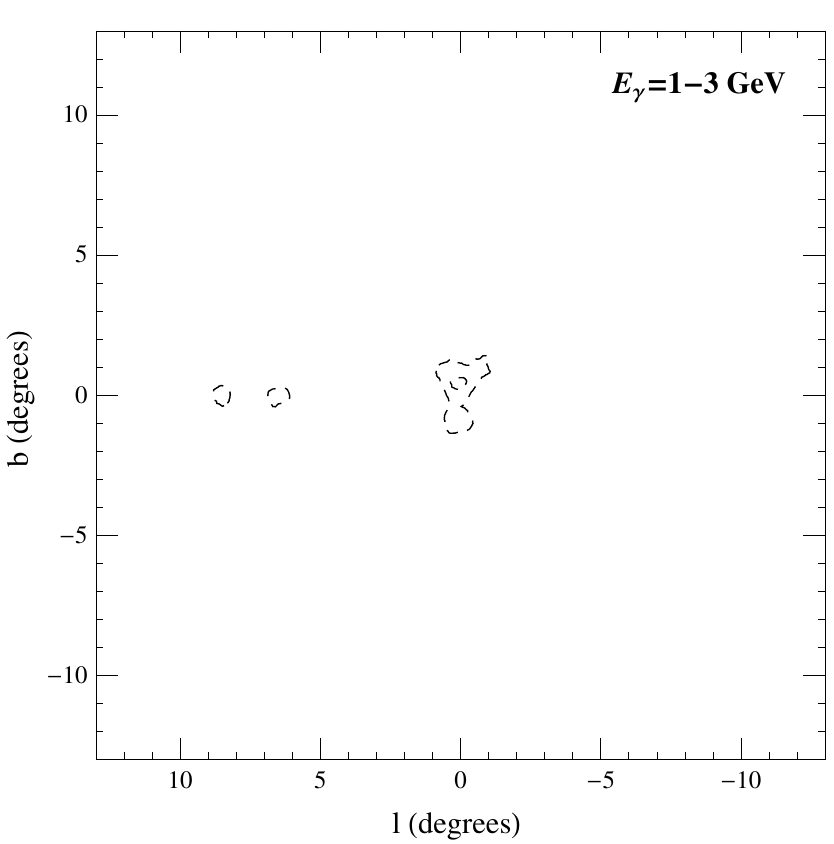}\\
\includegraphics[angle=0.0,width=1.86in]{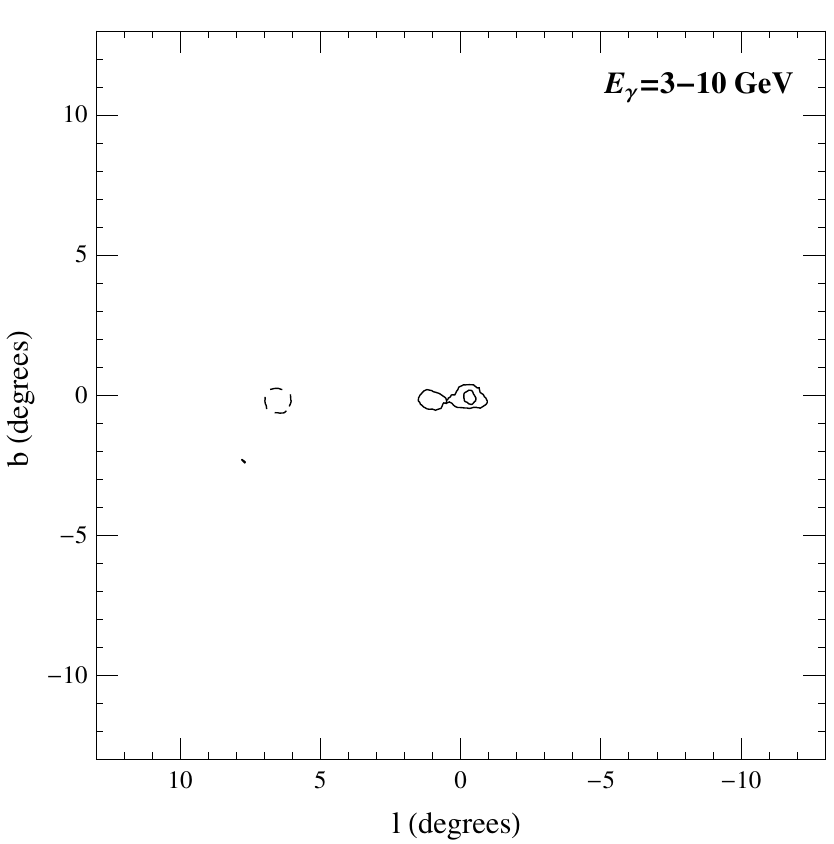}
\includegraphics[angle=0.0,width=1.86in]{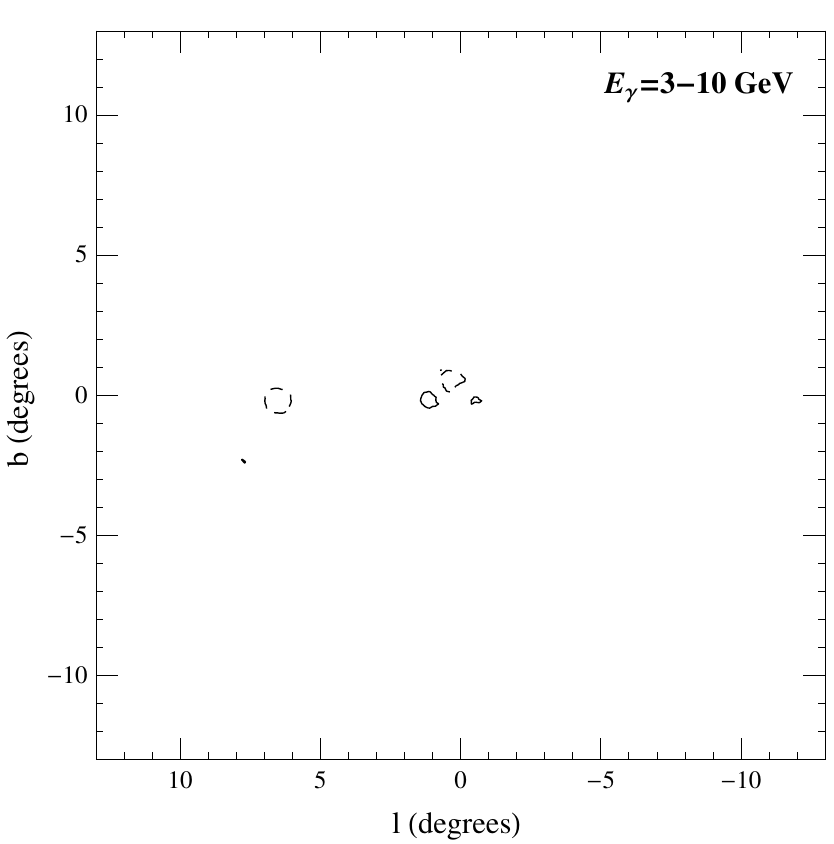}
\includegraphics[angle=0.0,width=1.86in]{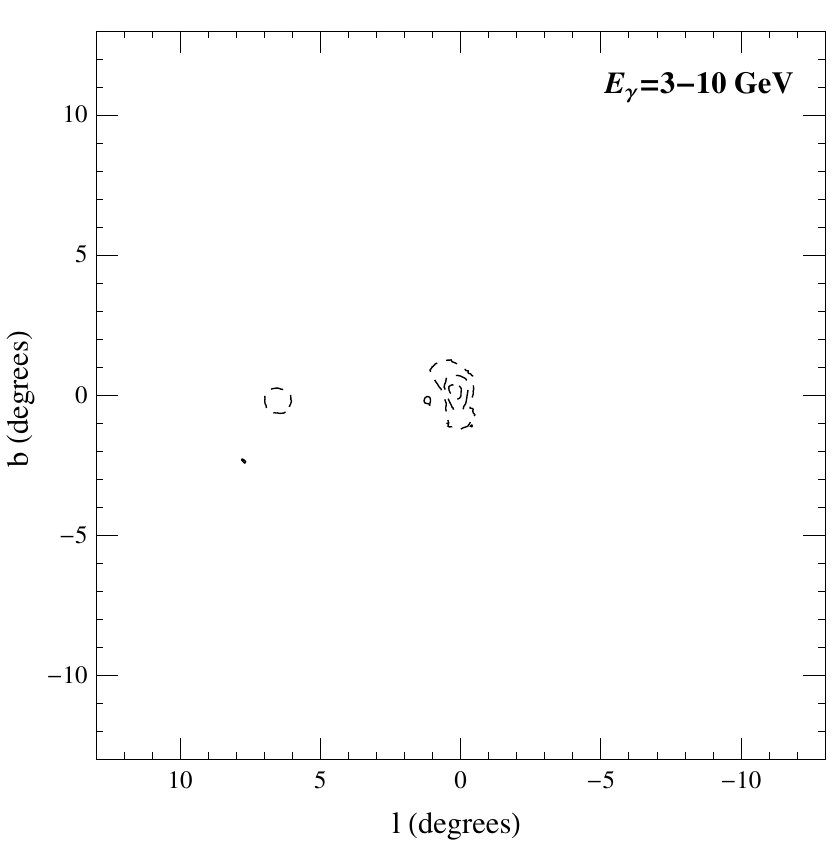}\\
\includegraphics[angle=0.0,width=1.86in]{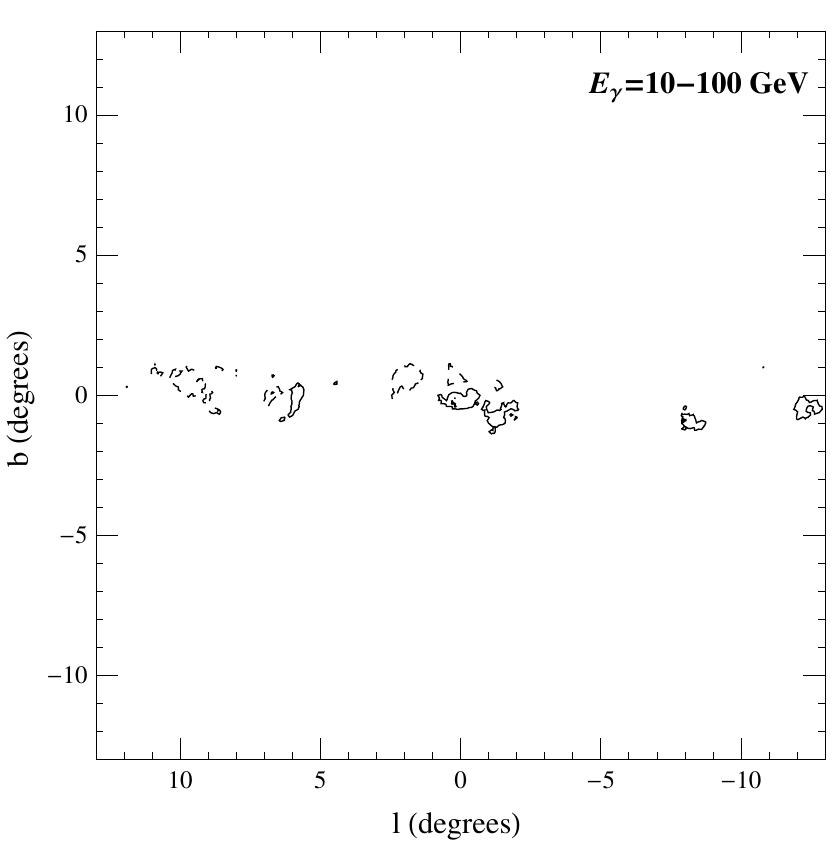}
\includegraphics[angle=0.0,width=1.86in]{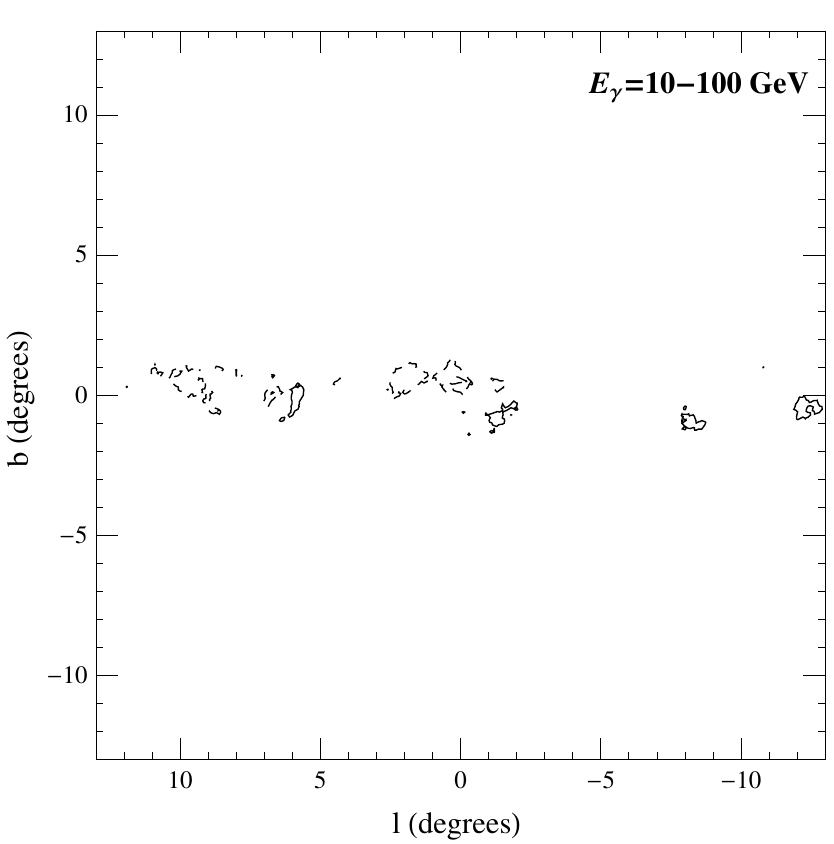}
\includegraphics[angle=0.0,width=1.86in]{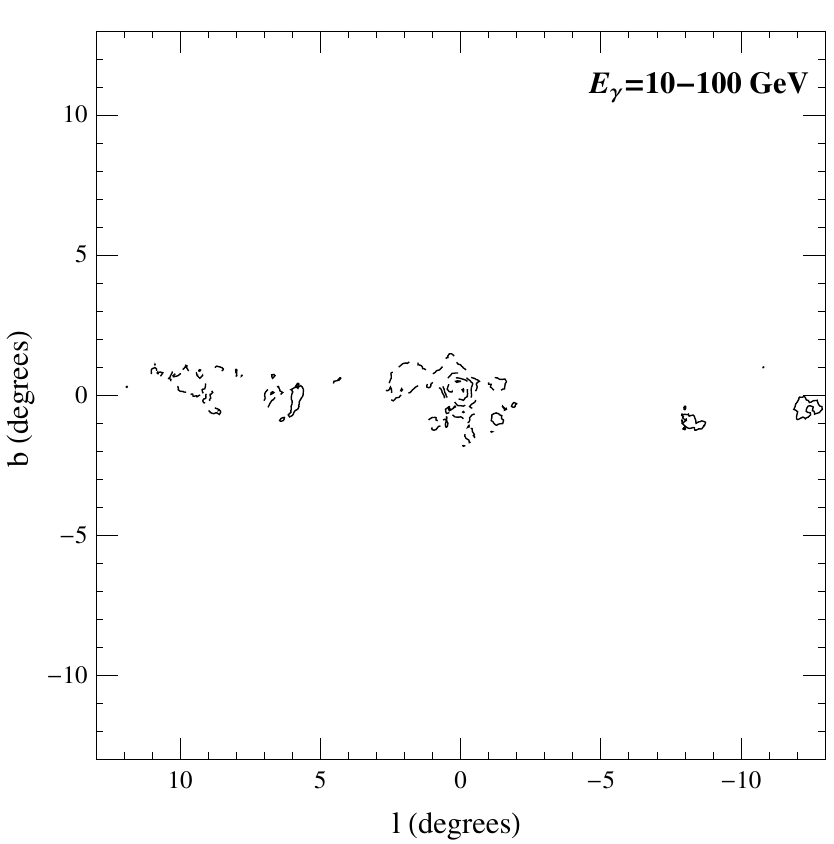}
\caption{Contour maps of the gamma-ray flux from the region surrounding the Galactic Center, after subtracting varying degrees of emission from dark matter with a mildly contracted (generalized NFW, $\gamma=1.2$) distribution. As the flux of dark matter annihilation products is increased (moving from left-to-right), regions of the maps become increasingly oversubtracted (denoted by dashed contours).}
\label{maps1pt2}
\end{figure*}

\begin{figure*}[t]
\centering
\includegraphics[angle=0.0,width=1.86in]{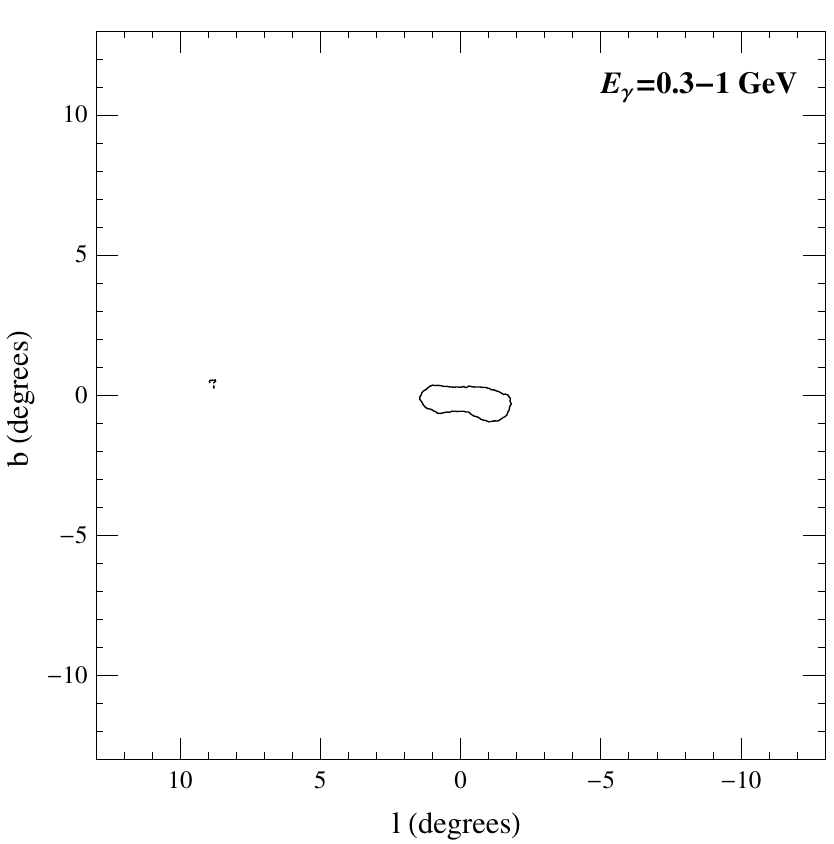}
\includegraphics[angle=0.0,width=1.86in]{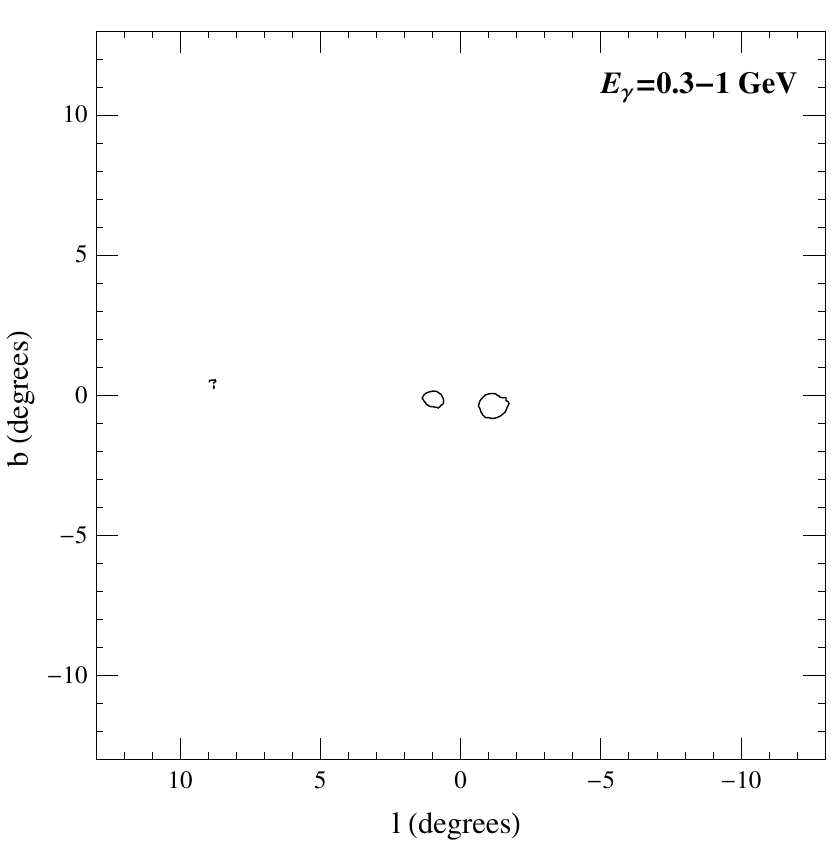}
\includegraphics[angle=0.0,width=1.86in]{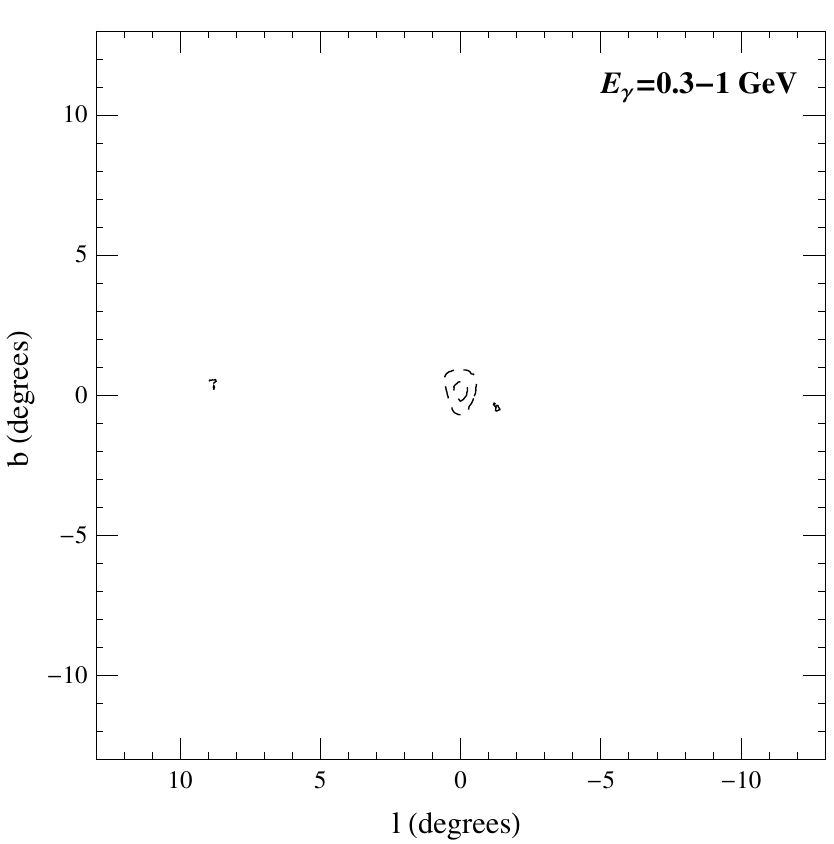}\\
\includegraphics[angle=0.0,width=1.86in]{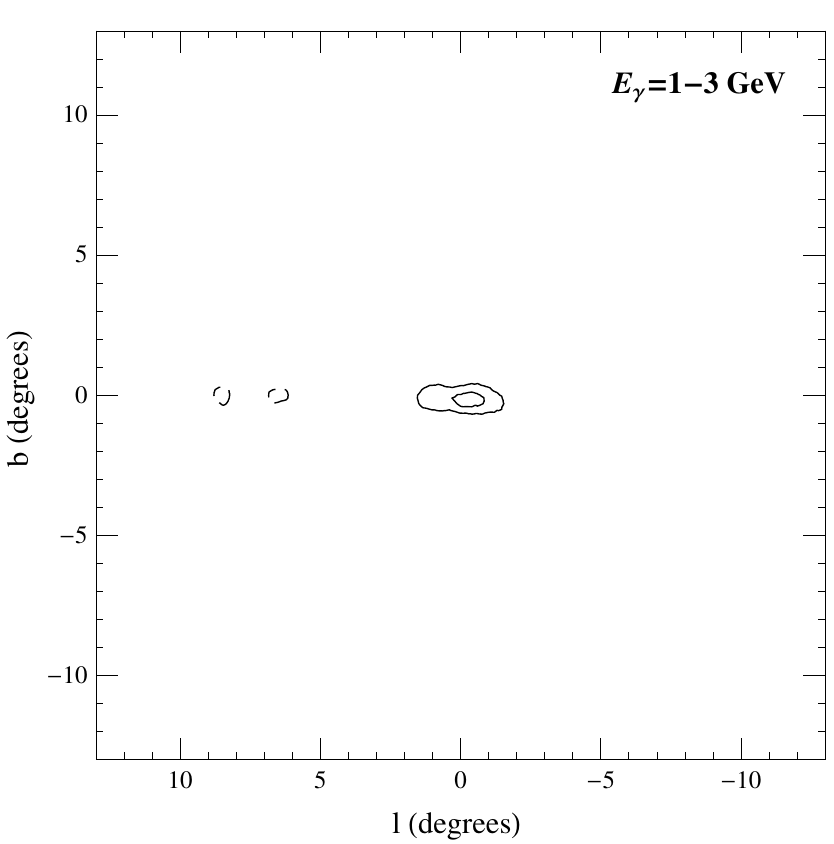}
\includegraphics[angle=0.0,width=1.86in]{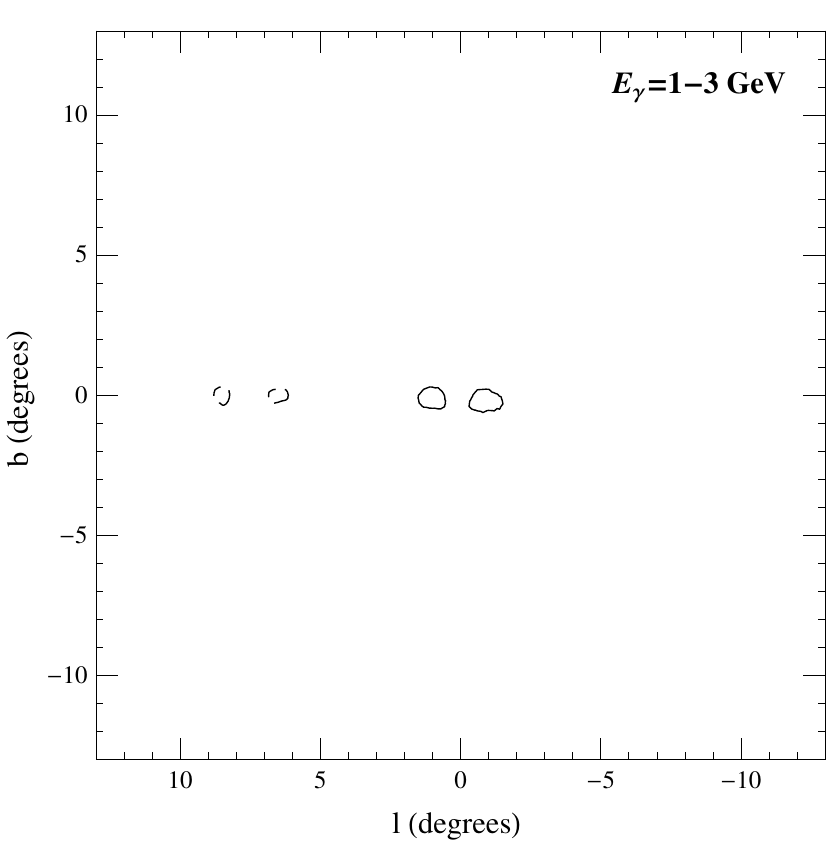}
\includegraphics[angle=0.0,width=1.86in]{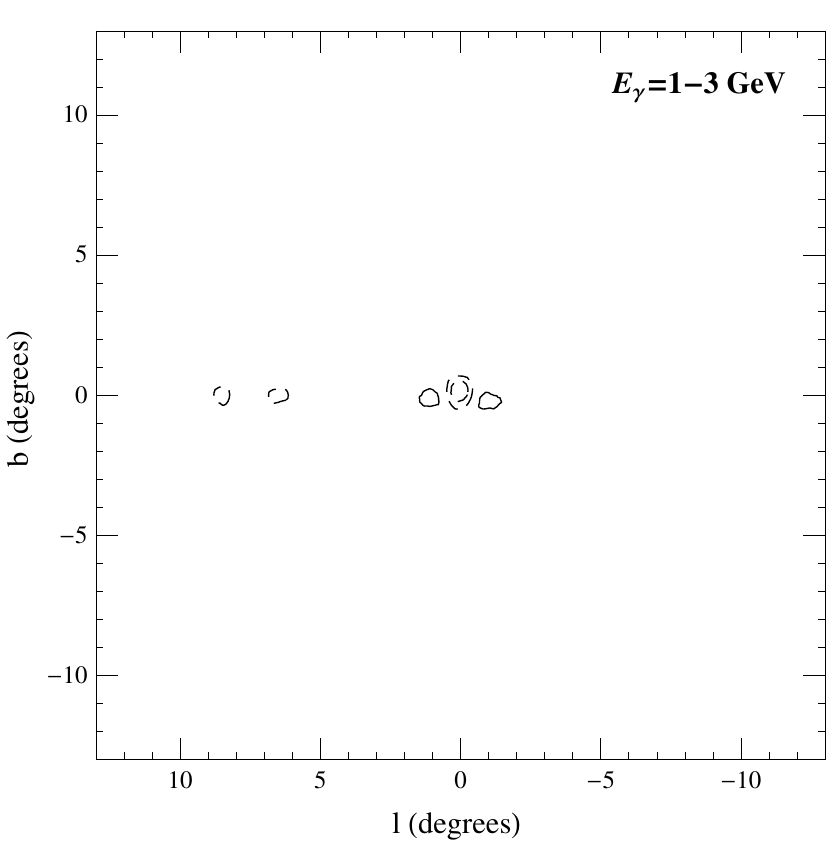}\\
\includegraphics[angle=0.0,width=1.86in]{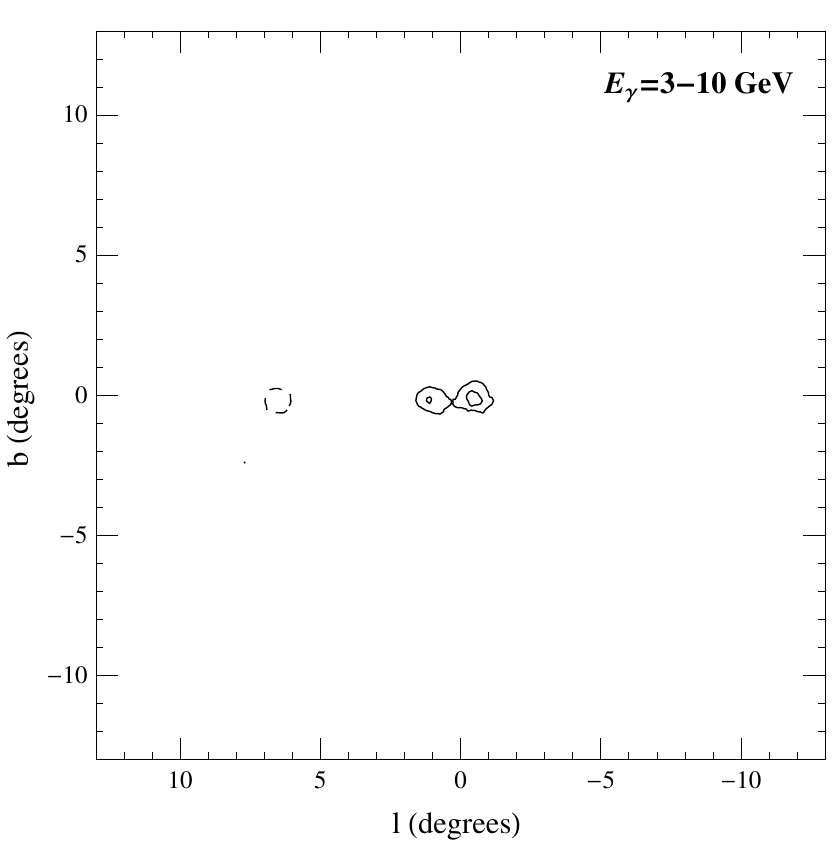}
\includegraphics[angle=0.0,width=1.86in]{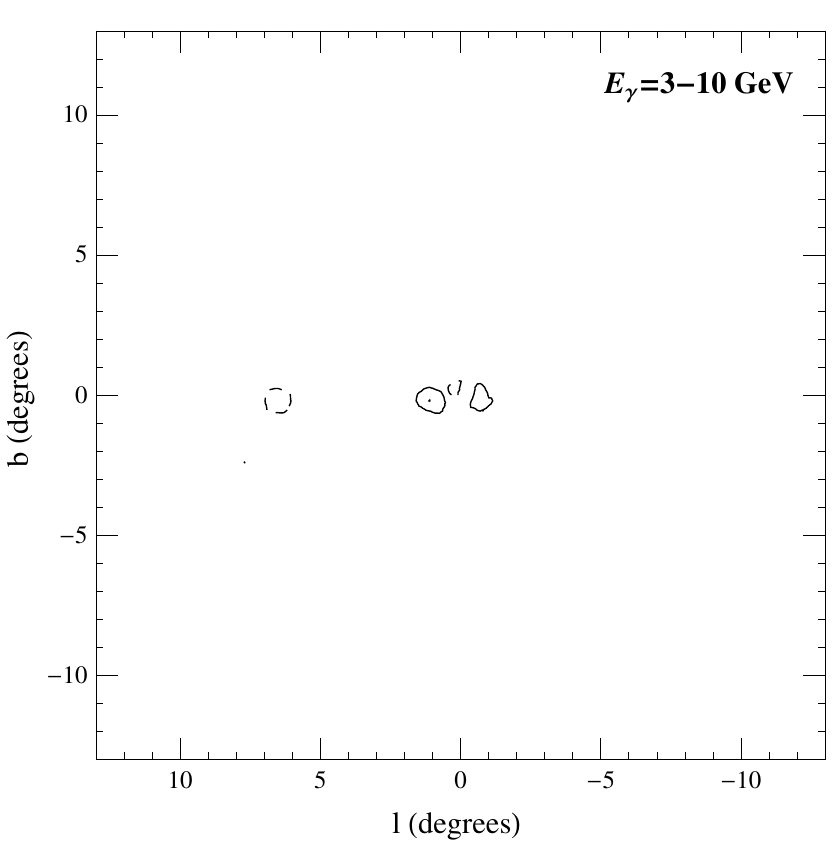}
\includegraphics[angle=0.0,width=1.86in]{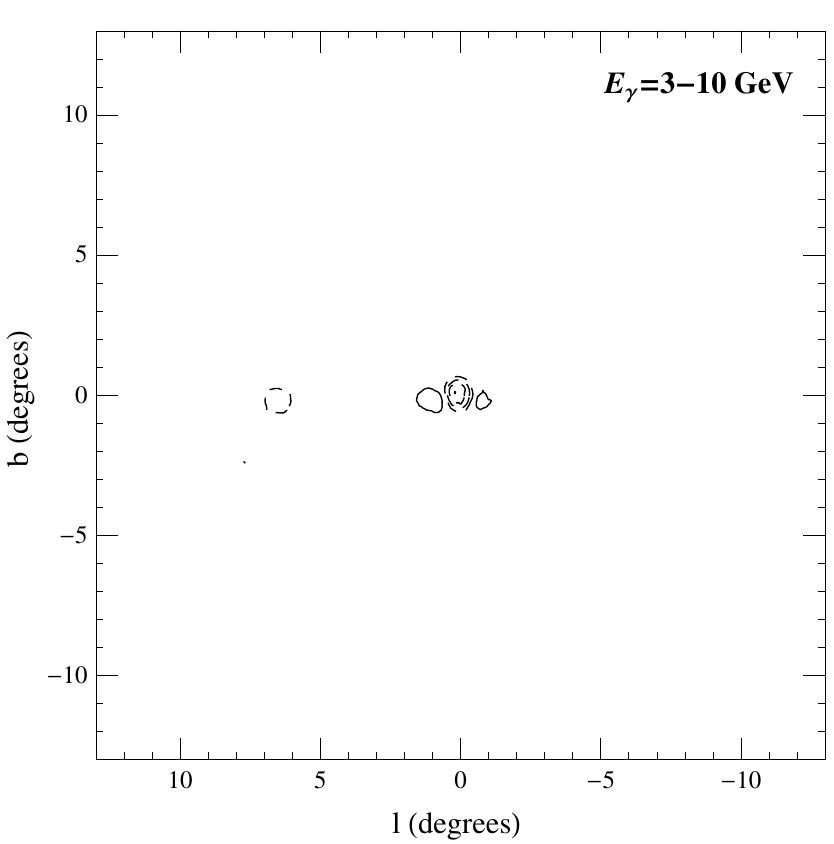}\\
\includegraphics[angle=0.0,width=1.86in]{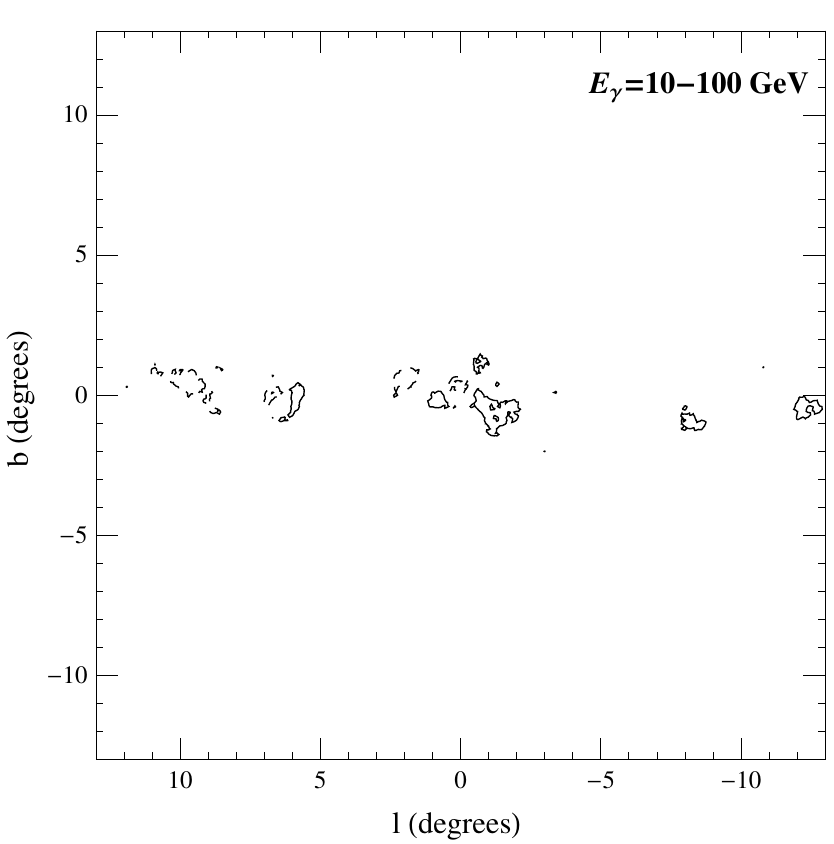}
\includegraphics[angle=0.0,width=1.86in]{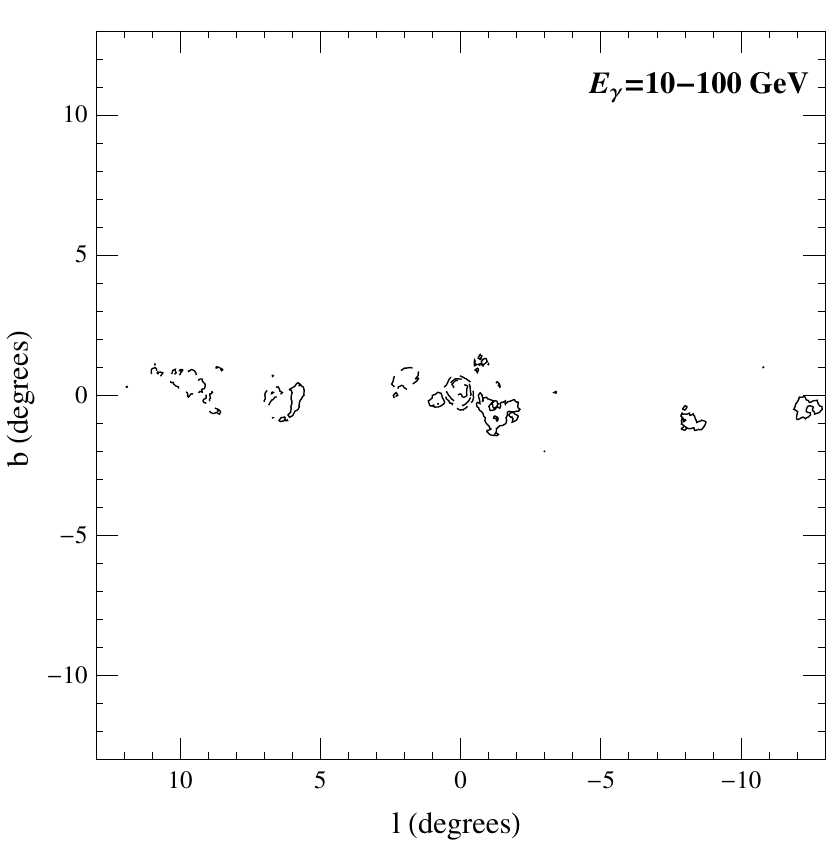}
\includegraphics[angle=0.0,width=1.86in]{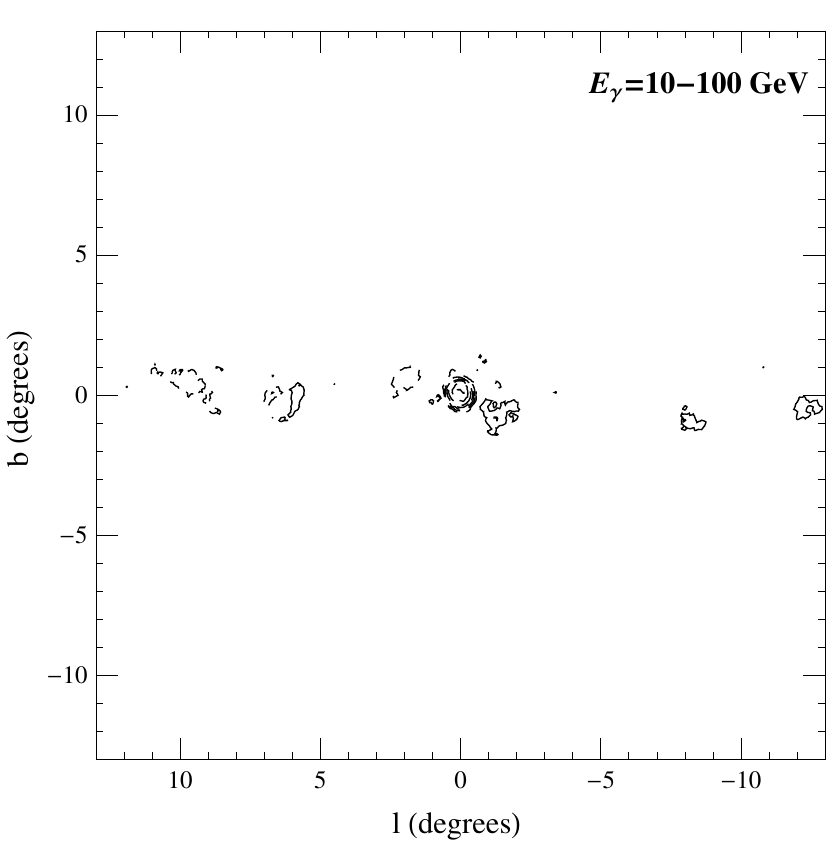}
\caption{Contour maps of the gamma-ray flux from the region surrounding the Galactic Center, after subtracting varying degrees of emission from dark matter with a significantly contracted (generalized NFW, $\gamma=1.4$) distribution. As the flux of dark matter annihilation products is increased (moving from left-to-right), regions of the maps become increasingly oversubtracted (denoted by dashed contours).}
\label{maps1pt4}
\end{figure*}

\begin{figure*}[t]
\centering
\includegraphics[angle=0.0,width=1.86in]{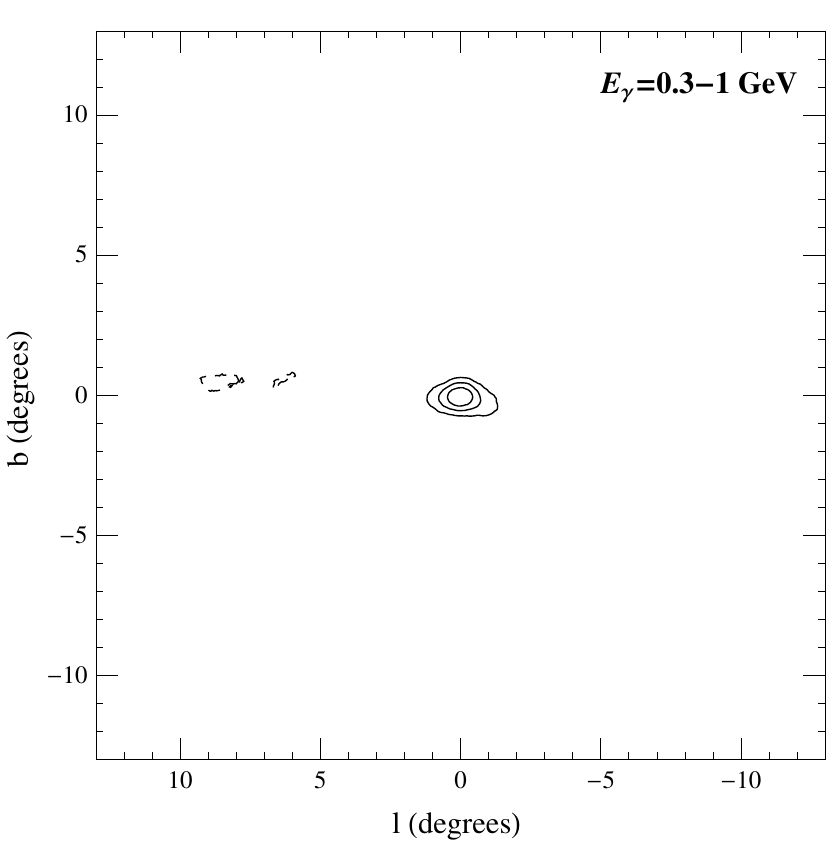}
\includegraphics[angle=0.0,width=1.86in]{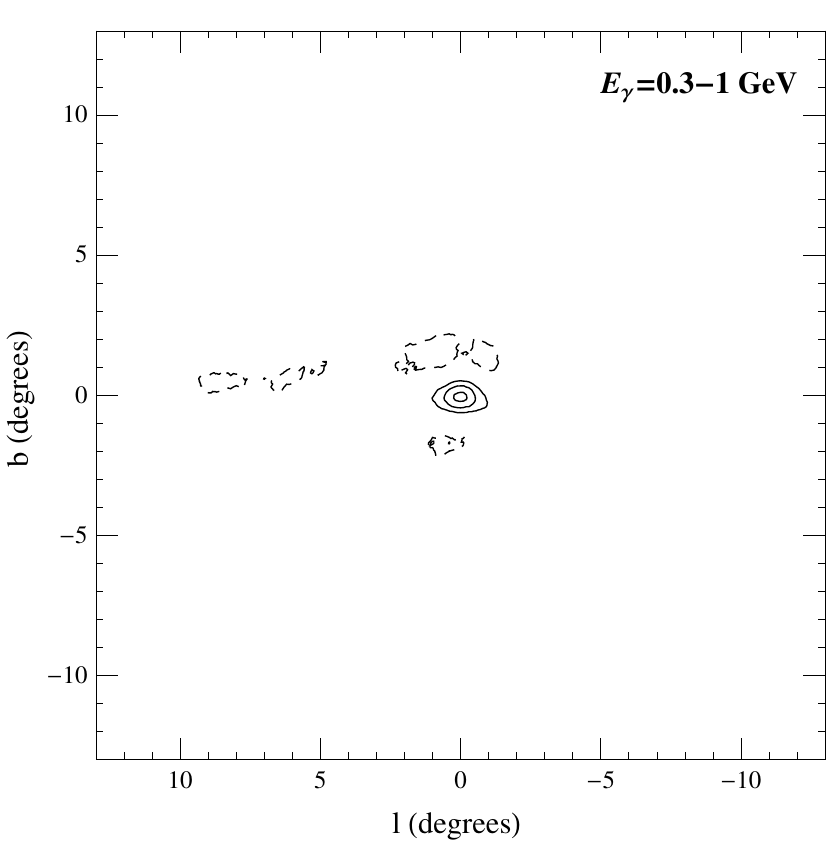}
\includegraphics[angle=0.0,width=1.86in]{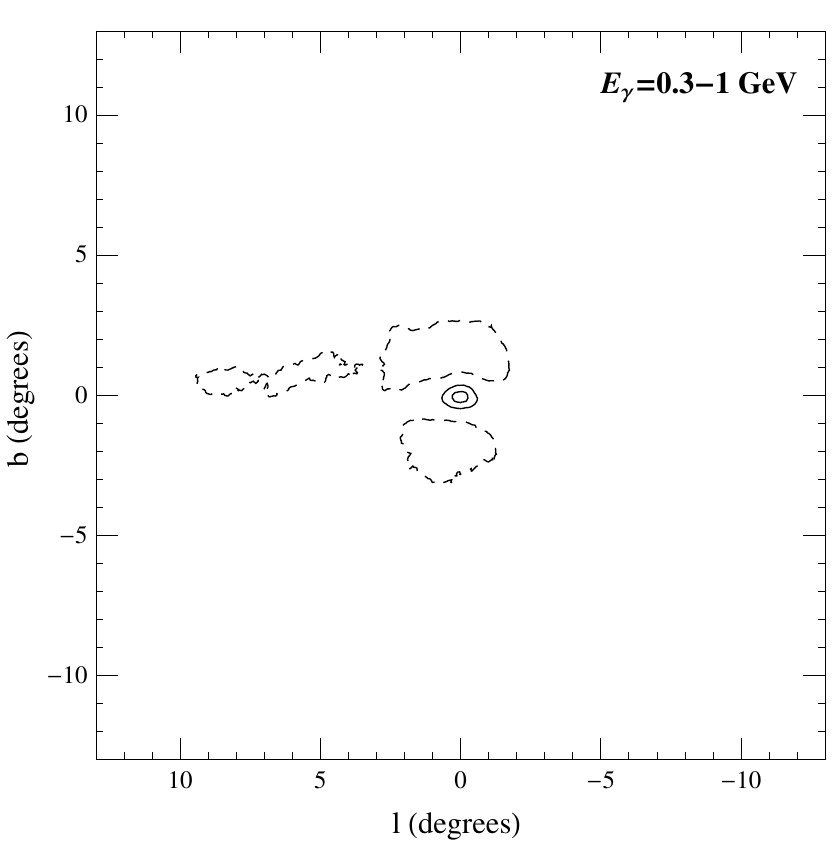}\\
\includegraphics[angle=0.0,width=1.86in]{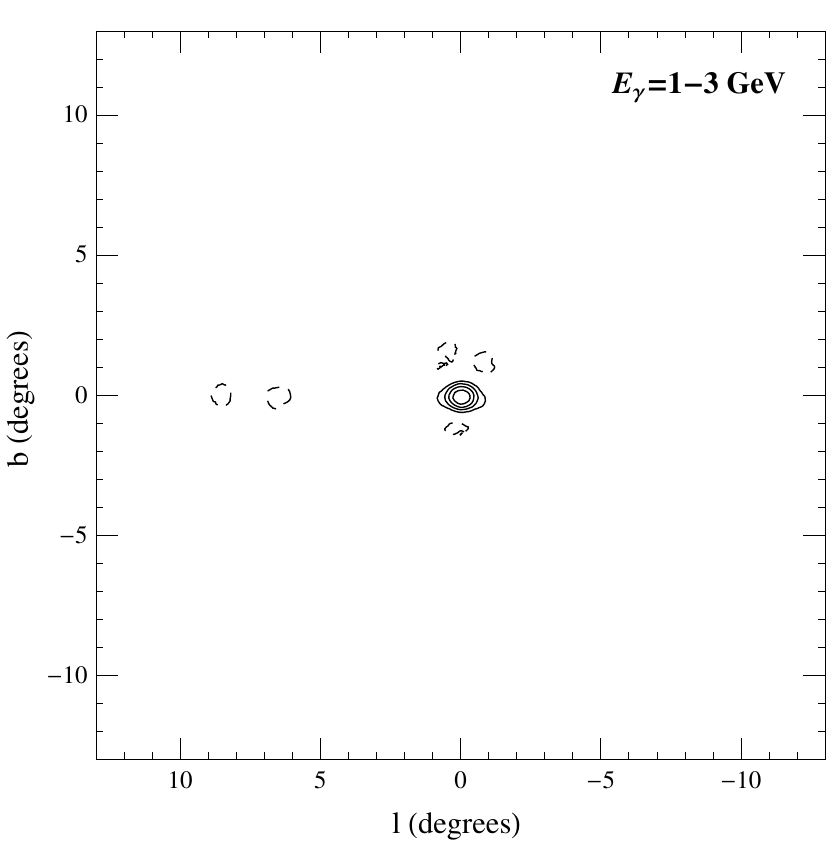}
\includegraphics[angle=0.0,width=1.86in]{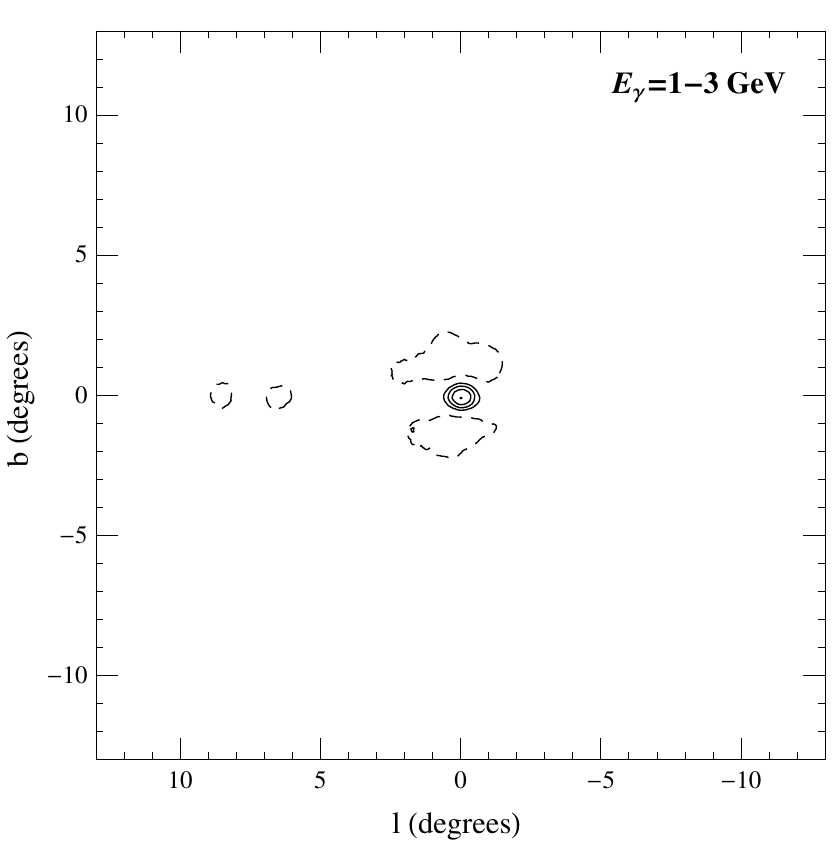}
\includegraphics[angle=0.0,width=1.86in]{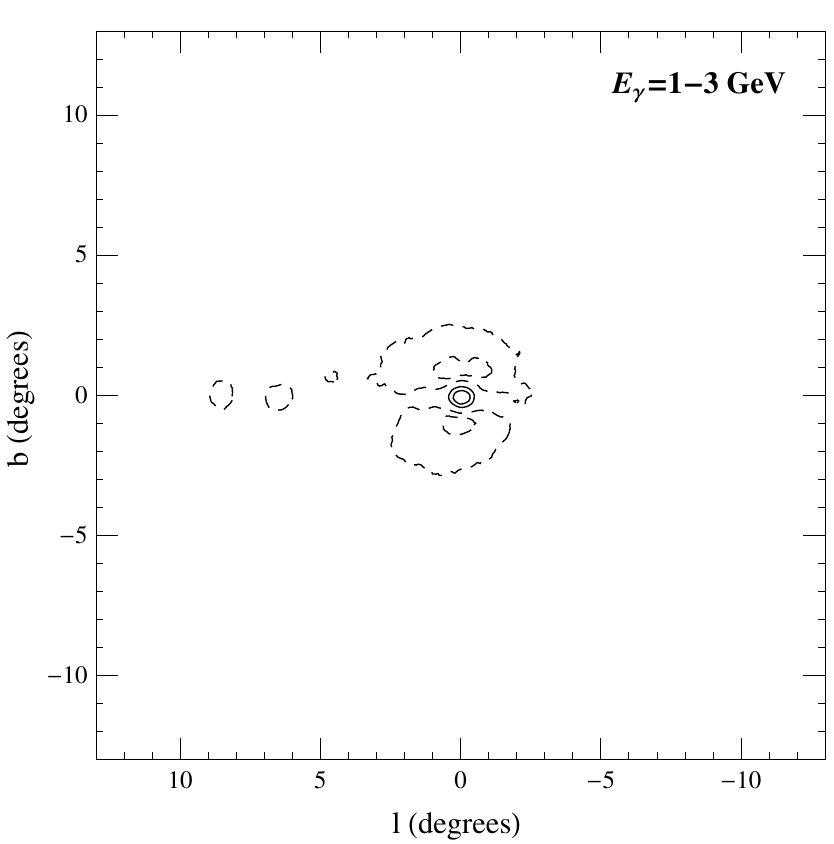}\\
\includegraphics[angle=0.0,width=1.86in]{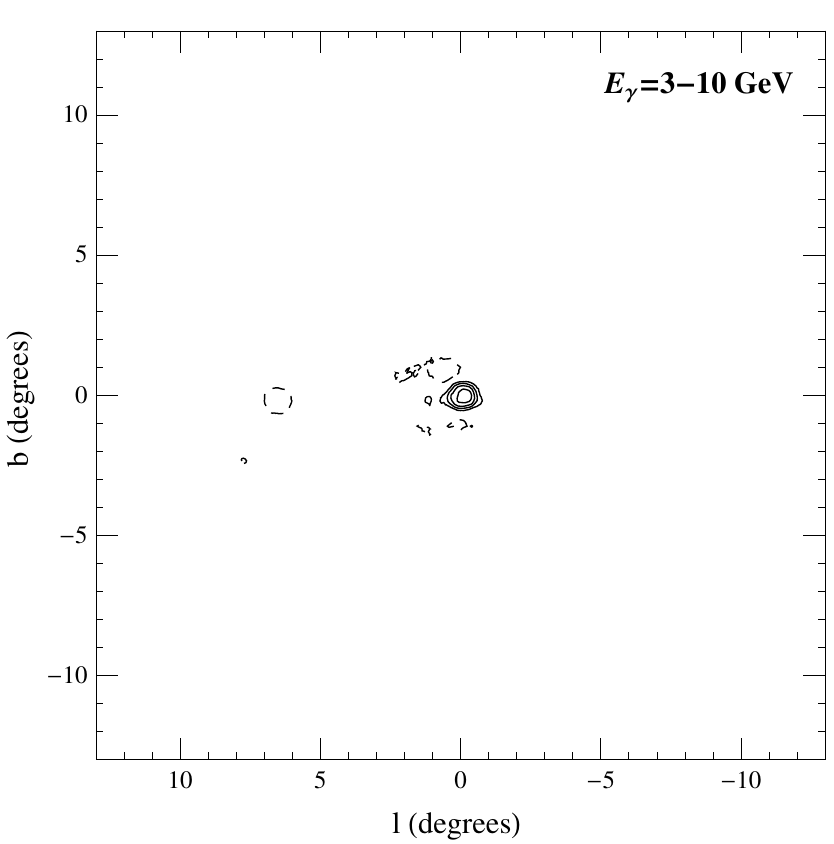}
\includegraphics[angle=0.0,width=1.86in]{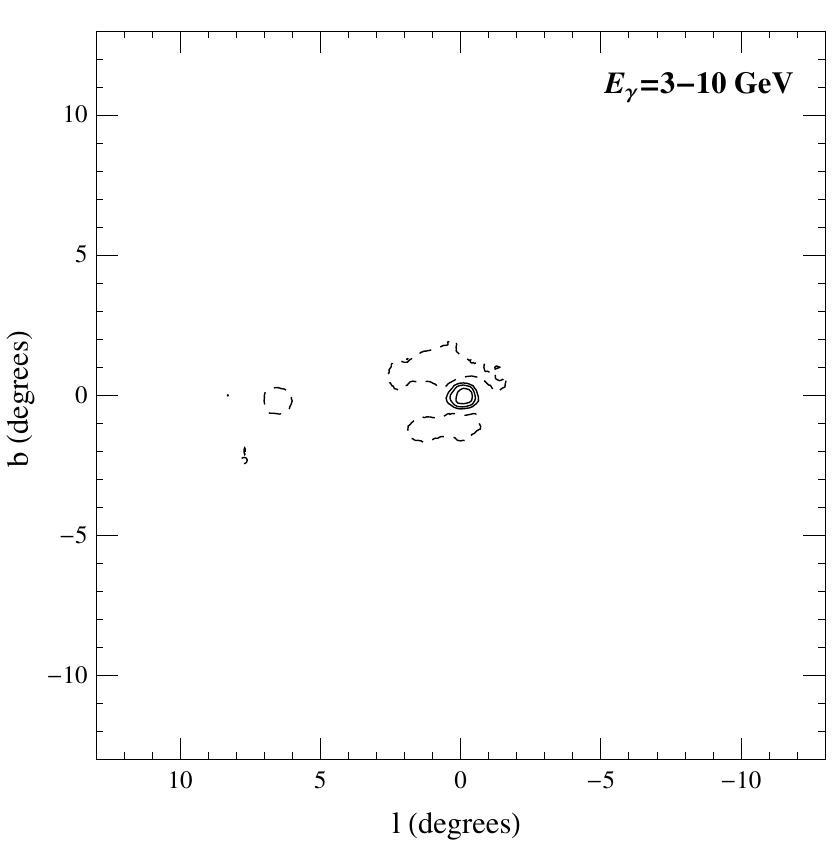}
\includegraphics[angle=0.0,width=1.86in]{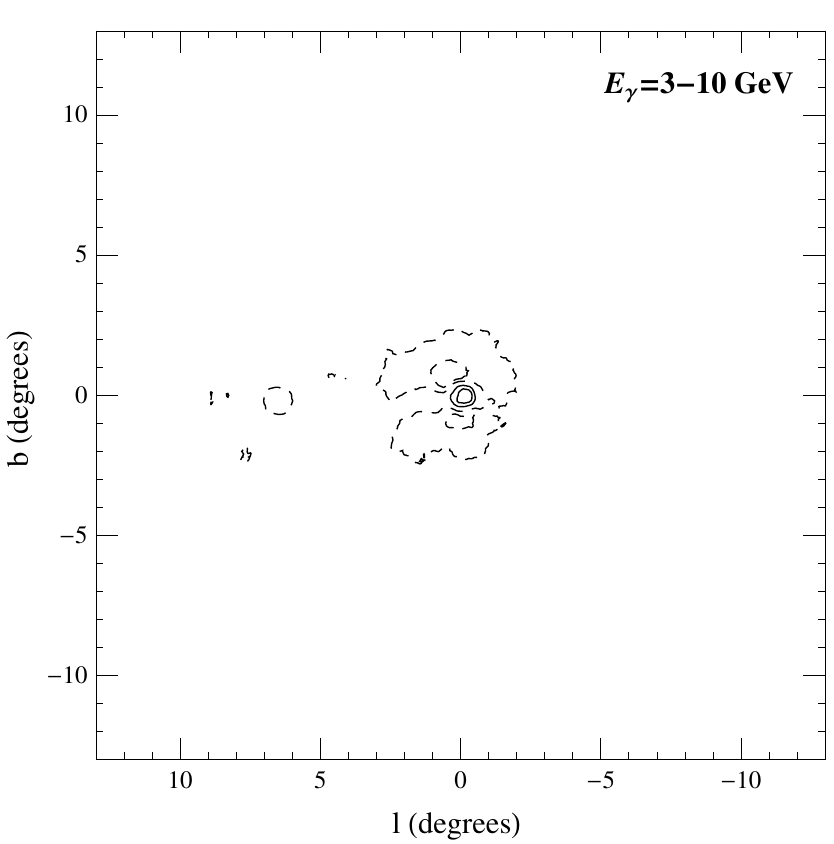}\\
\includegraphics[angle=0.0,width=1.86in]{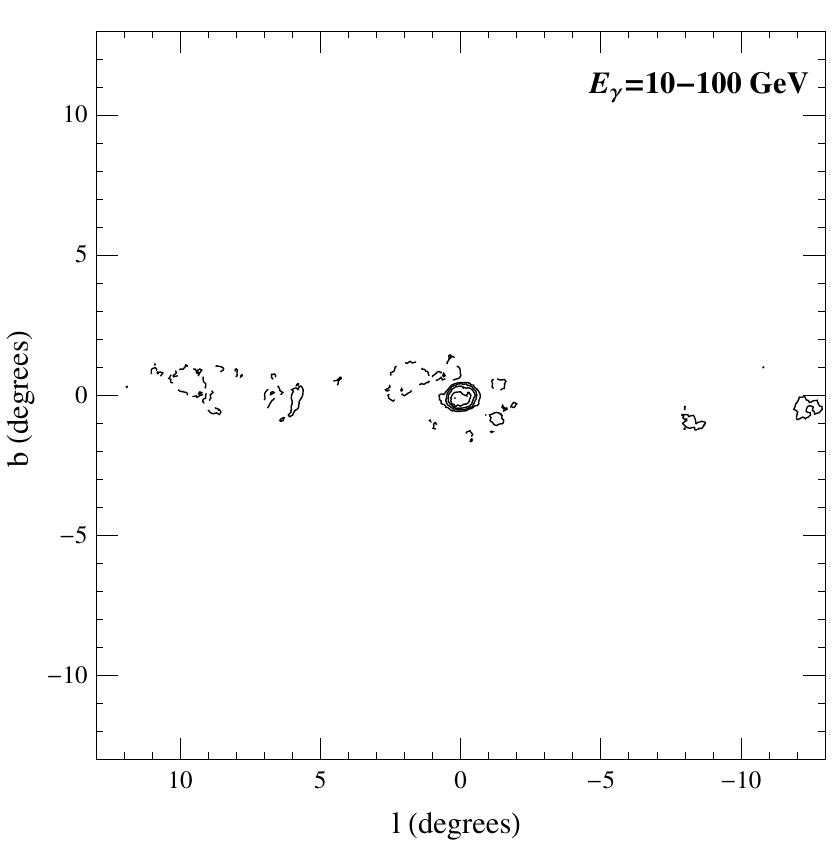}
\includegraphics[angle=0.0,width=1.86in]{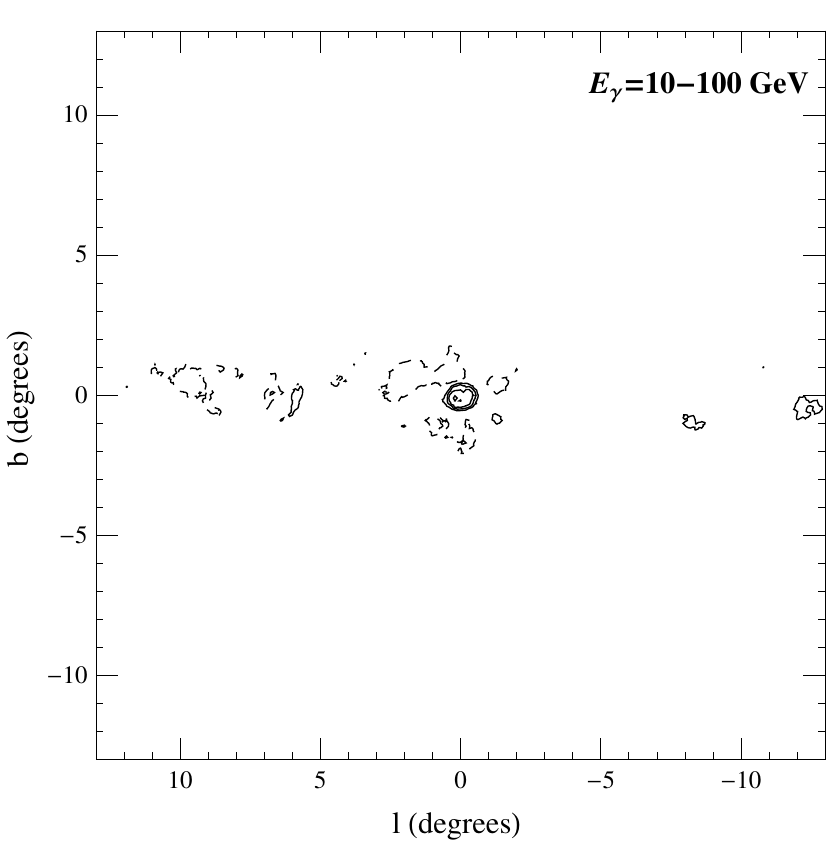}
\includegraphics[angle=0.0,width=1.86in]{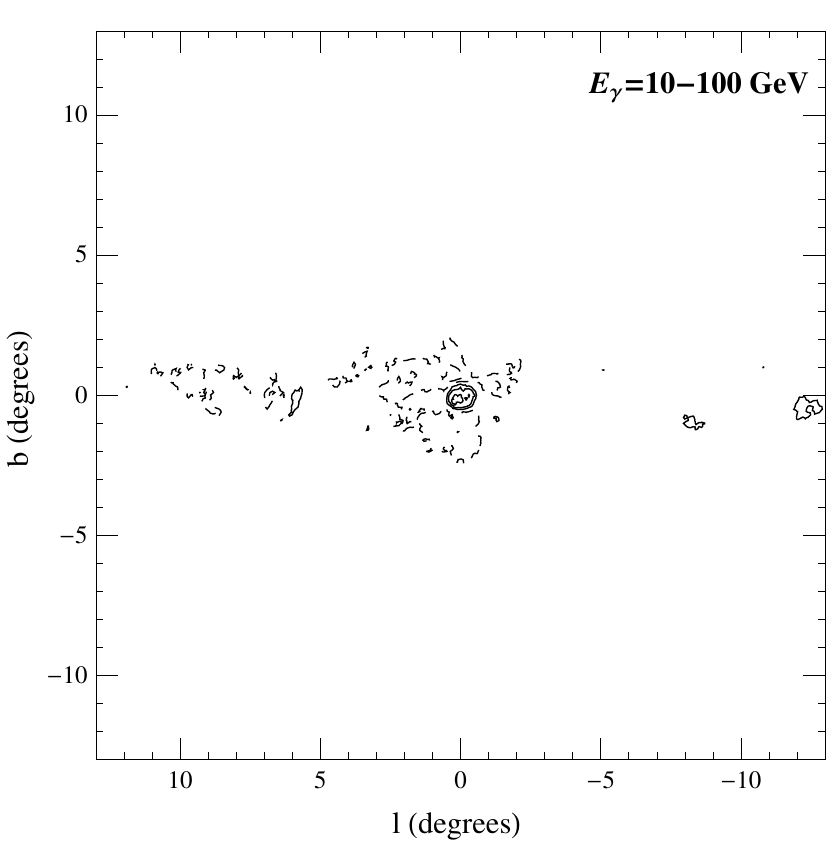}
\caption{Contour maps of the gamma-ray flux from the region surrounding the Galactic Center, after subtracting varying degrees of emission from dark matter distributed according to an NFW profile with a 100 parsec, constant-density core. As the flux of dark matter annihilation products is increased (moving from left-to-right), regions of the maps become increasingly oversubtracted (denoted by dashed contours).}
\label{maps100pccore}
\end{figure*}

\begin{figure*}[t]
\centering
\includegraphics[angle=0.0,width=1.86in]{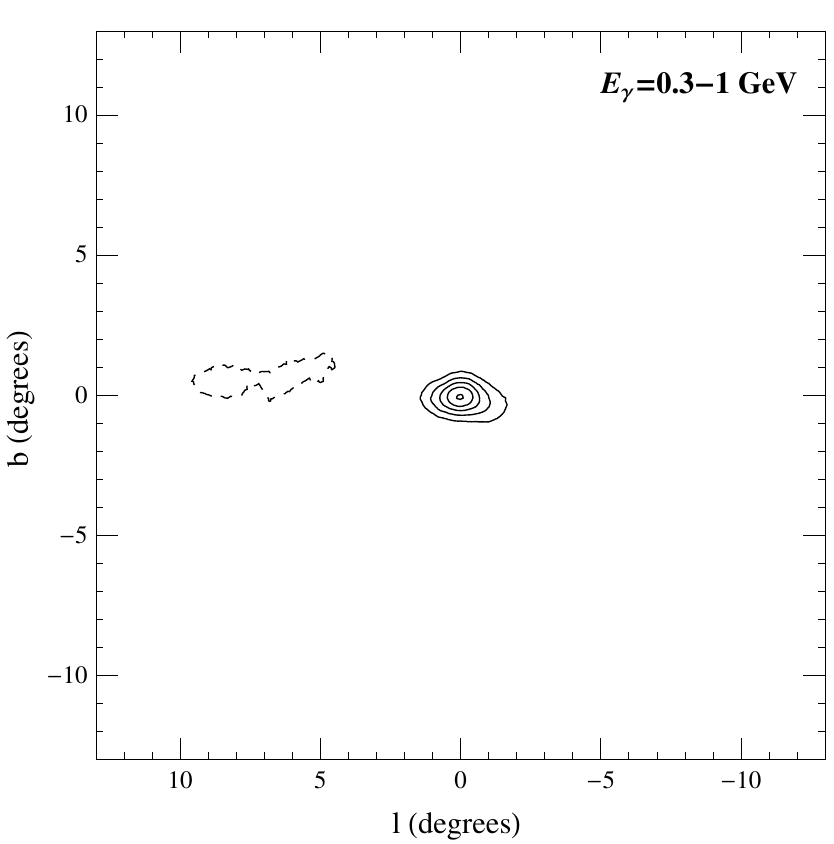}
\includegraphics[angle=0.0,width=1.86in]{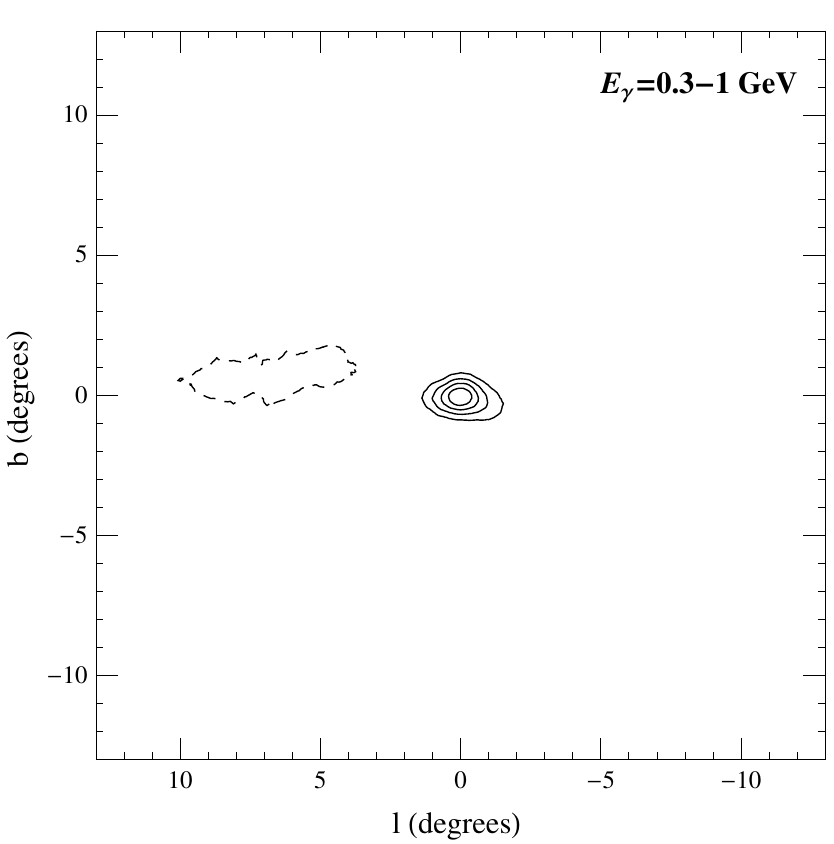}
\includegraphics[angle=0.0,width=1.86in]{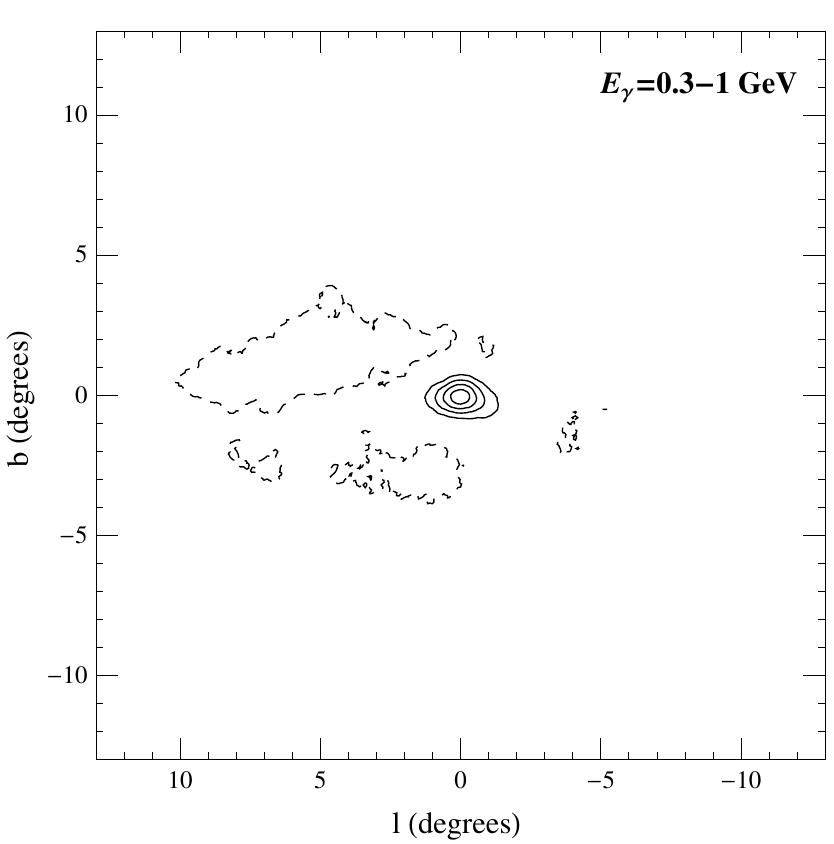}\\
\includegraphics[angle=0.0,width=1.86in]{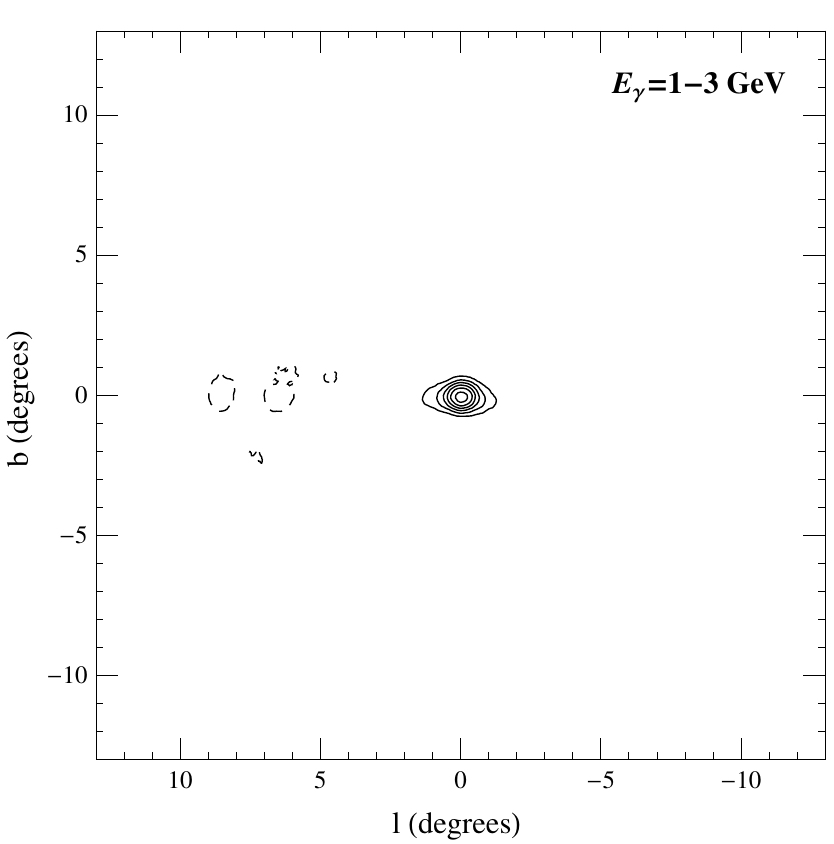}
\includegraphics[angle=0.0,width=1.86in]{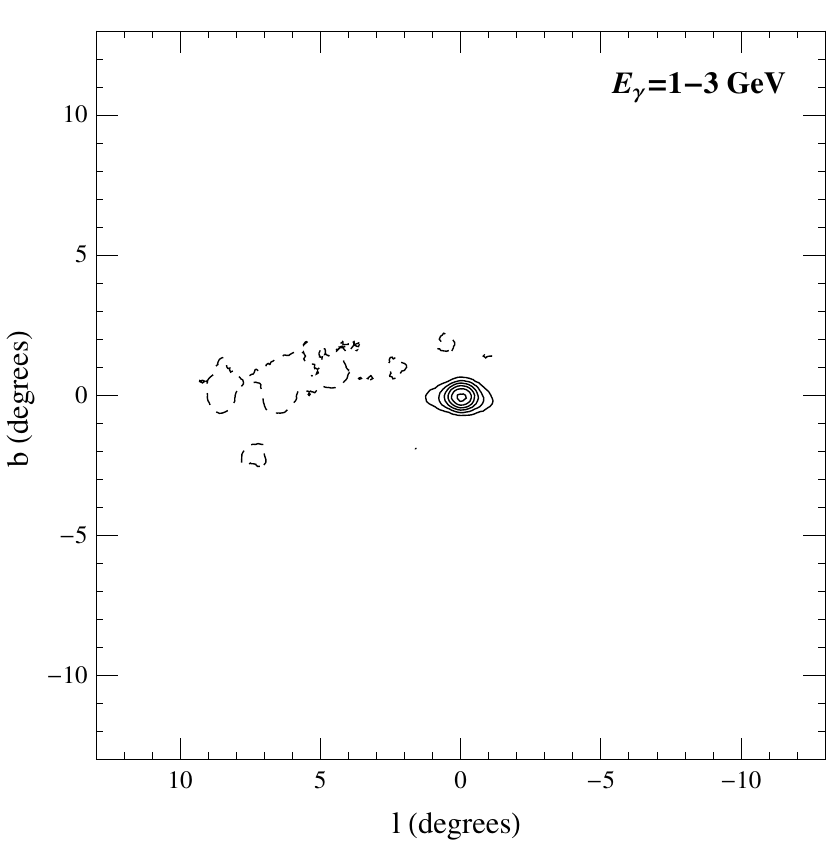}
\includegraphics[angle=0.0,width=1.86in]{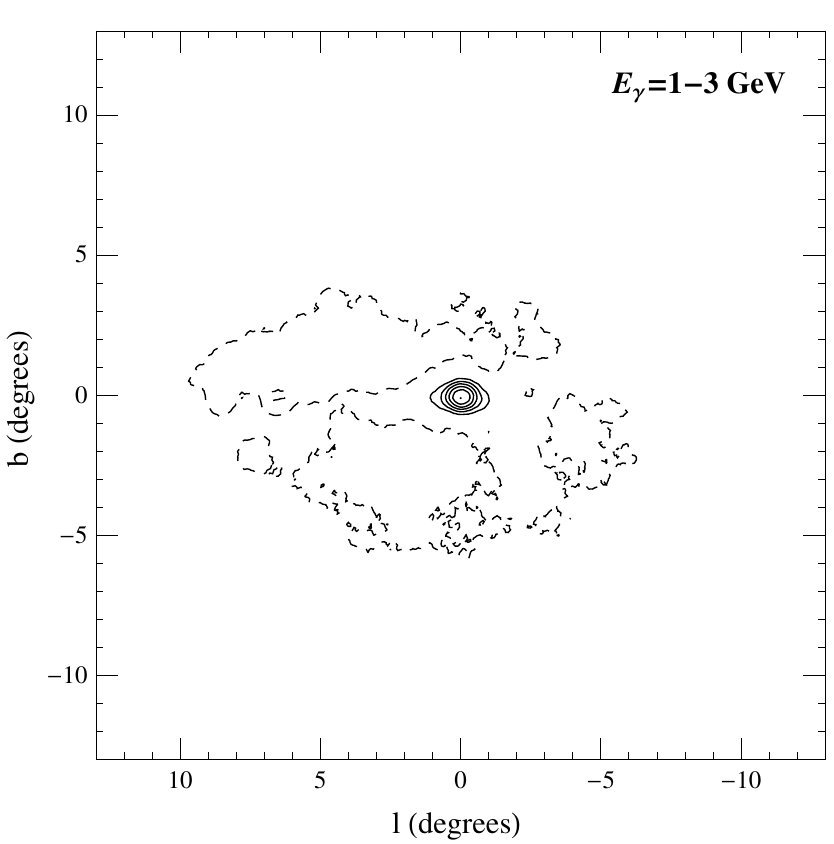}\\
\includegraphics[angle=0.0,width=1.86in]{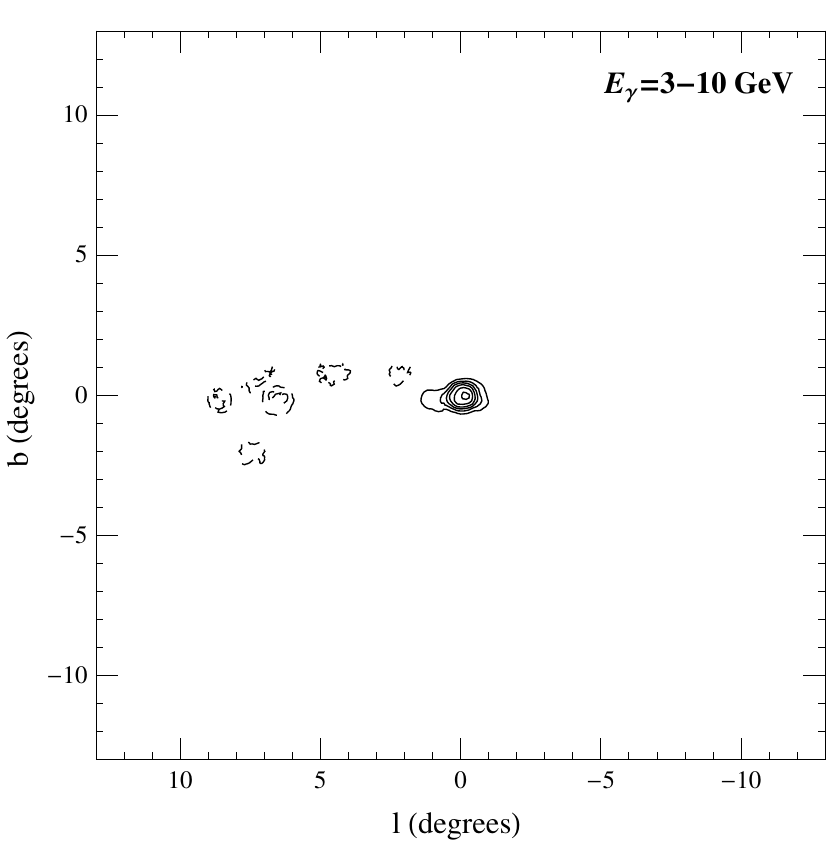}
\includegraphics[angle=0.0,width=1.86in]{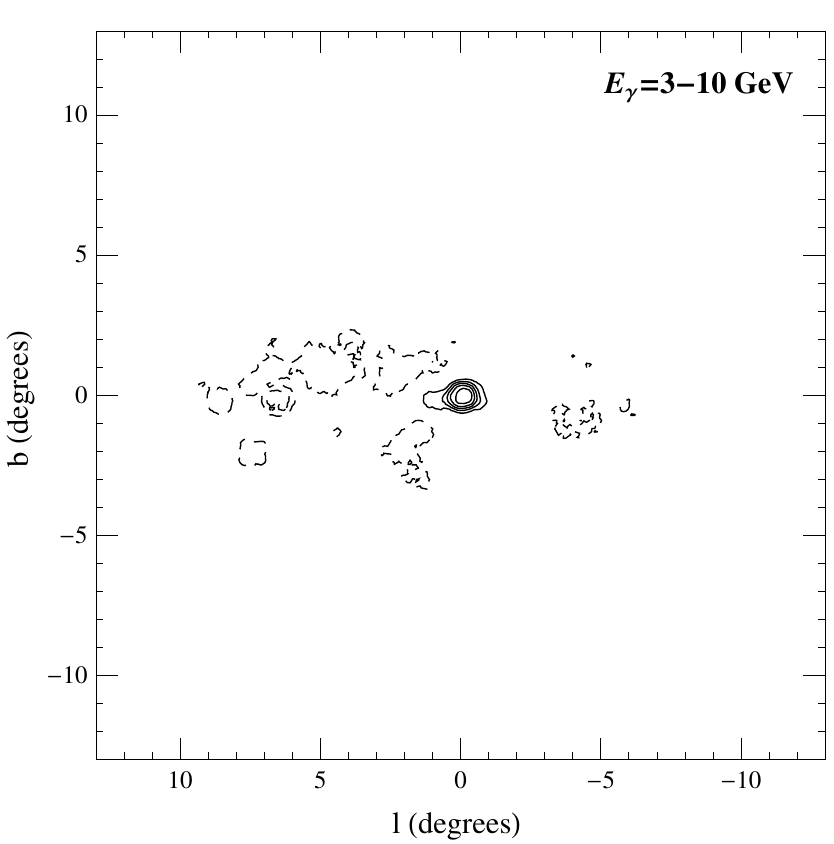}
\includegraphics[angle=0.0,width=1.86in]{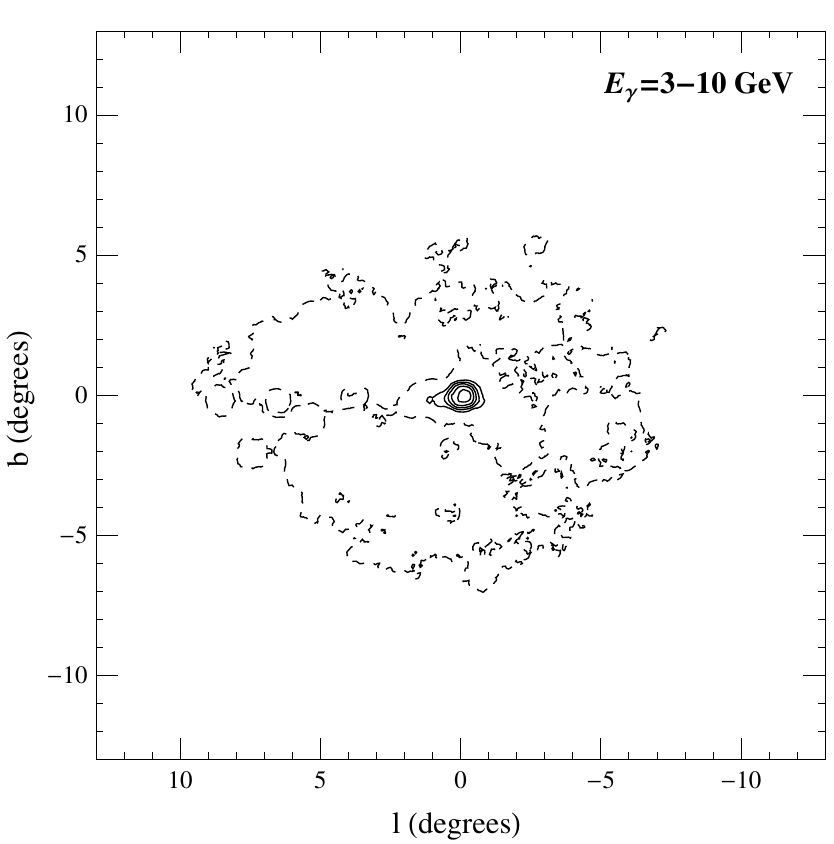}\\
\includegraphics[angle=0.0,width=1.86in]{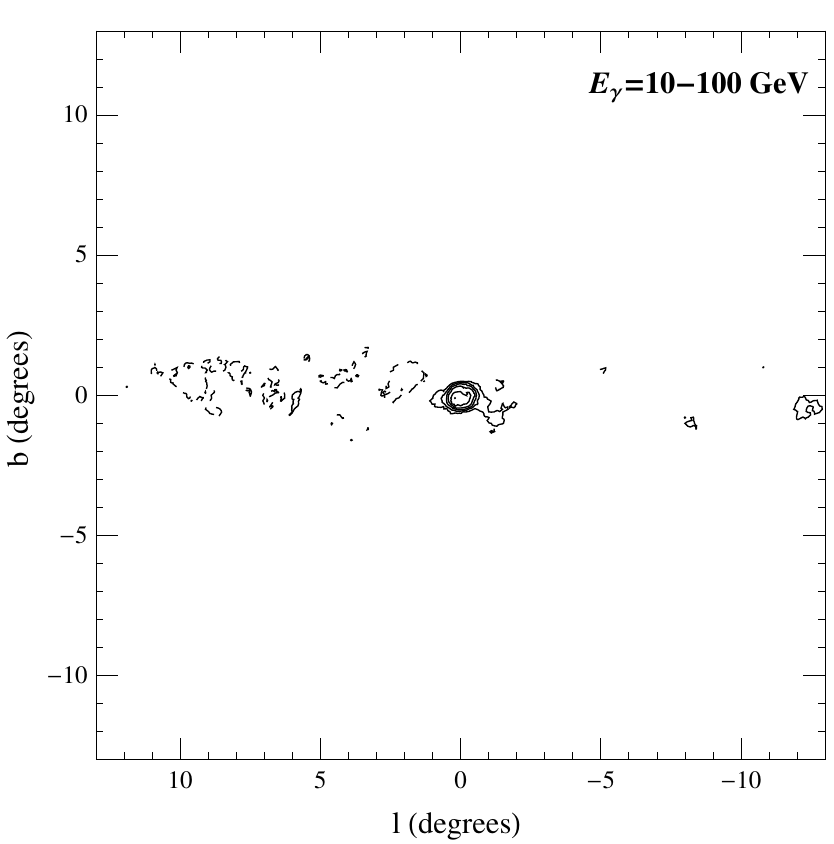}
\includegraphics[angle=0.0,width=1.86in]{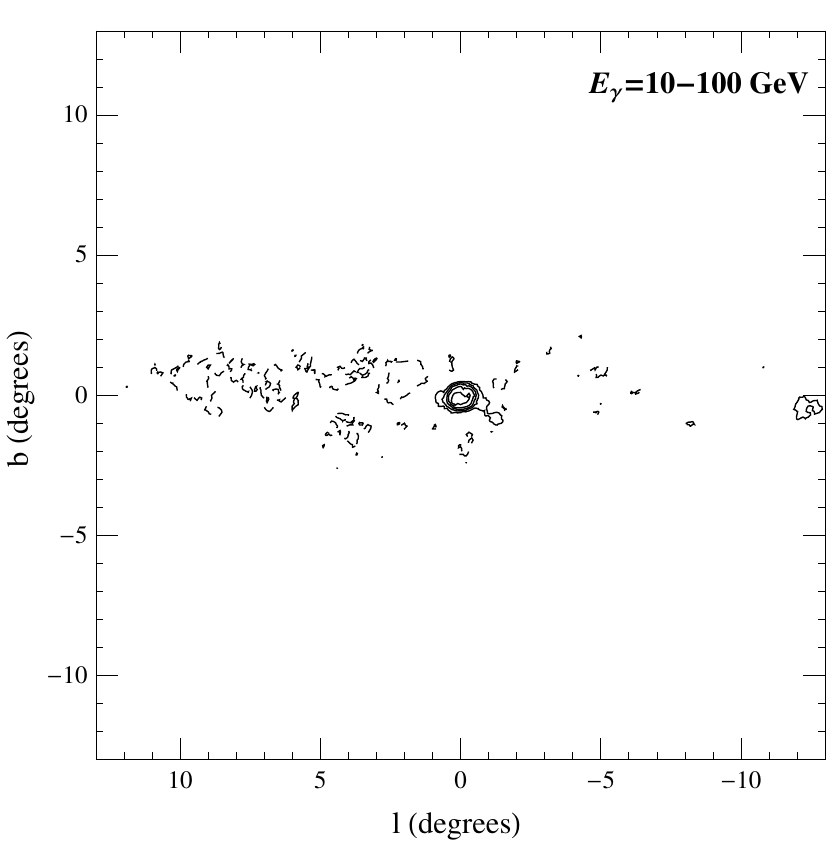}
\includegraphics[angle=0.0,width=1.86in]{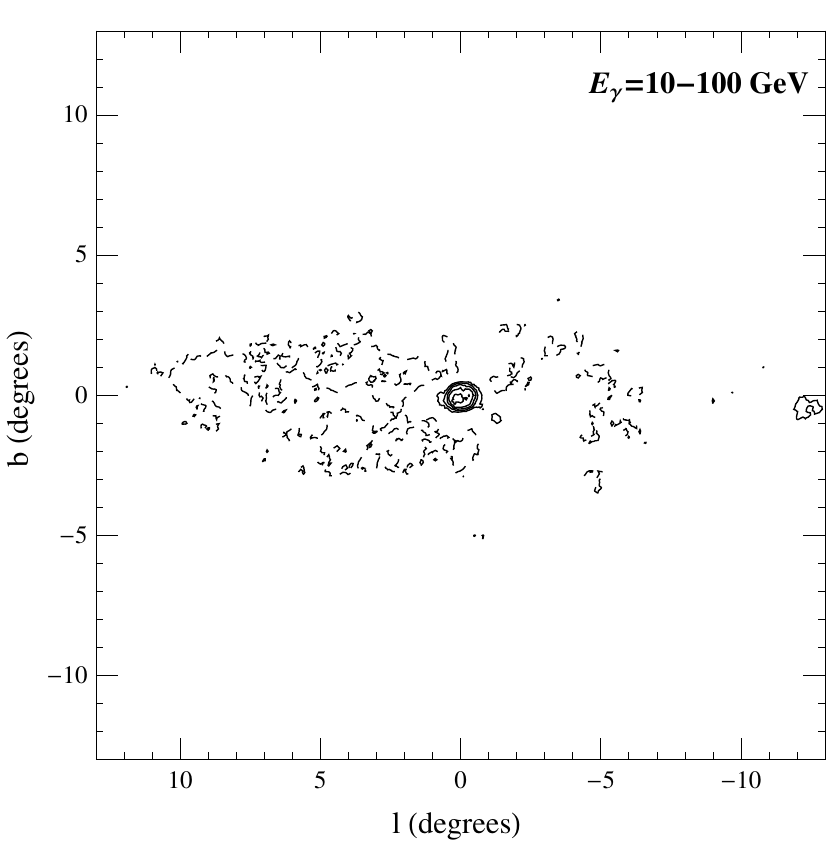}
\caption{Contour maps of the gamma-ray flux from the region surrounding the Galactic Center, after subtracting varying degrees of emission from dark matter distributed according to an NFW profile with a 1 kiloparsec, constant-density core. As the flux of dark matter annihilation products is increased (moving from left-to-right), regions of the maps become increasingly oversubtracted (denoted by dashed contours).}
\label{mapskpccore}
\end{figure*}

In this section, we derive constraints on the dark matter's annihilation cross section for each of the halo profiles described in the previous section. To do this, we generate maps of the quantity $J$, convolved with the point spread function of Fermi, and again smoothed at the scale of 0.5$^{\circ}$. We then subtract these templates from the maps shown in the right frames of Fig.~\ref{maps} with varying degrees of normalization. The results are shown in Figs.~\ref{maps2}-\ref{mapskpccore}. In each case, the left and right frames include a contribution from dark matter annihilations which is 40\% less or more than in the center frame. 

As we increase the flux of dark matter annihilation products in a given energy range, eventually regions of the map begin to be oversubtracted (fluxes with negative values, as denoted by dashed contours). In the 300-1000 MeV range shown in the top frames of Fig.~\ref{maps2}, for example, we see significantly oversubtracted regions in both the middle and right frames. In either of these two maps, the oversubtracted regions are too large to be statistical fluctuations (no variations nearly this large are observed over the rest of the 40$^{\circ}\times 40^{\circ}$ region we have studied). From this procedure, we can determine the maximum flux of gamma-rays from dark matter annihilations in each energy range, and convert this into an upper limit on the dark matter annihilation cross section, as a function of mass and dominant annihilation channel.\footnote{Quantitatively, we determine the upper limit by finding the normalization at which we oversubtract a region that exceeds the scale of fluctuations along the disk at more than the 95\% confidence level. For example, if the residual found in a region near the Galactic Center after subtracting a dark matter template is more negative than 95\% of all regions of the same angular size along the plane (over the range $|l|=5^{\circ}-20^{\circ}$), we consider it to be oversubtracted at the 95\% confidence level. This allows us to empirically take into account the frequency at which fluctuations of a given magnitude occur among the relevant backgrounds.}

\begin{figure*}[t]
\centering
\includegraphics[angle=0.0,width=7.in]{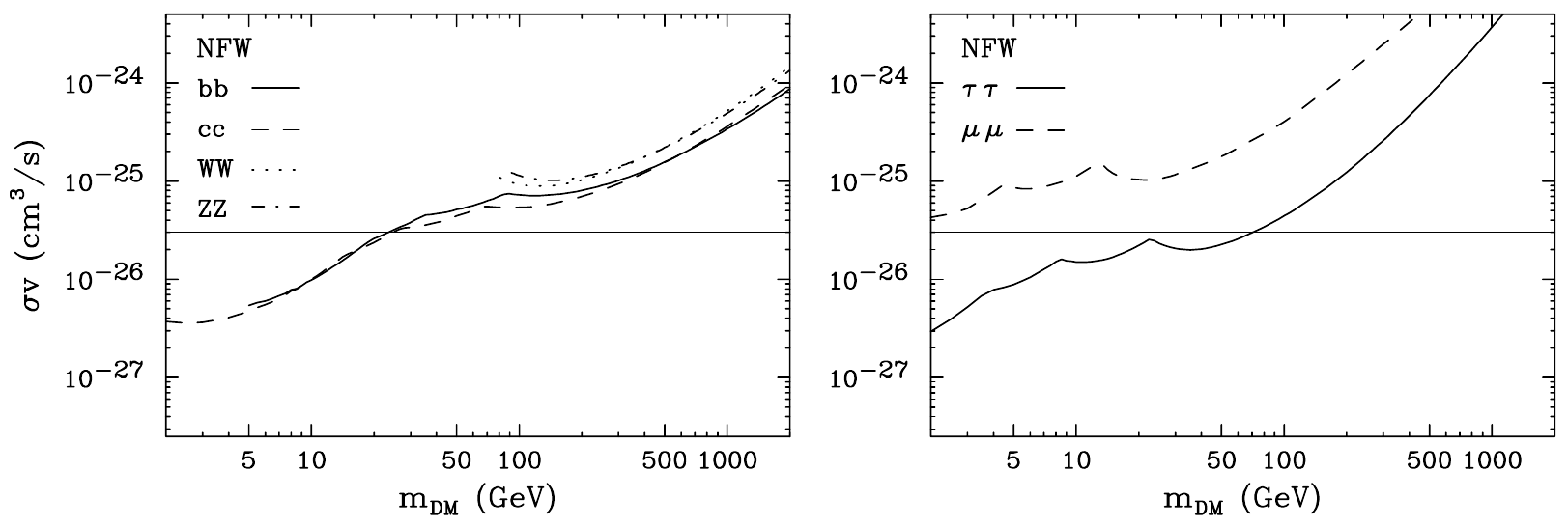}
\includegraphics[angle=0.0,width=7.in]{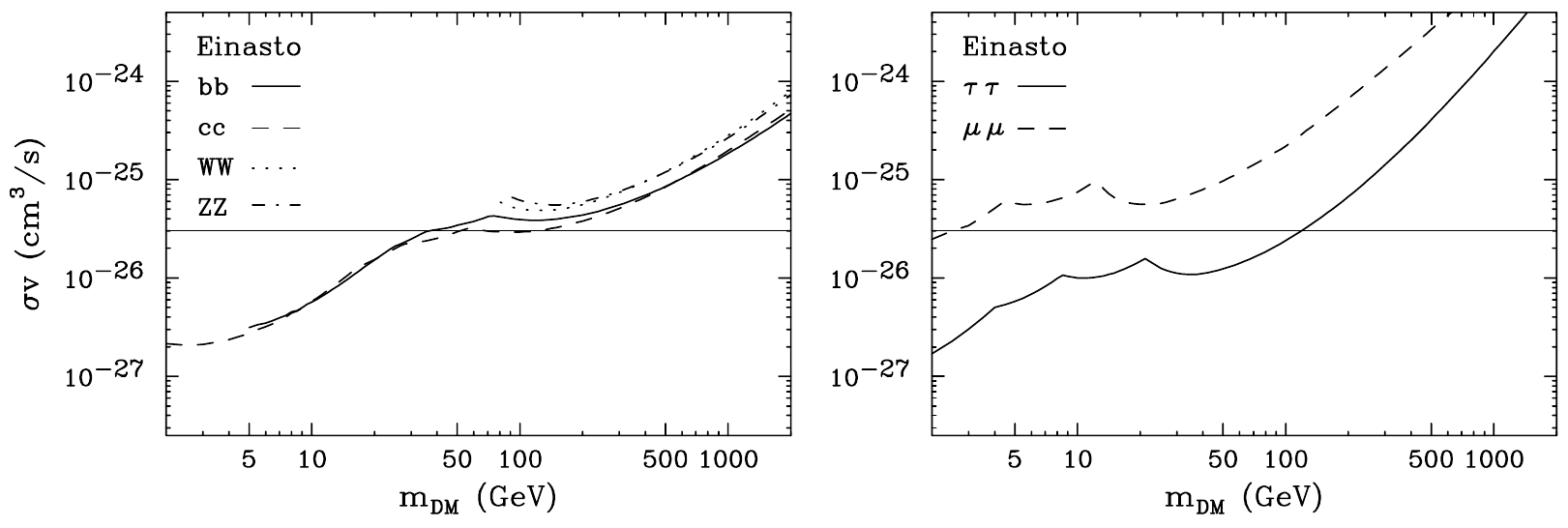}
\caption{The 95\% confidence level upper limits on the dark matter annihilation cross section, for various annihilation channels, assuming a distribution which follows an NFW (upper frames) or Einasto (lower frames) halo profile. To be conservative, we have normalized the halo profile to the minimum value capable of providing a good fit to the combination of the Milky Way's measured rotation curve and microlensing constraints~\cite{local} (corresponding to a local density of $\rho \approx 0.28$ GeV/cm$^3$ or 0.25 GeV/cm$^3$ in the upper and lower frames, respectively). For comparison, the horizontal line denotes the estimate for a simple thermal relic ($\sigma v\approx 3\times 10^{-26}$ cm$^3$/s).}
\label{nfw}
\end{figure*}

\begin{figure*}[t]
\centering
\includegraphics[angle=0.0,width=7.in]{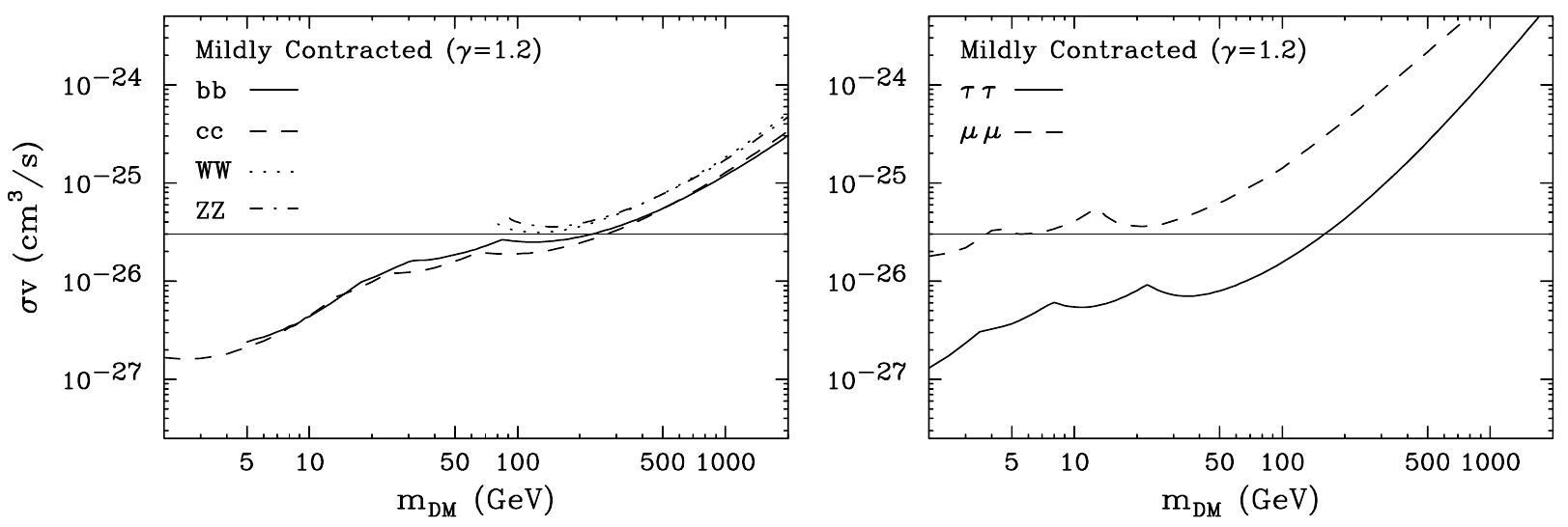}
\includegraphics[angle=0.0,width=7.in]{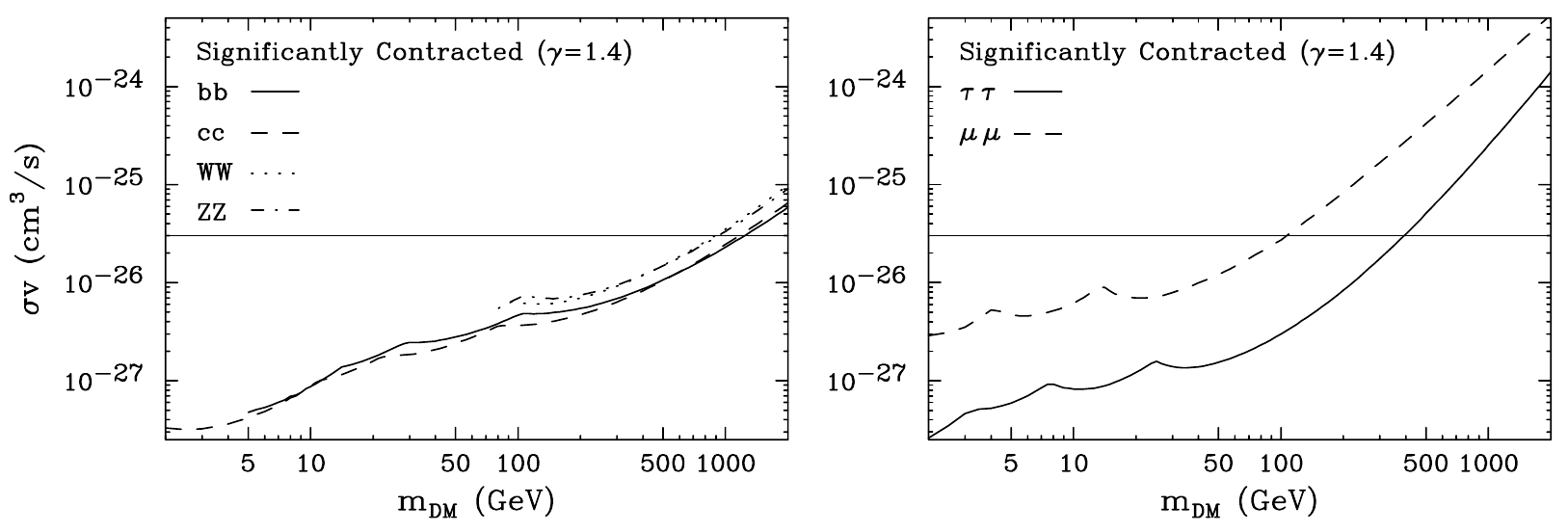}
\caption{The 95\% confidence level upper limits on the dark matter annihilation cross section, for various annihilation channels, assuming a distribution which follows a mildly contracted (upper frames) or a significantly contracted (lower frames) halo profile. To be conservative, we have normalized the halo profile to the minimum value capable of providing a good fit to the combination of the Milky Way's measured rotation curve and microlensing constraints~\cite{local} (corresponding to a local density of $\rho \approx 0.25$ GeV/cm$^3$ or 0.22 GeV/cm$^3$ in the upper and lower frames, respectively). For comparison, the horizontal line denotes the estimate for a simple thermal relic ($\sigma v\approx 3\times 10^{-26}$ cm$^3$/s).}
\label{con}
\end{figure*}

\begin{figure*}[t]
\centering
\includegraphics[angle=0.0,width=7.in]{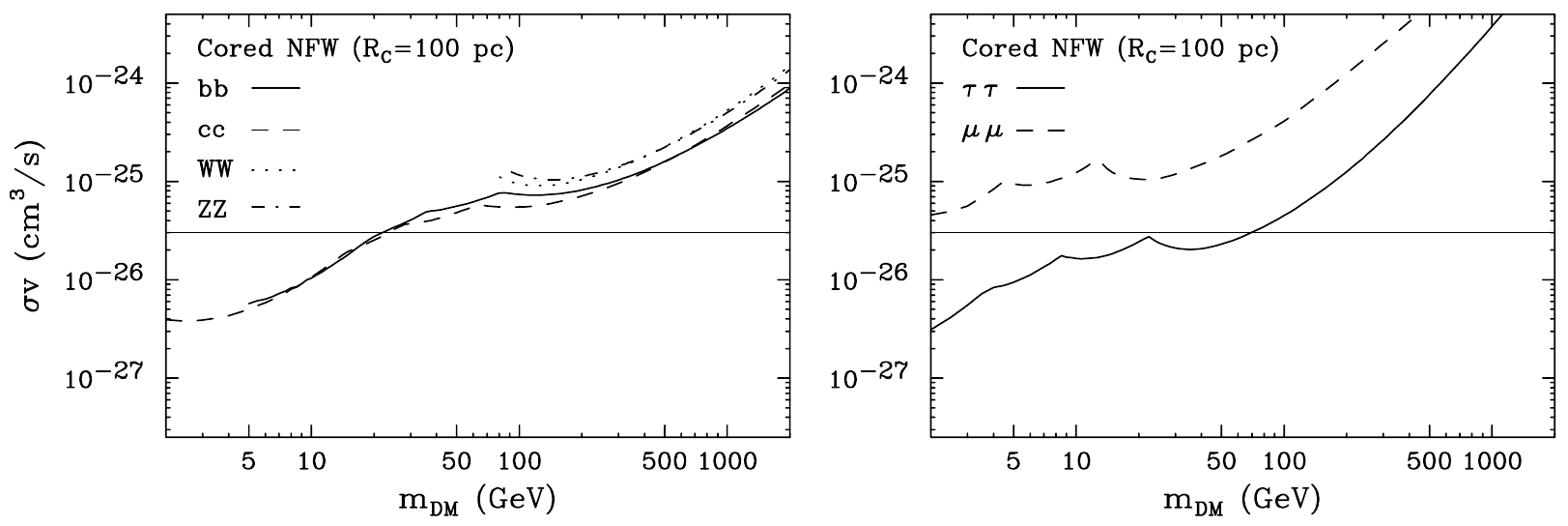}
\includegraphics[angle=0.0,width=7.in]{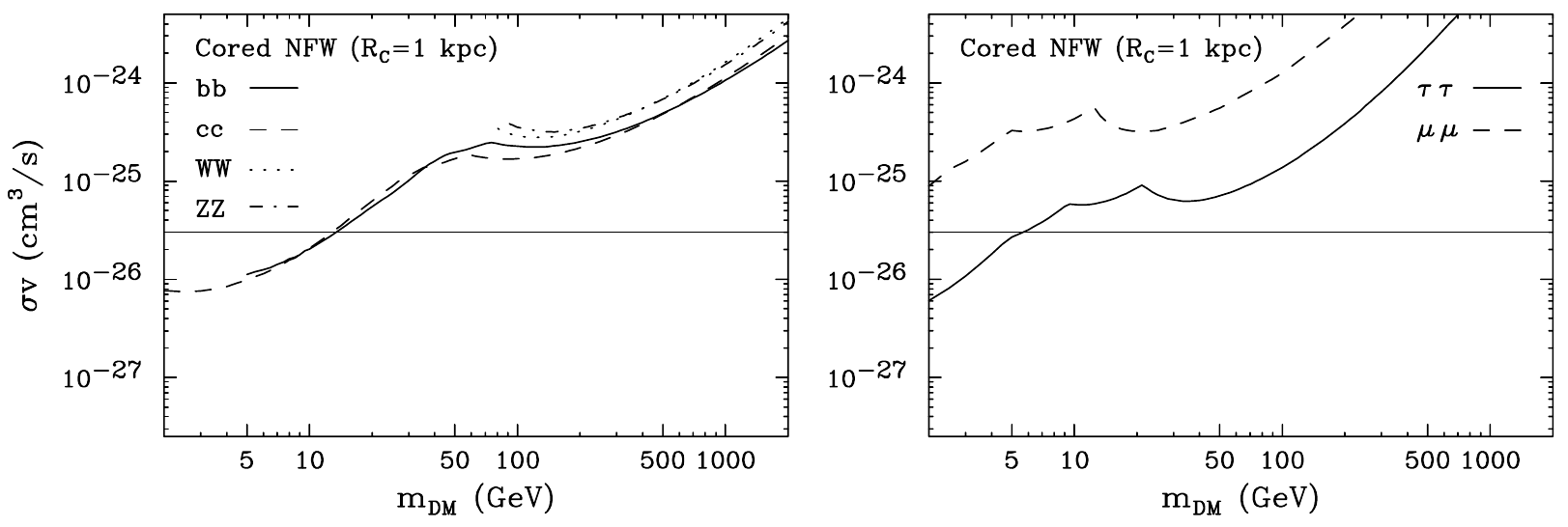}
\caption{The 95\% confidence level upper limits on the dark matter annihilation cross section, for various annihilation channels, assuming a distribution which follows an NFW distribution with a constant-density 100 pc (upper frames) or 1 kpc (lower frames) radius core. To be conservative, we have normalized the halo profile to the minimum value capable of providing a good fit to the combination of the Milky Way's measured rotation curve and microlensing constraints~\cite{local} (corresponding to a local density of $\rho \approx 0.28$ GeV/cm$^3$). Notice that the limits derived for a profile with a 100 pc core are nearly indistinguishable from those derived in the NFW case. For comparison, the horizontal line denotes the estimate for a simple thermal relic ($\sigma v\approx 3\times 10^{-26}$ cm$^3$/s).}
\label{core}
\end{figure*}

In Fig.~\ref{nfw}, we show the resulting constraints under the assumption of either an NFW or Einasto profile. In Figs.~\ref{con} and~\ref{core}, we show the constraints derived for contracted or cored distributions, respectively. In each case, we have conservatively adopted the minimum normalization for a given profile consistent with the Milky Way's rotation curve and microlensing constraints (see Fig.~\ref{bertone}). The constraints on the dark matter annihilation cross section would be about twice as stringent if we had adopted the central value for this normalization, as is often done.

Although the strength of the resulting constraints depend on the dark matter distribution that is assumed, the variation from profile-to-profile is more modest than might be expected. For example, the constraints derived using a profile with a 1 kpc core are only a factor of a few weaker than those derived in the NFW case. The presence of a 100 pc core has almost no effect on the resulting limits. The main reason for this is the limited angular resolution of Fermi (see the right frame of Fig.~\ref{profiles}). A second reason that limits the impact of the choice of halo profile is that the profiles which predict the highest annihilation rates (contracted profiles) also predict much of their signal to appear within the inner degree or so of the Galactic Center, where there is a significant gamma-ray flux observed. In contrast, profiles with a significant core are constrained primarily in the region 2-5$^{\circ}$ away from the Galactic Center, where the gamma-ray flux is lower.

In order to present our results in a way that can be easily applied to dark matter models that are not explicitly presented here (such as models which annihilate to combinations of different final states, or to final states that are not shown in our figures), we have present model independent limits in Table~I.

\begin{table}[t]
\centering
\begin{tabular}{|c|c|c|}
	\hline
Halo Model &  Energy Range    & $\frac{\sigma v}{m^2_{\rm DM}} \int dE_{\gamma} \frac{dN_{\gamma}}{dE_{\gamma}}$ (U.L.)      \\
	\hline
NFW   & 0.3-1 GeV & 1.8$\times 10^{-28}$ cm$^3$\,s$^{-1}$\,GeV$^{-2}$  \\
NFW   & 1-3 GeV   & 6.4$\times 10^{-29}$ cm$^3$\,s$^{-1}$\,GeV$^{-2}$  \\
NFW   & 3-10 GeV &  1.4$\times 10^{-29}$ cm$^3$\,s$^{-1}$\,GeV$^{-2}$ \\
NFW   & 10-100 GeV  & 1.7$\times 10^{-30}$ cm$^3$\,s$^{-1}$\,GeV$^{-2}$  \\
	\hline
Einasto   & 0.3-1 GeV &    1.0$\times 10^{-28}$ cm$^3$\,s$^{-1}$\,GeV$^{-2}$  \\
Einasto   & 1-3 GeV   &    4.1$\times 10^{-29}$ cm$^3$\,s$^{-1}$\,GeV$^{-2}$  \\
Einasto   & 3-10 GeV &     9.2$\times 10^{-30}$ cm$^3$\,s$^{-1}$\,GeV$^{-2}$  \\
Einasto   & 10-100 GeV   & 9.2$\times 10^{-31}$ cm$^3$\,s$^{-1}$\,GeV$^{-2}$  \\
	\hline
Contracted ($\gamma$=$1.2$)    & 0.3-1 GeV &    8.2$\times 10^{-29}$ cm$^3$\,s$^{-1}$\,GeV$^{-2}$  \\
Contracted ($\gamma$=$1.2$)    & 1-3 GeV   &    2.8$\times 10^{-29}$ cm$^3$\,s$^{-1}$\,GeV$^{-2}$  \\
Contracted ($\gamma$=$1.2$)   & 3-10 GeV &     5.0$\times 10^{-30}$ cm$^3$\,s$^{-1}$\,GeV$^{-2}$  \\
Contracted ($\gamma$=$1.2$)    & 10-100 GeV   & 6.0$\times 10^{-31}$ cm$^3$\,s$^{-1}$\,GeV$^{-2}$  \\
	\hline
Contracted ($\gamma$=$1.4$)   & 0.3-1 GeV &   1.6$\times 10^{-29}$ cm$^3$\,s$^{-1}$\,GeV$^{-2}$   \\
Contracted ($\gamma$=$1.4$)   & 1-3 GeV   &   4.3$\times 10^{-30}$ cm$^3$\,s$^{-1}$\,GeV$^{-2}$   \\
Contracted ($\gamma$=$1.4$)   & 3-10 GeV &   7.5$\times 10^{-31}$ cm$^3$\,s$^{-1}$\,GeV$^{-2}$   \\
Contracted ($\gamma$=$1.4$)   & 10-100 GeV &   1.1$\times 10^{-31}$ cm$^3$\,s$^{-1}$\,GeV$^{-2}$   \\
	\hline
Cored ($R_C$=$100$\,pc)   & 0.3-1 GeV & 2.0$\times 10^{-28}$ cm$^3$\,s$^{-1}$\,GeV$^{-2}$  \\
Cored ($R_C$=$100$\,pc)    & 1-3 GeV   & 6.8$\times 10^{-29}$ cm$^3$\,s$^{-1}$\,GeV$^{-2}$  \\
Cored ($R_C$=$100$\,pc)    & 3-10 GeV &   1.5$\times 10^{-29}$ cm$^3$\,s$^{-1}$\,GeV$^{-2}$\\
Cored ($R_C$=$100$\,pc)    & 10-100 GeV   &  1.7$\times 10^{-30}$ cm$^3$\,s$^{-1}$\,GeV$^{-2}$ \\
	\hline
Cored ($R_C$=$1$\,kpc)   & 0.3-1 GeV &  3.7$\times 10^{-28}$ cm$^3$\,s$^{-1}$\,GeV$^{-2}$   \\
Cored ($R_C$=$1$\,kpc)    & 1-3 GeV   &  2.0$\times 10^{-28}$ cm$^3$\,s$^{-1}$\,GeV$^{-2}$   \\
Cored ($R_C$=$1$\,kpc)    & 3-10 GeV &  5.3$\times 10^{-29}$ cm$^3$\,s$^{-1}$\,GeV$^{-2}$   \\
Cored ($R_C$=$1$\,kpc)    & 10-100 GeV   &  5.3$\times 10^{-29}$ cm$^3$\,s$^{-1}$\,GeV$^{-2}$   \\
	\hline
\end{tabular}
\caption{The 95\% confidence level upper limits on the quantity $(\sigma v /m^2_{\rm DM}) \int dE_{\gamma} dN_{\gamma}/dE_{\gamma}$ for various halo profiles and integrated over four different energy ranges. This table is intended to make it possible to derive limits for dark matter models which are not explicitly considered in this paper, such as those in which the dark matter annihilates into combinations of different final states, or to any final states which are not considered here. For such models, one can use PYTHIA~\cite{pythia} to determine the spectrum of gamma-rays per annihilation, $dN_{\gamma}/dE_{\gamma}$, and then apply the constraints as presented in this table. In each case, we have conservatively normalized the halo profile to the minimum value capable of providing a good fit to the combination of the Milky Way's measured rotation curve and microlensing constraints~\cite{local}.}
\label{tab}
\end{table}

The constraints derived in this paper are somewhat more stringent than those presented in Ref.~\cite{Hooper:2011ti}. This is largely due to the details of the dark matter distributions being considered. In Ref.~\cite{Hooper:2011ti}, halo profiles with a single power-law ($\rho \propto r^{-\gamma}$) within the solar circle were adopted, recovering the NFW form in the $R_s \rightarrow \infty$ limit. If $R_s$ is taken to have the conventional value used here ($R_s \approx 20$ kpc), the results of Ref.~\cite{Hooper:2011ti} are in good agreement with those found in this study.

\begin{figure*}[t]
\centering
\includegraphics[angle=0.0,width=1.86in]{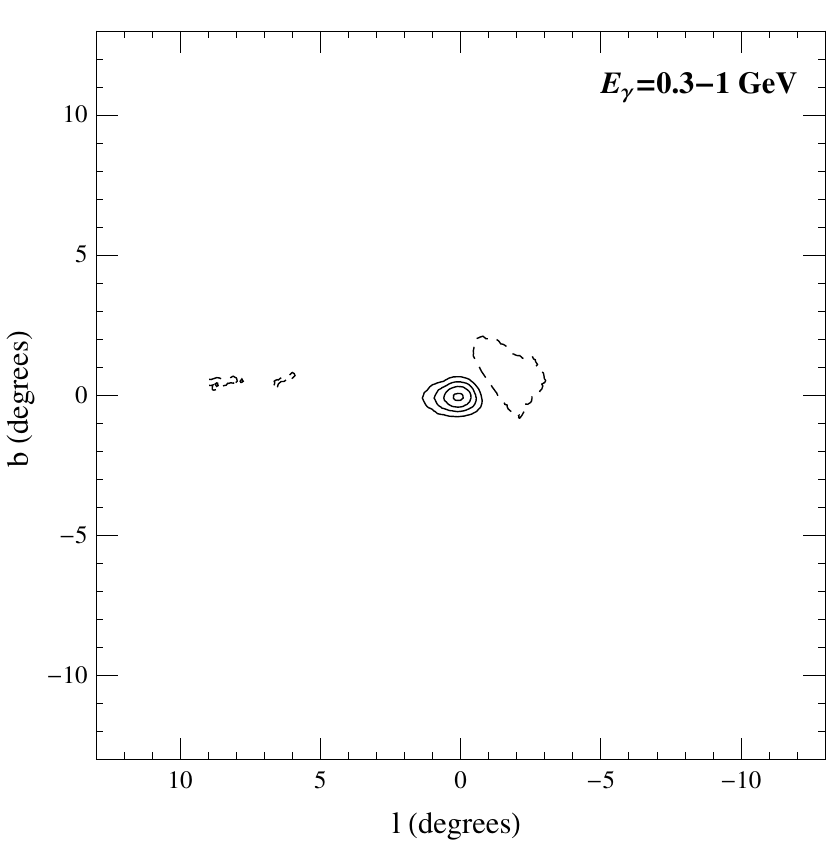}
\includegraphics[angle=0.0,width=1.86in]{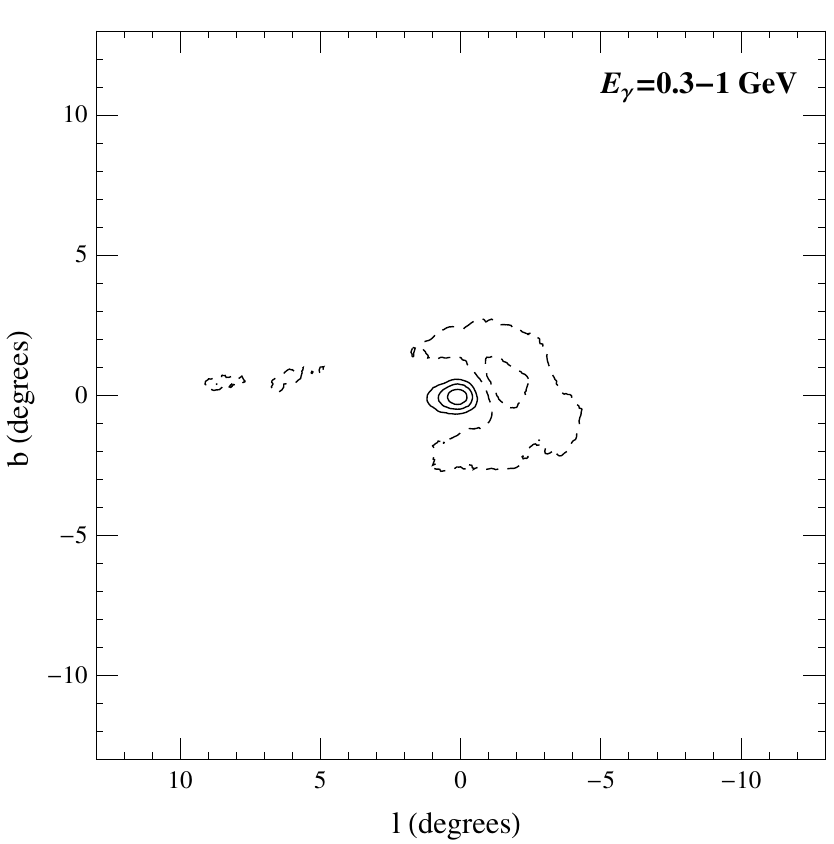}
\includegraphics[angle=0.0,width=1.86in]{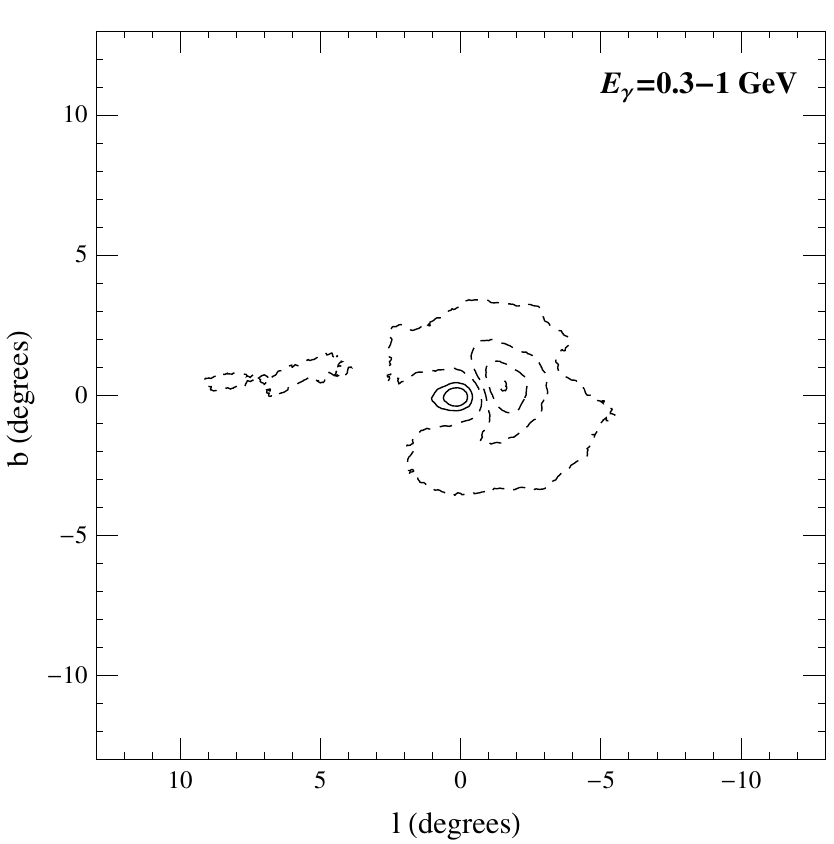}\\
\includegraphics[angle=0.0,width=1.86in]{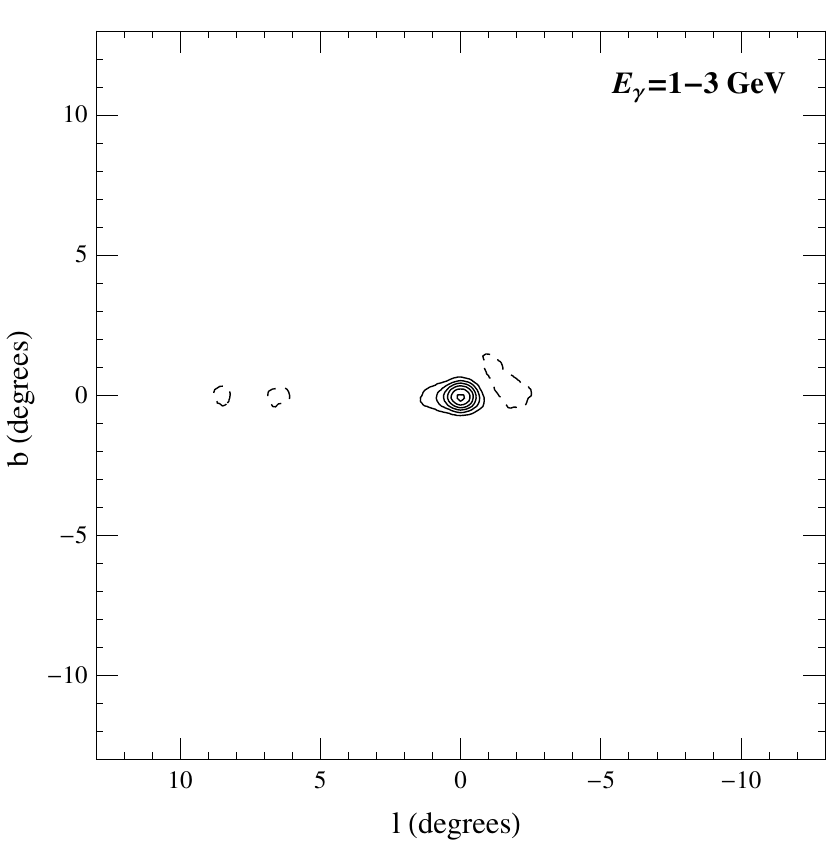}
\includegraphics[angle=0.0,width=1.86in]{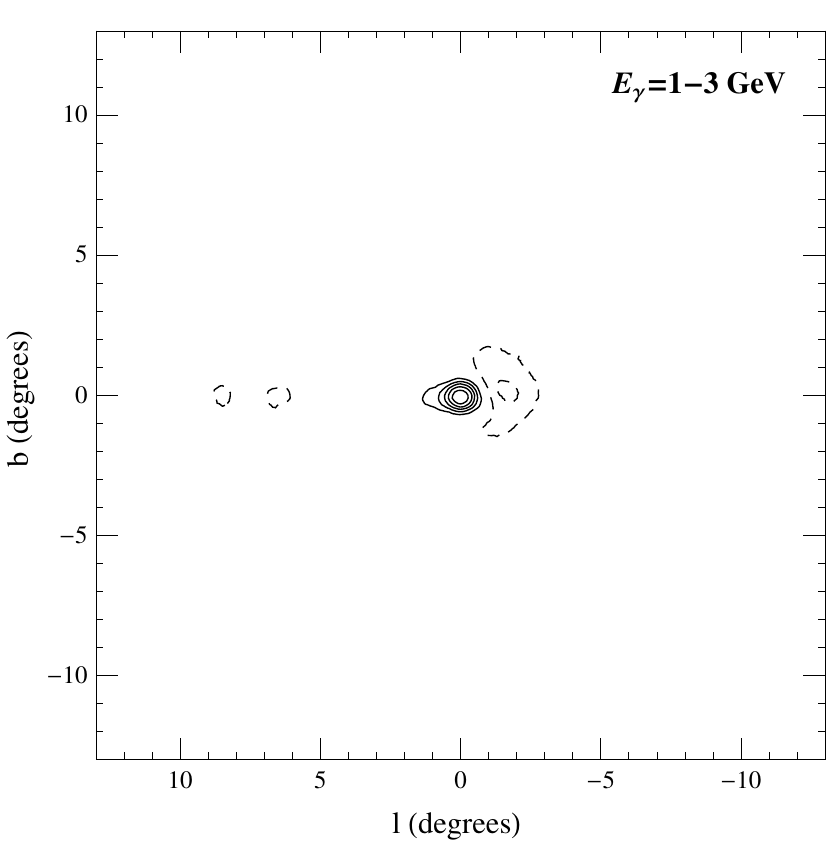}
\includegraphics[angle=0.0,width=1.86in]{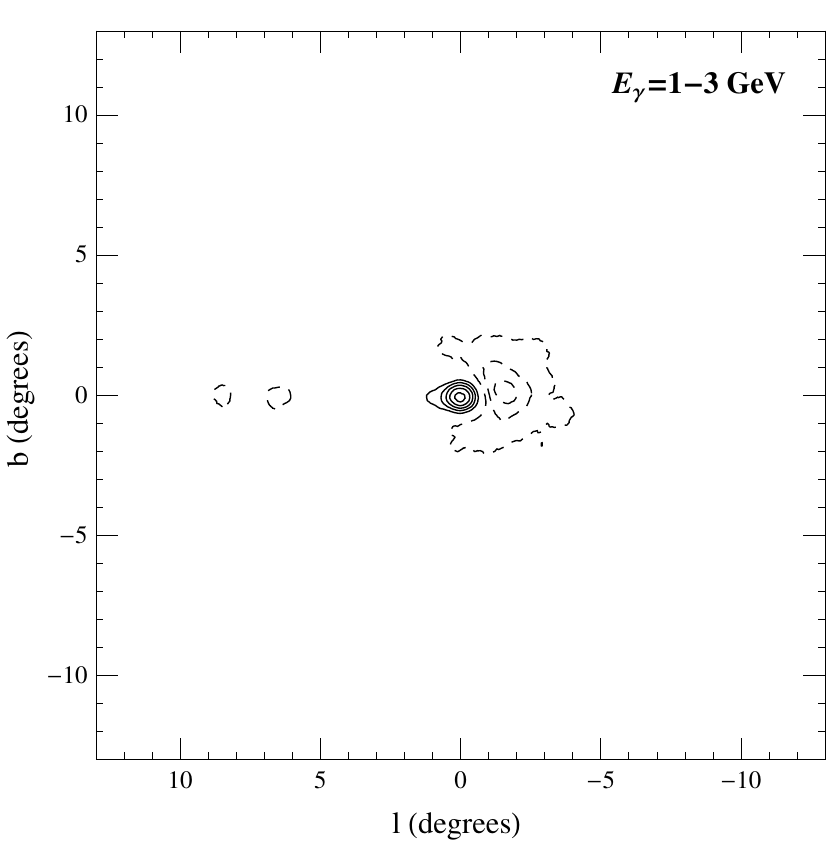}\\
\includegraphics[angle=0.0,width=1.86in]{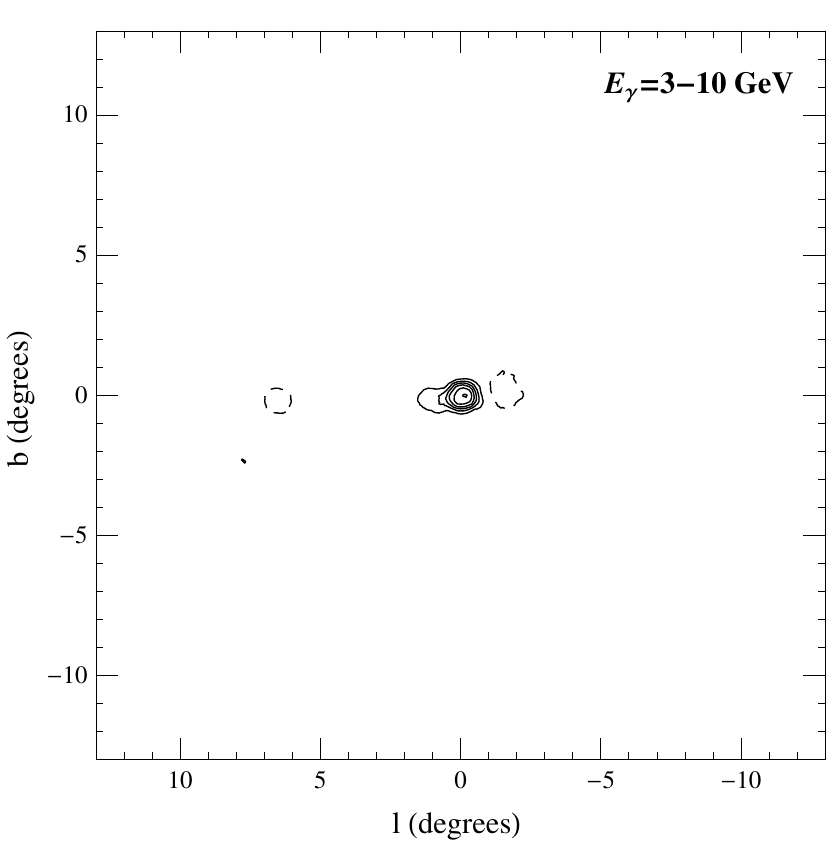}
\includegraphics[angle=0.0,width=1.86in]{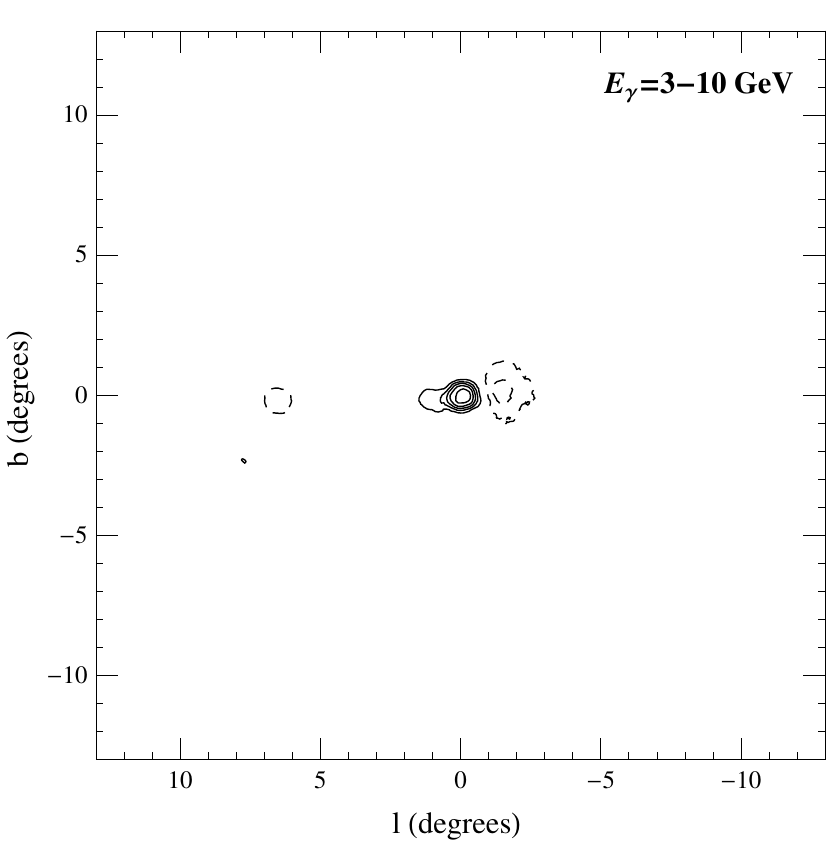}
\includegraphics[angle=0.0,width=1.86in]{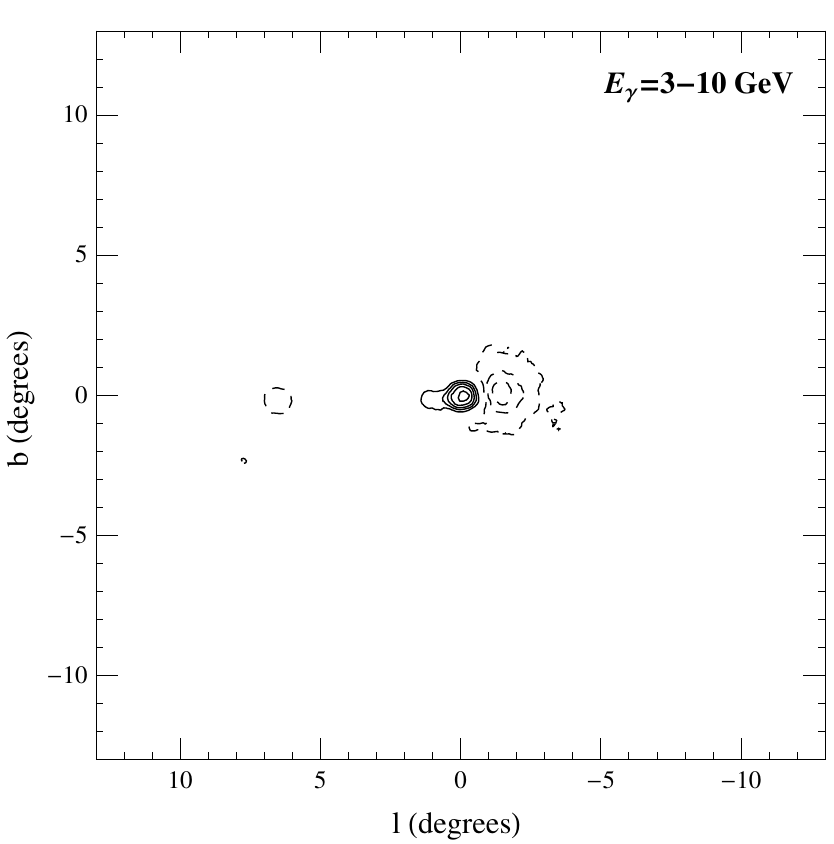}\\
\includegraphics[angle=0.0,width=1.86in]{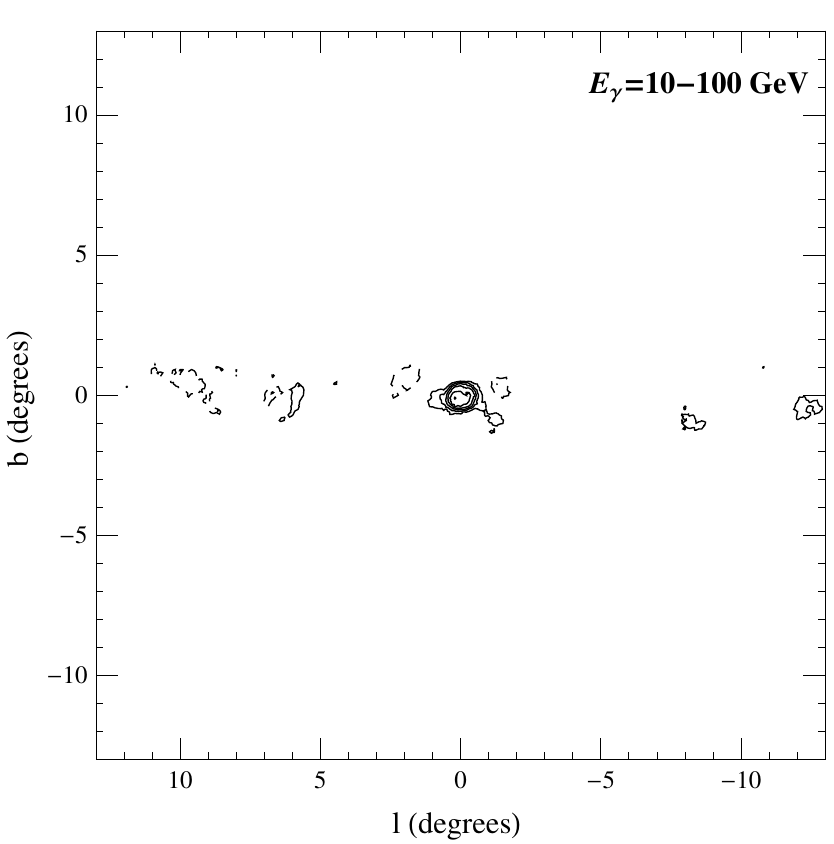}
\includegraphics[angle=0.0,width=1.86in]{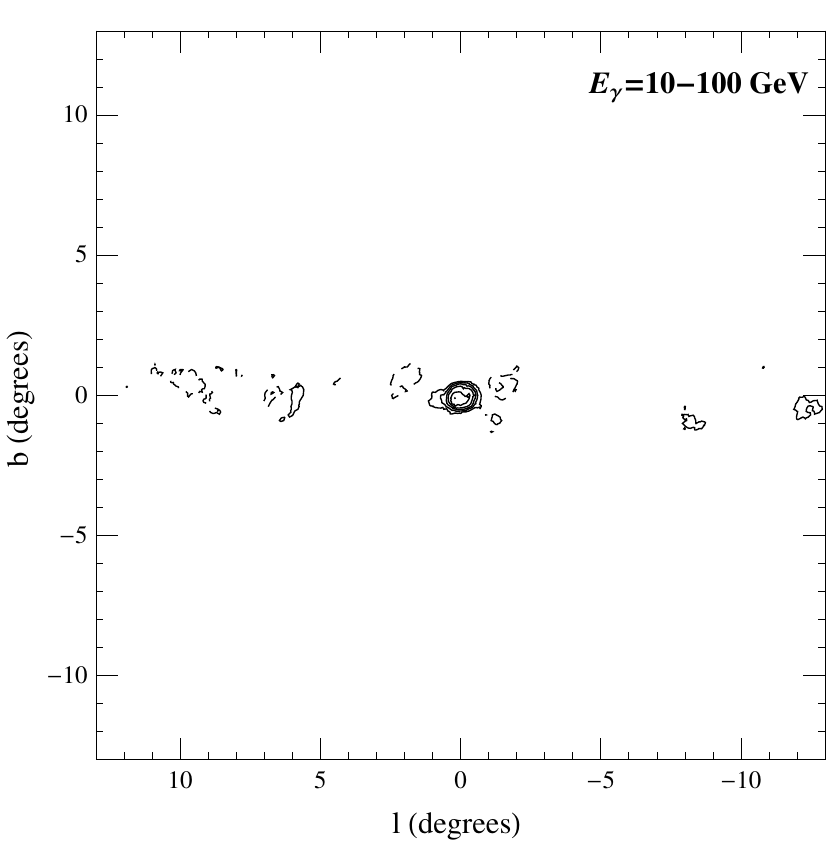}
\includegraphics[angle=0.0,width=1.86in]{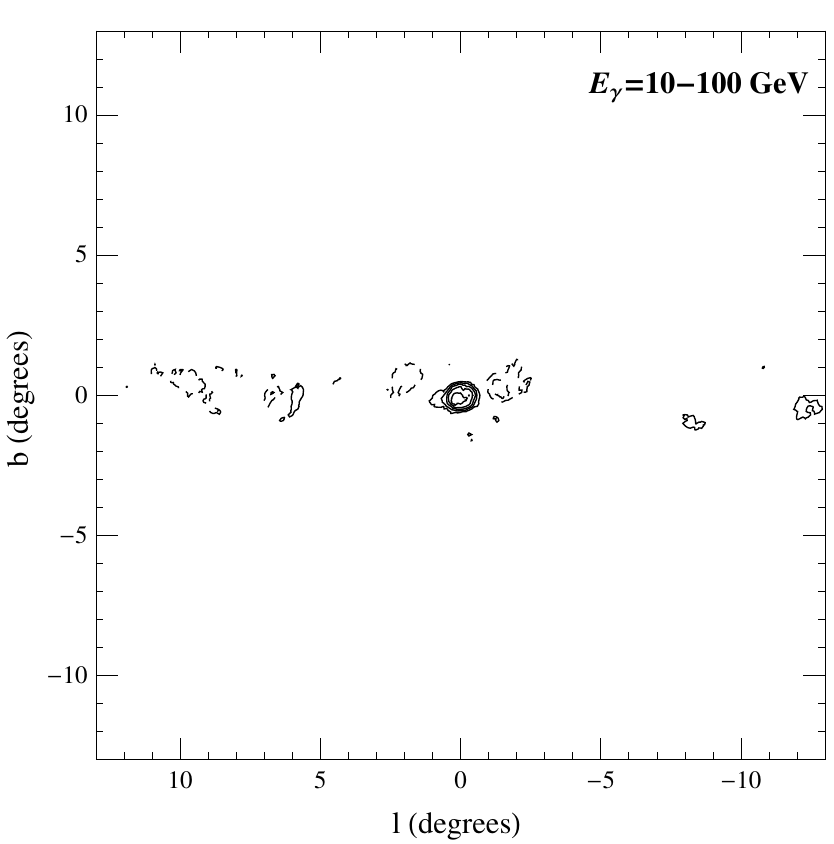}
\caption{Contour maps of the gamma-ray flux from the region surrounding the Galactic Center, after subtracting varying degrees of emission from dark matter distributed according to a Einasto profile centered around the point $(l,b)=(-1.5^{\circ},0)$, as motivated by the morphology of Fermi's 130 GeV line~\cite{Su:2012ft}. As the flux of dark matter annihilation products is increased (moving from left-to-right), regions of the maps become increasingly oversubtracted (denoted by dashed contours).}
\label{mapsshifted}
\end{figure*}

Lastly, we note that some recent interest has been given to the possibility that the Milky Way's dark matter halo profile may peak at a location not coincident with the Galactic Center. This could be the case if our halo profile exhibits a large, nearly constant-density core~\cite{Kuhlen:2012qw}. In such a case, the resulting constraints would be only slightly different than those we have derived for the cored profile cases. On the other hand, it has been shown that the morphology of the 130 GeV line present within the FGST data~\cite{line} is best fit by a roughly Einasto-like distribution, but centered around a point along the Galactic Plane approximately 1.5$^{\circ}$ away from the Galactic Center ($l=-1.5^{\circ}$)~\cite{Su:2012ft} (see also, however, Fig.~3 of Ref.~\cite{oncenter}). In Fig.~\ref{mapsshifted}, we consider this case and find a resulting constraint of $(\sigma v /m^2_{\rm DM}) \int^{100 {\rm GeV}}_{10 {\rm GeV}} dE_{\gamma} dN_{\gamma}/dE_{\gamma}< 3.6\times 10^{-31}$ cm$^3$ s$^{-1}$ GeV$^{-2}$, which is more stringent than in the on-center Einasto case by a factor of approximately 2.5. This makes a dark matter annihilation explanations of the observed 130 GeV gamma-ray line very difficult to accommodate with dark matter candidates which possess any significant annihilation cross section to non-line final states~\cite{Buckley:2012ws,Cohen:2012me,Huang:2012yf}.


\clearpage

\begin{figure*}[t]
\centering
\includegraphics[angle=0.0,width=5.in]{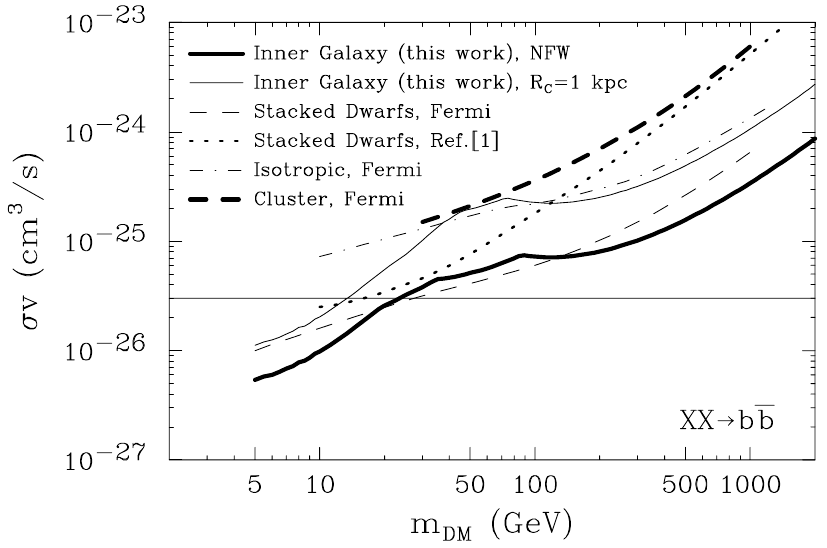}
\caption{A comparison of the upper limits on the dark matter annihilation cross section derived in this work to those from other gamma-ray observations. In particular, we show the constraints derived from the observations of dwarf spheroidal galaxies~\cite{dwarffirst,dwarf}, the isotropic gamma-ray background~\cite{cosmo}, and from the Fornax galaxy cluster~\cite{clusters}. If we adopt an NFW halo profile (or an Einasto or contracted profile), the constraints derived from the Galactic Center are the most stringent. Only if the dark matter halo profile of the Milky Way has a significant core (while dwarf galaxies retain their cusps) are constraints from dwarfs more stringent. The constraints from the Galactic Center are, for all dark matter masses, more stringent than those reliably extracted from the isotropic gamma-ray background or from galaxy clusters.}
\label{comparison}
\end{figure*}

\section{Comparisons With Other Dark Matter Annihilation Constraints}

In this section, we compare the constraints derived in this study to those from other gamma-ray observations. The most stringent of these constraints result from observations of:
\begin{itemize}
\item{Dwarf Spheroidal Galaxies: Geringer-Sameth \& Koushiappas~\cite{dwarffirst}, and the Fermi Collaboration~\cite{dwarf} have each published constraints on the dark matter annihilation cross section from the stacked observations of dwarf spheroidal galaxies. Both of these studies assume that the dark matter distribution in each dwarf is described by an NFW profile (in contrast, there is evidence from observations~\cite{BoylanKolchin:2011dk,Wolf:2012wp,Hayashi:2012si} and simulations~\cite{Brooks:2012vi,dwarfcores}  that favor cored profiles among dwarf galaxies). Relaxing this assumption may weaken the resulting constraints, although only moderately~\cite{salucci,dwarf}. With this exception, the constraints derived from dwarfs are robust and not subject to significant astrophysical uncertainties. In Fig.~\ref{comparison}, we show these constraints, as derived from each of these groups~\cite{dwarffirst,dwarf} (to present a fair comparison with our results, we have shown the constraint from Ref.~\cite{dwarffirst} as derived using the minimum $J$ factor, as we have normalized to the minimum acceptable halo density).}
\item{The Isotropic Gamma-Ray Background: The contribution to the isotropic gamma-ray background from cosmological dark matter annihilations could be significant. Depending on what one assumes about the quantity of substructure present in dark matter halos, constraints from the isotropic background can be stringent. The uncertainties associated with the issue of substructure are large, however. For example, in Fermi's analysis, results are presented for a range substructure models that produce annihilation constraints which vary by approximately three orders of magnitude~\cite{cosmo}. Only for the most optimistic substructure models are the constraints from the isotropic gamma-ray background competitive with those presented in this study. In Fig.~\ref{comparison}, we include the constraint from the isotropic background as presented in Ref.~\cite{cosmo} as the ``stringent constraint'' for the case of substructure model MSII-Sub 1.}
\item{Galaxy Clusters: Much like those derived from the isotropic gamma-ray background, dark matter annihilation constraints from galaxy clusters depend very strongly on the quantities of substructures assumed to be present in such objects. The Fermi Collaboration's cluster analysis~\cite{clusters} was able to place only fairly weak constraints on the dark matter annihilation cross section (their most optimistic substructure model yielded $\sigma v \lsim 3 \times 10^{-25}$ cm$^3$/s for $m_{\rm DM}=100$ GeV and annihilating to $b\bar{b}$), see Fig.~\ref{comparison}. In contrast, the more recent analysis of Ref.~\cite{clusters3} used a much more optimistic substructure model to produce constraints $\sim$$10^2$ times more stringent. Given these enormous uncertainties, it can be somewhat difficult to interpret the constraints derived from clusters.}
\end{itemize}

In Fig.~\ref{comparison}, we compare the constraints derived in this study to those derived from dwarf spheroidals, from the isotropic gamma-ray background, and from galaxy clusters. If we adopt an NFW halo profile (or an Einasto or contracted profile), the constraints derived from the Galactic Center are the most stringent. Only if the dark matter halo profile of the Milky Way has a significant core (while dwarf galaxies retain their cusps) are constraints from dwarfs more stringent. The constraints from the Galactic Center are, for all dark matter masses, more stringent than those reliably extracted from the isotropic gamma-ray background or from galaxy clusters.

\section{Summary and Conclusions}

In this article, we have used data from the Fermi Gamma-Ray Space Telescope to place constraints on the dark matter annihilation cross section. In doing so, we have considered a wide range of dark matter distributions, each consistent with rotation curves and microlensing data. And while the strength of our constraints vary somewhat depending on which halo profile is adopted, we find that even in the most conservative cases (such as profiles with a significant, flat-density core) our constraints are comparably stringent to those derived from observations of dwarf spheroidal galaxies (which were derived assuming NFW distributions). In most other cases, our constraints are stronger than those derived from any other region of the sky.

In an effort to be conservative in our analysis, we have chosen not to subtract a number of astrophysical components from the gamma-ray map of the inner Galaxy. In particular, we have not subtracted any component of the emission associated with the Galactic Ridge (as observed by HESS) or from the central gamma-ray point source (as observed by HESS, VERITAS, and MAGIC). Subtracting these components would have improved on the limits presented here by a factor of approximately two. Furthermore, we have in each case adopted normalizations for the dark matter halo profile which are the ($2\sigma$) minimum value consistent with dynamical and microlensing constraints. If we had instead used the central values for the overall dark matter density, our constraints would be more stringent by another factor of two. We have also not applied any boost factors or other enhancements due to dark matter substructure in deriving our results. The existence of any such substructures would only strengthen our constraints further. 
 
Throughout this study, we have remained agnostic about the origin of the gamma-ray flux observed from the inner Galaxy. We only briefly mention here that the results presented here are in no way in conflict with those presented previously which find that annihilating dark matter can provide a good fit to the observed emission~\cite{Hooper:2011ti,HG2,Abazajian:2012pn,HG1}. In particular, Fermi's Galactic Center observations, coupled with observations of the Milky Way's radio filaments, are most easily explained by a dark matter particle with a mass of $m_{\rm DM} \approx 7-10$ GeV, an annihilation cross section of $\sigma v \sim 10^{-26}$ cm$^3$/s to charged leptons, and distributed in a somewhat contracted profile ($\rho \propto r^{-1.3}$).

Looking toward the future, we find very promising the possibility of the post-Fermi gamma-ray satellite, GAMMA-400~\cite{Bergstrom:2012vd}. As GAMMA-400's overall effective area and acceptance will be comparable to that of Fermi, it will likely not be more sensitive to dark matter annihilations from flux-limited sources, such as dwarf galaxies. With considerable improvements in both angular and energy resolution relative to Fermi, however, GAMMA-400 should be able to much better separate astrophysical backgrounds in the inner Galaxy from any dark matter annihilation signal that is present. Furthermore, multi-wavelength studies of the Galactic Center, and progress from hydrodynamical simulations of dark matter in Milky Way-like galaxies, could further strengthen the dark matter constraints that can be derived from the inner Galaxy.

\bigskip
\bigskip

{\it Acknowledgements}: We would like to thank Tim Linden, Mariangela Lisanti, and Keith Bechtol for insightful comments as well as Gianfranco Bertone, Miguel Pato, Fabio Iocco and Philippe Jetzer for providing the contours used in our Fig.~\ref{bertone}. DH is supported by the US Department of Energy. CK is supported by a Fermilab Fellowship in Theoretical Physics. FSQ is supported by Coordenacao de Aperfeisoamento de Pessoal de Nivel Superior (CAPES). This work was supported in part by the National Science Foundation under Grant No.~PHY-1066293. We thank the Aspen Center for Physics for their hospitality.


\begin{thebibliography}{}

\bibitem{dwarffirst}
  A.~Geringer-Sameth, S.~M.~Koushiappas
  Phys.\ Rev.\ Lett.\  {\bf 107}, 241303 (2011)
  [arXiv:1108.2914 [astro-ph.CO]].

\bibitem{dwarf}
  The Fermi-LAT Collaboration,
Phys.\ Rev.\ Lett.\  {\bf 107}, 241302 (2011)
  [arXiv:1108.3546 [astro-ph.HE]];
  C.~Farnier {\it et al.}  [Fermi-LAT Collaboration],
  Nucl.\ Instrum.\ Meth.\ A {\bf 630}, 143 (2011).

\bibitem{clusters}
  M.~Ackermann, M.~Ajello, A.~Allafort {\it et al.},
  JCAP {\bf 1005}, 025 (2010)
  [arXiv:1002.2239 [astro-ph.CO]].

\bibitem{clusters2}
  L.~Dugger, T.~E.~Jeltema, S.~Profumo,
  JCAP {\bf 1012}, 015 (2010)
  [arXiv:1009.5988 [astro-ph.HE]].

\bibitem{clusters3}
  J.~Han, C.~S.~Frenk, V.~R.~Eke, L.~Gao, S.~D.~M.~White, A.~Boyarsky, D.~Malyshev and O.~Ruchayskiy,
  arXiv:1207.6749 [astro-ph.CO].


\bibitem{Anderson:2010hh} 
  B.~Anderson [Fermi-LAT Collaboration],
  PoS IDM {\bf 2010}, 113 (2011)
  [arXiv:1012.0863 [hep-ph]];
  T.~F.~-: M.~Ackermann {\it et al.}  [LAT Collaboration],
  arXiv:1205.6474 [astro-ph.CO];
  G.~Zaharijas {\it et al.}  [Fermi-LAT Collaboration],
  PoS IDM {\bf 2010}, 111 (2011)
  [arXiv:1012.0588 [astro-ph.HE]].


\bibitem{Ackermann:2012nb} 
  M.~Ackermann {\it et al.}  [Fermi LAT Collaboration],
  Astrophys.\ J.\  {\bf 747}, 121 (2012)
  [arXiv:1201.2691 [astro-ph.HE]];
  A.~Drlica-Wagner {\it et al.}  [Fermi-LAT Collaboration],
  arXiv:1111.3358 [astro-ph.HE];
  A.~V.~Belikov, D.~Hooper and M.~R.~Buckley,
  arXiv:1111.2613 [hep-ph];
  N.~Mirabal, V.~Frias-Martinez, T.~Hassan and E.~Frias-Martinez,
  Mon.\ Not.\ Roy.\ Astron.\ Soc.\  {\bf 424}, L64 (2012)
  [arXiv:1205.4825 [astro-ph.HE]];
  M.~R.~Buckley, D.~Hooper,
  Phys.\ Rev.\  {\bf D82}, 063501 (2010)
  [arXiv:1004.1644 [hep-ph]].



\bibitem{cosmo}
  A.~A.~Abdo {\it et al.} [Fermi-LAT Collaboration],
  JCAP {\bf 1004}, 014 (2010)
  [arXiv:1002.4415 [astro-ph.CO]].



\bibitem{cosmo2}
  K.~N.~Abazajian, P.~Agrawal, Z.~Chacko, C.~Kilic,
  JCAP {\bf 1011}, 041 (2010)
  [arXiv:1002.3820 [astro-ph.HE]]. 


\bibitem{Steigman:2012nb} 
  G.~Steigman, B.~Dasgupta and J.~F.~Beacom,
  Phys.\ Rev.\ D {\bf 86}, 023506 (2012)
  [arXiv:1204.3622 [hep-ph]].




\bibitem{Hooper:2011ti} 
  D.~Hooper and T.~Linden,
  Phys.\ Rev.\ D {\bf 84}, 123005 (2011)
  [arXiv:1110.0006 [astro-ph.HE]].

\bibitem{HG2}
  D.~Hooper, L.~Goodenough,
  Phys.\ Lett.\  {\bf B697}, 412-428 (2011)
  [arXiv:1010.2752 [hep-ph]].

\bibitem{Abazajian:2012pn} 
  K.~N.~Abazajian and M.~Kaplinghat,
  arXiv:1207.6047 [astro-ph.HE].

\bibitem{aharonian}
M.~Chernyakova, D.~Malyshev, F.~A.~Aharonian, R.~M.~Crocker, and D.~I.~Jones,
Ap.~J. 726, 60C (2011).



\bibitem{Boyarsky:2010dr}
  A.~Boyarsky, D.~Malyshev and O.~Ruchayskiy,
  Phys.\ Lett.\ B {\bf 705}, 165 (2011)
  [arXiv:1012.5839 [hep-ph]].


\bibitem{HG1}
  L.~Goodenough, D.~Hooper,
  arXiv:0910.2998 [hep-ph].

\bibitem{Vitale:2011zz} 
  V.~Vitale {\it et al.}  [Fermi-LAT Collaboration],
  Nucl.\ Instrum.\ Meth.\ A {\bf 630}, 147 (2011).

\bibitem{Morselli:2010ty} 
  A.~Morselli {\it et al.}  [Fermi-LAT Collaboration],
  arXiv:1012.2292 [astro-ph.HE].

\bibitem{Vitale:2008zz} 
  V.~Vitale {\it et al.}  [Fermi LAT Collaboration],
  PoS IDM {\bf 2008}, 115 (2008).


\bibitem{Linden:2012iv} 
  T.~Linden, E.~Lovegrove and S.~Profumo,
  Astrophys.\ J.\  {\bf 753}, 41 (2012)
  [arXiv:1203.3539 [astro-ph.HE]].


\bibitem{Linden:2012bp} 
  T.~Linden and S.~Profumo,
  arXiv:1206.4308 [astro-ph.HE].

\bibitem{pulsars}
  K.~N.~Abazajian,
  JCAP {\bf 1103}, 010 (2011).
  [arXiv:1011.4275 [astro-ph.HE]].

\bibitem{Wharton:2011dv} 
  R.~S.~Wharton, S.~Chatterjee, J.~M.~Cordes, J.~S.~Deneva and T.~J.~W.~Lazio,
  Astrophys.\ J.\  {\bf 753}, 108 (2012)
  [arXiv:1111.4216 [astro-ph.HE]].



\bibitem{catalog}
  The Fermi-LAT Collaboration,
  arXiv:1108.1435 [astro-ph.HE].

\bibitem{gas}
P.~M.~W.~Kalberla and J.~Kerp, 
Ann.~Rev.~AA. {\bf 47}, 27-61 (2009).

\bibitem{gas2}
  H.~Nakanishi and Y.~Sofue,
  Publ.\ Astron.\ Soc.\ Jap.\  {\bf 55}, 191 (2003)
  [arXiv:astro-ph/0304338].



\bibitem{hesspoint}
F.~Aharonian {\it et al.}  [The HESS Collaboration],
  Astron.\ Astrophys.\  {\bf 425}, L13-L17 (2004).
  [astro-ph/0408145]; 
  C.~van Eldik {\it et al.} [HESS Collaboration],
  J.\ Phys.\ Conf.\ Ser.\  {\bf 110}, 062003 (2008)
  [arXiv:0709.3729 [astro-ph]].


\bibitem{others}
 K.~Kosack {\it et al.}  [The VERITAS Collaboration],
  Astrophys.\ J.\  {\bf 608}, L97 (2004)
  [arXiv:astro-ph/0403422];
 J.~Albert {\it et al.}  [MAGIC Collaboration],
  Astrophys.\ J.\  {\bf 638}, L101 (2006)
  [arXiv:astro-ph/0512469].



\bibitem{ridge}
  F.~Aharonian {\it et al.}  [H.E.S.S. Collaboration],
  Nature {\bf 439}, 695 (2006)
  [arXiv:astro-ph/0603021].

\bibitem{pythia}
T.~Sjostrand,~{\it et al.}, 
Comput.~Phys.~Commun., 135, 238 (2001).



\bibitem{nfw}
 J.~F.~Navarro, C.~S.~Frenk and S.~D.~M.~White,
  Astrophys.\ J.\  {\bf 462}, 563 (1996)
  [arXiv:astro-ph/9508025];
  J.~F.~Navarro, C.~S.~Frenk and S.~D.~M.~White,
  Astrophys.\ J.\  {\bf 490}, 493 (1997).




\bibitem{local}
  F.~Iocco, M.~Pato, G.~Bertone, P.~Jetzer,
  [arXiv:1107.5810 [astro-ph.GA]].

\bibitem{local2}
  R.~Catena, P.~Ullio,
  JCAP {\bf 1008}, 004 (2010).
  [arXiv:0907.0018 [astro-ph.CO]].











\bibitem{Navarro:2008kc}
  J.~F.~Navarro, A.~Ludlow, V.~Springel, J.~Wang, M.~Vogelsberger, S.~D.~M.~White, A.~Jenkins, C.~S.~Frenk {\it et al.},
  [arXiv:0810.1522 [astro-ph]].


\bibitem{vialactea}
  J.~Diemand, M.~Kuhlen, P.~Madau, M.~Zemp, B.~Moore, D.~Potter, J.~Stadel,
  Nature {\bf 454}, 735-738 (2008)
  [arXiv:0805.1244 [astro-ph]];
 J.~Diemand, M.~Zemp, B.~Moore, J.~Stadel, M.~Carollo,
  Mon.\ Not.\ Roy.\ Astron.\ Soc.\  {\bf 364}, 665 (2005)
  [astro-ph/0504215].

\bibitem{aquarius}
  J.~F.~Navarro {\it et al.},
  Mon.\ Not.\ Roy.\ Astron.\ Soc.\  {\bf 349}, 1039 (2004)
  [arXiv:astro-ph/0311231].
  V.~Springel {\it et al.};
  Mon.\ Not.\ Roy.\ Astron.\ Soc.\  {\bf 391}, 1685 (2008)
  [arXiv:0809.0898 [astro-ph]].

\bibitem{ac}
  G.~R.~Blumenthal, S.~M.~Faber, R.~Flores, J.~R.~Primack,
  Astrophys.\ J.\  {\bf 301}, 27 (1986);
B.~S.~Ryden and J.~E.~Gunn,
Astrophys.\ J.\  {\bf 318}, 15 (1987).
  

\bibitem{Gnedin:2011uj}
  O.~Y.~Gnedin, D.~Ceverino, N.~Y.~Gnedin, A.~A.~Klypin, A.~V.~Kravtsov, R.~Levine, D.~Nagai, G.~Yepes
  [arXiv:1108.5736 [astro-ph.CO]].




\bibitem{mac}
  O.~Y.~Gnedin, A.~V.~Kravtsov, A.~A.~Klypin and D.~Nagai,
  Astrophys.\ J.\  {\bf 616}, 16 (2004)
  [arXiv:astro-ph/0406247].

\bibitem{100pc}
  R.~Levine, N.~Y.~Gnedin, A.~J.~S.~Hamilton and A.~V.~Kravtsov,
  Astrophys.\ J.\  {\bf 678}, 154 (2008)
  [arXiv:0711.3478 [astro-ph]].

\bibitem{governato} 
  F.~Governato, A.~Zolotov, A.~Pontzen, C.~Christensen, S.~H.~Oh, A.~M.~Brooks, T.~Quinn and S.~Shen {\it et al.},
  Mon.\ Not.\ Roy.\ Astron.\ Soc.\  {\bf 422}, 1231 (2012)
  [arXiv:1202.0554 [astro-ph.CO]].

\bibitem{Kuhlen:2012qw} 
  M.~Kuhlen, J.~Guedes, A.~Pillepich, P.~Madau and L.~Mayer,
  arXiv:1208.4844 [astro-ph.GA].

\bibitem{Weinberg:2001gm} 
  M.~D.~Weinberg and N.~Katz,
  Astrophys.\ J.\  {\bf 580}, 627 (2002)
  [astro-ph/0110632];
  M.~D.~Weinberg and N.~Katz,
  Mon.\ Not.\ Roy.\ Astron.\ Soc.\  {\bf 375}, 460 (2007)
  [astro-ph/0601138].

\bibitem{Sellwood:2002vb} 
  J.~A.~Sellwood,
  Astrophys.\ J.\  {\bf 587}, 638 (2003)
  [astro-ph/0210079].

\bibitem{Valenzuela:2002np} 
  O.~Valenzuela and A.~Klypin,
  Mon.\ Not.\ Roy.\ Astron.\ Soc.\  {\bf 345}, 406 (2003)
  [astro-ph/0204028].

\bibitem{Colin:2005rr} 
  P.~Colin, O.~Valenzuela and A.~Klypin,
  Astrophys.\ J.\  {\bf 644}, 687 (2006)
  [astro-ph/0506627].


\bibitem{line}
  T.~Bringmann, X.~Huang, A.~Ibarra, S.~Vogl and C.~Weniger,
  JCAP {\bf 1207}, 054 (2012)
  [arXiv:1203.1312 [hep-ph]];
  C.~Weniger,
  JCAP {\bf 1208}, 007 (2012)
  [arXiv:1204.2797 [hep-ph]].

\bibitem{Su:2012ft} 
  M.~Su and D.~P.~Finkbeiner,
  arXiv:1206.1616 [astro-ph.HE];
  E.~Tempel, A.~Hektor and M.~Raidal,
  JCAP {\bf 1209}, 032 (2012)
  [arXiv:1205.1045 [hep-ph]].

\bibitem{oncenter}
  T.~Bringmann and C.~Weniger,
  arXiv:1208.5481 [hep-ph].


\bibitem{Buckley:2012ws} 
  M.~R.~Buckley and D.~Hooper,
  Phys.\ Rev.\ D {\bf 86}, 043524 (2012)
  [arXiv:1205.6811 [hep-ph]].

\bibitem{Cohen:2012me} 
  T.~Cohen, M.~Lisanti, T.~R.~Slatyer and J.~G.~Wacker,
  arXiv:1207.0800 [hep-ph].

\bibitem{Huang:2012yf} 
  X.~-Y.~Huang, Q.~Yuan, P.~-F.~Yin, X.~-J.~Bi and X.~-L.~Chen,
  arXiv:1208.0267 [astro-ph.HE].



\bibitem{BoylanKolchin:2011dk} 
  M.~Boylan-Kolchin, J.~S.~Bullock and M.~Kaplinghat,
  Mon.\ Not.\ Roy.\ Astron.\ Soc.\  {\bf 422}, 1203 (2012)
  [arXiv:1111.2048 [astro-ph.CO]];
  M.~Boylan-Kolchin, J.~S.~Bullock and M.~Kaplinghat,
  Mon.\ Not.\ Roy.\ Astron.\ Soc.\  {\bf 415}, L40 (2011)
  [arXiv:1103.0007 [astro-ph.CO]].

\bibitem{Wolf:2012wp} 
  J.~Wolf and J.~S.~Bullock,
  arXiv:1203.4240 [astro-ph.CO].

\bibitem{Hayashi:2012si} 
  K.~Hayashi and M.~Chiba,
  Astrophys.\ J.\  {\bf 755}, 145 (2012)
  [arXiv:1206.3888 [astro-ph.CO]].

\bibitem{Brooks:2012vi} 
  A.~M.~Brooks and A.~Zolotov,
  arXiv:1207.2468 [astro-ph.CO].


\bibitem{dwarfcores}
F.~Governato {\it et al.},
Nature, 463, 203 (2010);
  S.~-H.~Oh, C.~Brook, F.~Governato, E.~Brinks, L.~Mayer, W.~J.~G.~de Blok, A.~Brooks, F.~Walter,
arXiv:1011.2777 [astro-ph.CO].


\bibitem{salucci} 
  I.~Cholis and P.~Salucci,
  Phys.\ Rev.\ D {\bf 86}, 023528 (2012)
  [arXiv:1203.2954 [astro-ph.HE]].


\bibitem{Bergstrom:2012vd} 
  L.~Bergstrom, G.~Bertone, J.~Conrad, C.~Farnier and C.~Weniger,
  arXiv:1207.6773 [hep-ph].


\end{thebibliography}
\end{document}